%% file: m-joyce.tex
\documentstyle[11pt]{article}

\input amssym 
\input psfig-97   
\psfiginit

 1
 1
\font\footnotesizebbfont=msbm9 scaled \magstep 0
\font\smallbbfont=msbm7 scaled \magstep 2
\font\bbfont=msbm9 scaled \magstep1  
 1
\font\Largebbfont=msbm10 scaled \magstep 2
 3
 4

\def\footnotesizeBbb#1{\hbox{\footnotesizebbfont #1}}
\def\smallBbb#1{\hbox{\smallbbfont #1}}
\def\Bbb#1{\hbox{\bbfont #1}}

\def\LargeBbb#1{\hbox{\Largebbfont #1}}

\newcommand{\Aut}{\mbox{\it Aut}\,}
\newcommand{\Id}{\mbox{\it Id}\,}

\newcommand{\pr}{\mbox{\rm pr}}

\newcommand{\Spin}{\mbox{\it Spin}\,}
\newcommand{\Stab}{\mbox{\it Stab}\,}

\begin{document}

\enlargethispage{23cm}

\begin{titlepage}

$ $

\vspace{-1.5cm} 

\noindent\hspace{-1cm}
\parbox{6cm}{\footnotesize August 1998\newline
              Micron MPC P166 by Marco Monti}\
   \hspace{7.5cm}\
   \parbox{5cm}{ {\tt hep-th/9809007}\newline
                 {ut-ma/980011}}

\vspace{2cm}     

\centerline{\large\bf
 On the global structure of } 
\vspace{1ex}
\centerline{\large\bf some natural fibrations of Joyce manifolds}

\vspace{2cm}       

\centerline{\large Chien-Hao Liu\footnotemark}
\vspace{1.1em}
\centerline{\it Department of Mathematics}
\centerline{\it University of Texas at Austin}
\centerline{\it Austin, Texas 78712-1082}


\vspace{2em}

\begin{quotation}
\centerline{\bf Abstract}
\vspace{0.3cm}

\baselineskip 12pt  
{\small
 The study of fibrations of the target manifolds of
 string/M/F-theories has provided many insights to the dualities
 among these theories or even as a tool to build up dualities
 since the work of Strominger, Yau, and Zaslow on the Calabi-Yau
 case. For M-theory compactified on a Joyce manifold $M^7$, the
 fact that $M^7$ is constructed via a generalized Kummer
 construction on a $7$-torus ${\smallBbb T}^7$ with a torsion-free
 $G_2$-structure $\varphi$ suggests that there are natural
 fibrations of $M^7$ by ${\smallBbb T}^3$, ${\smallBbb T}^4$, and
 K3 surfaces in a way governed by $\varphi$. The local picture of
 some of these fibrations and their roles in dualities between
 string/M-theory have been studied intensively in the work of
 Acharya. In this present work, we explain how one can understand
 their global and topological details in terms of bundles over
 orbifolds. After the essential background is provided in Sec.\ 1,
 we give general discussions in Sec.\ 2 about these fibrations,
 their generic and exceptional fibers, their monodromy, and the
 base orbifolds. Based on these, one obtains a 5-step-routine
 to understand the fibrations, which we illustrate by examples
 in Sec.\ 3. In Sec.\ 4, we turn to another kind of fibrations for
 Joyce manifolds, namely the fibrations by the Calabi-Yau
 threefolds constructed by Borcea and Voisin. All these fibrations
 arise freely and naturally from the work of Joyce. Understanding
 how the global structure of these fibrations may play roles in
 string/M-theory duality is one of the major issues for further
 pursuit.
} 
\end{quotation}

\bigskip

\baselineskip 12pt


\bigskip

\baselineskip 11pt

{\footnotesize
\noindent{\bf Acknowledgements.}
I would like to thank 
 Orlando Alvarez, Philip Candelas, Jacques Distler, and
 William Thurston
for educations/group-meetings/discussions that strongly influence
the work; to 
 Hung-Wen Chang, Chong-Sun Chu, Daniel Freed, Robert Gompf,
 Cameron Gordon, Pei-Ming Ho, Se\'{a}n Keel, Vinek Narayanan,
 Rafael Nepomechie, Xenia de la Ossa, Alan Reid, Lorenzo Sadun,
 and Margaret Symington
for valuable conversations/discussions/correspondences/references
of a mother project that leads to this work; to
 Rev.\ Cam and Barbara Willman and Ann
for warm hospitality while the draft is in progress, 
and to Ling-Miao Chou for the indispensable moral support.
} 

\noindent
\underline{\hspace{20em}}

$^1${\footnotesize E-mail: chienliu@math.utexas.edu}

\end{titlepage}

\newpage
$ $

\vspace{-4em}  

\centerline{\sc
 Fibrations of Joyce Manifolds}

\vspace{4em}

\baselineskip 14pt  

\begin{flushleft}
{\Large\bf 0. Introduction and outline.}
\end{flushleft}

\begin{flushleft}
{\bf Introduction.}
\end{flushleft}
The study of fibrations of the target manifolds of
string/M/F-theories has provided many insights to the dualities
among these theories or even as a tool to build up dualities
since the work of Strominger, Yau, and Zaslow on the Calabi-Yau
case. For M-theory compactified on a Joyce manifold $M^7$, the
fact that $M^7$ is constructed via a generalized Kummer
construction on a $7$-torus ${\smallBbb T}^7$ with a torsion-free
$G_2$-structure $\varphi$ suggests that there are natural
fibrations of $M^7$ by ${\smallBbb T}^3$, ${\smallBbb T}^4$, and
K3 surfaces in a way governed by $\varphi$. The local picture of
these fibrations and their roles in dualities between
string/M-theory have been studied intensively in the work of
Acharya. In this present work, based on the work of Joyce, we set
up a $5$-step-routine which provides the global and topological
details to these fibrations in the terms of flat bundles over
orbifold. After illuminating this routine by several examples, we
turn to a discussion on some natural fibrations of Joyce manifolds
of the first kind by the Calabi-Yau threefolds constructed by
Borcea and Voisin. This renders Joyce manifolds of the first kind
to be similar to the $7$-spaces obtained by rolling Calabi-Yau
threefolds, as discussed in [Li].

Since all these fibrations arise freely and naturally from the
work of Joyce, one is surely very curious about how such
fibrations, particularly their global structures, should enter
string/M-theory dualities. This will have to await a future
pursuit.

\bigskip

\noindent
{\bf Convention.}
Since both real and complex manifolds are involved in this article,
to avoid confusion, a {\it real} $n$-dimensional manifold will be
called an {\it $n$-manifold} while a {\it complex} $n$-dimensional
manifold an {\it $n$-fold}. Also, a real $n$-dimensional orbifold
will be called an {\it $n$-orbifold}.

\bigskip

\begin{flushleft}
{\bf Outline.}
\end{flushleft}
{\small
\baselineskip 11pt  

\begin{quote}
 1. Essential mathematical backgrounds for physicists.

 2. \parbox[t]{11cm}{
    Fibrations of Joyce manifolds from decomposition of
    ${\Bbb R}^7$ governed by the $G_2$-structure.}
    \begin{quote}
     \hspace{-1.3em}
     2.1 Fibrations in terms of bundles over orbifolds.
         
     \hspace{-1.3em}
     2.2 Adjustment to the fibration and the monodromy after
         resolving $S$.
    \end{quote}

 3. Examples in the 5-step-routine.

 4. Fibrations of Joyce manifolds by Borcea-Voisin threefolds.

 5. Remarks on further questions/works.
\end{quote}
} 

\bigskip

\newpage
\section{Essential mathematical background for physicists.}

To set up some notations and to provide some essential background
for physicists, three key ingredients of the work: orbifolds,
Joyce manifolds, and Borcea-Voisin threefolds, are concisely
explained in this section. Readers are referred to the listed
references for more details.

\bigskip

\begin{flushleft}
{\bf Orbifolds {\rm ([Th1], also [B-S], [F-M], [Mo] and [Sc])}.}
  (Cf.\ Remark 2.1.4.)
\end{flushleft}
An $n$-dimensional {\it orbifold} $Q$ is a topological space
locally modelled by a quotient
$U=\widetilde{U}/\mbox{\raisebox{-.4ex}{$\Gamma_U$}}$, where
$\widetilde{U}$ is a connected open set in ${\Bbb R}^n$ and
$\Gamma_U$ is a finite group that acts on $\widetilde{U}$
effectively. Associated to each point $p$ in $Q$ is a group
$\Gamma_p$ that is isomorphic to the stabilizer of any preimage
point of $p$ in a local model
$U=\widetilde{U}/\mbox{\raisebox{-.4ex}{$\Gamma_U$}}$
around $p$. The set $\Sigma_Q=\{p\in Q|\Gamma_p\ne\{1\}\}$ is
called the {\it singular locus} of $Q$. In general $\Sigma_Q$ is
stratified by manifolds of various dimensions
(cf.\ {\sc Figures 2-1, 3-1-1, 3-6-1}).
In particular, when a discrete group $\Gamma$ acts on a manifold
$X$ effectively and properly discontinuously but not necessarily
freely, then the quotient $X/\mbox{\raisebox{-.4ex}{$\Gamma$}}$
is an orbifold with $\Sigma_{X/\Gamma}$ descending from the set
of fixed points in $X$ of some element in $\Gamma$. The group
$\Gamma_p$ associated to each point $p$ in
$X/\mbox{\raisebox{-.4ex}{$\Gamma$}}$ is the stabilizer
$\Stab(\widetilde{p})$ of any preimage point $\widetilde{p}$ of
$p$ in $X$.

A {\it covering orbifold} of an orbifold $Q$ is an orbifold
$\widetilde{Q}$ with a projection
$\phi:X_{\widetilde{Q}}\rightarrow X_Q$ between the underlying
spaces such that each $x\in X_Q$ has a neighborhood, modelled by
$U=\widetilde{U}/\mbox{\raisebox{-.4ex}{$\Gamma_U$}}$, for which
each component $V_i$ of $\phi^{-1}(U)$ is isomorphic to
$\widetilde{U}/\mbox{\raisebox{-.4ex}{$\Gamma_i$}}$, where
$\Gamma_i$ is a subgroup in $\Gamma$ and the isomorphisms are
compatible with $\phi$.

\bigskip

\noindent
{\bf Fact 1.1 [universal covering orbifold].}
([Th1]: Proposition 13.2.4). {\it 
Any orbifold $Q$ has a universal covering orbifold
$\widetilde{Q}$. In other words, if $\ast\in X_Q-\Sigma_Q$ is a
base point for $Q$, then $\phi:\widetilde{Q}\rightarrow Q$ is a
connected covering orbifold with base point
$\widetilde{\ast}\in\phi^{-1}(\ast)$ such that, for any 
covering orbifold
$\phi^{\prime}:\widetilde{Q}^{\prime}\rightarrow Q$ with base
point $\widetilde{\ast}^{\prime}\in{\phi^{\prime}}^{-1}(\ast)$,
there is an orbifold covering map
$\chi:\widetilde{Q}\rightarrow\widetilde{Q}^{\prime}$
such that $\chi(\widetilde{\ast})=\widetilde{\ast}^{\prime}$
and $\phi=\phi^{\prime}\circ\chi$.
} 

\bigskip

\noindent
By definition, such $\widetilde{Q}$ is unique. The
{\it fundamental group} $\pi_1^{\rm orb}(Q)$ of an orbifold $Q$
is then defined to be the group of deck transformations of its
universal covering orbifold $\widetilde{Q}$.

The following example will be important to us:

\bigskip

\noindent
{\bf Example 1.2 [toroidal orbifold].} {
 Let ${\Bbb T}^n={\Bbb R}^n/\mbox{\raisebox{-.4ex}{$\Lambda$}}$
 be an Euclidean $n$-torus determined by a lattice $\Lambda$ in
 ${\Bbb R}^n$ and $Q$ be an $n$-orbifold obtained by the quotient
 of ${\Bbb T}^n$ by a discrete group $\Gamma$ of isometries. Then
 the universal covering orbifold of $Q$ is the same as the
 universal covering space of ${\Bbb T}^n$, which is ${\Bbb R}^n$.
 Let $\widetilde{\Gamma}$ be a lifting of the $\Gamma$-action on
 ${\Bbb T}^n$ to ${\Bbb R}^n$. Then
 $\pi_1^{\rm orb}(Q)=\langle\widetilde{\Gamma},\Lambda\rangle$,
 the group of isometries on ${\Bbb R}^n$ generated by
 $\widetilde{\Gamma}$ and the translation group determined by
 $\Lambda$.
} 

\bigskip

In the above example, let $\phi$ be the universal covering map,
which is the composition
${\Bbb R}^n\rightarrow{\Bbb T}^n\rightarrow Q$. Then
$\langle\widetilde{\Gamma},\Lambda\rangle$ acts on
$\phi^{-1}(Q-\Sigma_Q)$ freely. Hence, if $\ast\in Q-\Sigma_Q$ is
a base point for $Q$ and $\widetilde{\ast}\in\phi^{-1}(\ast)$ is
a base point for ${\Bbb R}^n$, then, for each $y$ in the orbit of
$\ast$, there is a unique deck transformation
$g\in\langle\widetilde{\Gamma},\Lambda\rangle$ that sends $\ast$
to $y$. Let $\widetilde{\gamma}$ be a path from $\widetilde{\ast}$
to $y$, then its projection $\gamma=\phi(\widetilde{\gamma})$
is a loop at $\ast$ in $Q$ that represents
$g\in\pi_1^{\rm orb}(Q)$. In terms of homotopy classes of loops at
$\ast$, one has a tautological group homomorphism
$\tau:\pi_1(Q-\Sigma_Q)\rightarrow\pi_1^{\rm orb}(Q)$.
(Cf.\ {\sc Figure 2-1}.)

\bigskip

\begin{flushleft}
{\bf Joyce manifolds {\rm ([Jo1])}.}
\end{flushleft}
Consider the standard oriented Euclidean space ${\Bbb R}^7$ and
the $3$-form
\begin{eqnarray*}
 \lefteqn{ \varphi_0\;=\;
   e^1\wedge e^2\wedge e^7\,
     +\, e^1\wedge e^3\wedge e^6
     +\, e^1\wedge e^4\wedge e^5 }\\
   & &  +\, e^2\wedge e^3\wedge e^5 
     -\, e^2\wedge e^4\wedge e^6\,
         +\, e^3\wedge e^4\wedge e^7\,
         +\, e^5\wedge e^6\wedge e^7\,.
\end{eqnarray*}
The group of orientation-preserving linear isomorphisms that
preserves $\varphi_0$ is the exceptional group $G_2$ and lies in
$\mbox{\it SO}\,(7)$. The same holds also for its Hodge dual 
\begin{eqnarray*}
 \lefteqn{ \ast\varphi_0\;
   =\; e^1\wedge e^2\wedge e^3\wedge e^4\,
      +\, e^1\wedge e^2\wedge e^5\wedge e^6\,
      -\, e^1\wedge e^3\wedge e^5\wedge e^7  }\\
   & &  +\, e^1\wedge e^4\wedge e^6\wedge e^7\,
      +\, e^2\wedge e^3\wedge e^6\wedge e^7\,
      +\, e^2\wedge e^4\wedge e^5\wedge e^7\,
      +\, e^3\wedge e^4\wedge e^5\wedge e^6\,.
\end{eqnarray*}                                                          
Thus a $3$-form $\varphi$ on an oriented $7$-manifold $M^7$ that
is pointwise oriented-isomorphic to $\varphi_0$ determines a
$G_2$-structure and, via which, a Riemannian metric $g_{\varphi}$
on $M^7$. When $\varphi$ is torsion-free, the holonomy of
$g_{\varphi}$ is contained in $G_2$, which implies that
$g_{\varphi}$ is Ricci-flat. This feature singles out the role in
string theory of closed $7$-manifolds with torsion-free
$G_2$-structures.

In [Jo1], Joyce has constructed numerous such examples from
desingularizations of the following two kinds of toroidal
orbifolds from the standard $({\Bbb R}^7,\varphi_0)$ above:
\begin{quote}
 \hspace{-1.9em}{\it Class (1)}$\,$:\hspace{1ex}
 The quotient ${\Bbb T}^7/\mbox{\raisebox{-.4ex}{$\Gamma$}}$,
 where
 ${\Bbb T}^7={\Bbb R}^7/
  \hspace{-.1ex}\mbox{\raisebox{-.4ex}{${\Bbb Z}^7$}}$
 with coordinates $(x_1,\cdots,x_7)$, where
 $x_i\in {\Bbb R}/\mbox{\raisebox{-.4ex}{${\Bbb Z}$}}$, and
 $\Gamma$ is the group
 ${\Bbb Z}_2\oplus{\Bbb Z}_2\oplus{\Bbb Z}_2$, generated by
 $$
 \begin{array}{rcl}
  \alpha(x_1,\cdots,x_7) & =
      & (-x_1,-x_2,-x_3,-x_4,x_5,x_6,x_7)\,, \\[.2ex]
  \beta(x_1,\cdots,x_7) & =
      & (b_1-x_1,b_2-x_2,x_3,x_4,-x_5,-x_6,x_7)\,, \\[.2ex]
  \gamma(x_1,\cdots,x_7) & =
      & (c_1-x_1,x_2,c_3-x_3,x_4,c_5-x_5,x_6,-x_7)\,,
 \end{array}
 $$
 where $b_1$, $b_2$, $c_1$, $c_3$, and $c_5$ are some constants
 in $\{0,\frac{1}{2}\}$.

 \hspace{-1.9em}{\it Class (2)}$\,$: \hspace{1ex}
 The quotient
 $({\Bbb C}^3\times{\Bbb R})/\mbox{\raisebox{-.4ex}{$\langle
                                      \Gamma,\Lambda\rangle$}}$,
 where ${\Bbb C}^3\times{\Bbb R}$ is parametrized by
 $(z_1,z_2,z_3,x)$, $\Gamma$ is the dihedral group $D_a$ of $2a$
 elements, generated by
 $$
 \begin{array}{rcl}
  \alpha(z_1,z_2,z_3,x) & =
   & (uz_1,\, vz_2,\, \overline{uv}z_3,\,x+\frac{1}{a})\,,\\[.4ex]
  \beta(z_1,z_2,z_3,x) & =
   & (-\overline{z_1},\, -\overline{z_2},\,
                              -\overline{z_3},\,-x)\,
 \end{array}
 $$
 with $u$, $v$ being unit complex numbers and $a$ being the
 smallest positive integer such that $u^a=v^a=1$,
 and $\Lambda$ is a lattice in ${\Bbb C}^3\times{\Bbb R}$ invariant
 under $\Gamma$.
\end{quote}

Let $W^7$ be a such quotient. Then its singular set $S$ arises
from the collection of all the fixed $3$-tori ${\Bbb T}^3$ of
some element in $\Gamma$. We require that the constants that appear
in the definition of $\Gamma$ are chosen so that the tubular
neighborhood $\nu(S_0)$ of a component $S_0$ of $S$ is modelled
by either (a) or (b) below:
\begin{quote}
 \hspace{-1.9em}{\it Model (a)}$\,$:
 ${\Bbb T}^3\times({\Bbb C}^2/
               \mbox{\raisebox{-.4ex}{$\langle -1\rangle$}})$,
 where $-1$ stands for negation of both coordinates.

 \hspace{-1.9em}{\it Model (b)}$\,$:
 $\{{\Bbb T}^3\times({\Bbb C}^2/
           \mbox{\raisebox{-.4ex}{$\langle -1\rangle$}})\}/
                         \mbox{\raisebox{-.4ex}{${\Bbb Z}_2$}}$,
 where ${\Bbb Z}_2$ acts on
 ${\Bbb T}^3\times({\Bbb C}^2/
           \mbox{\raisebox{-.4ex}{$\langle -1\rangle$}})$
 in Model (a) freely.
\end{quote}
We shall called these {\it transverse $A_1$-singularities} along
${\Bbb T}^3$ or
${\Bbb T}^3/\mbox{\raisebox{-.4ex}{${\Bbb Z}_2$}}$.

Both Model (a) and Model (b) appear for admissible $b_1$, $b_2$,
$c_1$, $c_3$, $c_5$ while, for all admissible $(u,v,\Lambda)$,
Model (a) is the only kind of singularity that appears.
They can be resolved as follows:
\begin{itemize}
 \item
 For Model (a), $S_0$ is a ${\Bbb T}^3$ and $\nu(S_0)$ can be
 resolved by the transverse blowup $\mbox{\it Id}\times\psi$ that
 resolves the $A_1$-singularity in each
 ${\Bbb C}^2/\mbox{\raisebox{-.4ex}{$\langle -1\rangle$}}$ along
 $S_0$:
 $$
  \mbox{\it Id}\,\times\,\psi\;:
   \;{\Bbb T}^3\times T^{\ast}{\Bbb C}{\rm P}^1\;
    \longrightarrow\;
     {\Bbb T}^3\times({\Bbb C}^2/
               \mbox{\raisebox{-.4ex}{$\langle -1\rangle$}})\,,
 $$
 where $\mbox{\it Id}$ is the identity map on ${\Bbb T}^3$ and
 $\psi:T^{\ast}{\Bbb C}{\rm P}^1\rightarrow
       {\Bbb C}^2/\mbox{\raisebox{-.4ex}{$\langle -1\rangle$}}$
 is the resolution of the isolated $A_1$-singularity of
 ${\Bbb C}^2/\mbox{\raisebox{-.4ex}{$\langle -1\rangle$}}$.
 Note that, since the exceptional locus for $\psi$ is the
 $0$-section of $T^{\ast}{\Bbb C}{\rm P}^1$, which is a
 ${\Bbb C}{\rm P}^1$, the exceptional locus of
 $\mbox{\it Id}\times\psi$ is ${\Bbb T}^3\times{\Bbb C}{\rm P}^1$.
 Note also that in this case the resolution is indifferent of how
 one identifies the fiber
 ${\Bbb R}^4/\mbox{\raisebox{-.4ex}{$\langle -1\rangle$}}$ of
 $\nu(S_0)$ with
 ${\Bbb C}^2/\mbox{\raisebox{-.4ex}{$\langle -1\rangle$}}$.

 \item
 For Model (b), $S_0$ is a free quotient
 ${\Bbb T}^3/\mbox{\raisebox{-.4ex}{${\Bbb Z}_2$}}$.
 Depending on the ways the fiber
 ${\Bbb R}^4/\mbox{\raisebox{-.4ex}{$\langle -1\rangle$}}$ of
 $\nu(S_0)$ is identified with
 ${\Bbb C}^2/\mbox{\raisebox{-.4ex}{$\langle -1\rangle$}}$,
 there are two inequivalent free ${\Bbb Z}_2$-actions on
 ${\Bbb T}^3\times({\Bbb C}^2/
            \mbox{\raisebox{-.4ex}{$\langle -1\rangle$}})$:
 one generated by a holomorphic map of the form
 $(\,\ast\, ,\,(z_1,z_2))\mapsto(\,\ast^{\prime}\,,(z_1,-z_2))$
 and the other generated by an antiholomorphic map of the form
 $(\,\ast\, ,\,(z_1,z_2))\mapsto
   (\,\ast^{\prime}\,,(\overline{z_1},\overline{z_2}))$.
 Each has a unique natural lifting to a free ${\Bbb Z}_2$-action
 on ${\Bbb T}^3\times T^{\ast}{\Bbb C}{\rm P}^1$ equivariant with
 the ${\Bbb Z}_2$-action on
 ${\Bbb T}^3\times({\Bbb C}^2/
            \mbox{\raisebox{-.4ex}{$\langle -1\rangle$}})$
 with respect to $\mbox{\it Id}\times \psi$ in Case (a).
 Each quotient
 $({\Bbb T}^3\times T^{\ast}{\Bbb C}{\rm P}^1)/
                          \mbox{\raisebox{-.4ex}{${\Bbb Z}_2$}}$
 gives then a different resolution for $\nu(S_0)$. The exceptional
 locus is
 $({\Bbb T}^3\times{\Bbb C}{\rm P}^1)/
                        \mbox{\raisebox{-.4ex}{${\Bbb Z}_2$}}$.
\end{itemize}
From a combination of three independent $1$-forms on ${\Bbb T}^3$
and three $2$-forms on $T^{\ast}{\Bbb C}{\rm P}^1$ from its hyper
K\"{a}hler structure, any above resolution $\widetilde{\nu(S_0)}$ 
of $\nu(S_0)$ admits a family of torsion-free $G_2$-structures
depending on a real parameter $t$. The singular $\nu(S_0)$ is
recovered as $t\rightarrow 0$. 

Hence, after resolving the singular set $S$ of $W^7$ as above,
the resulting smooth closed $7$-manifold $M^7$ admits a
$G_2$-structure $\varphi_t$ by sewing the $G_2$-structure
of $W^7$ inherited from $({\Bbb R}^7,\varphi_0)$ and
that of $\widetilde{\nu(S)}$ via a partition of unity. By
construction, $\varphi_t$ is torsion-free outside a ring domain
$A$ around the boundary $\partial\nu(S)$ but has torsion in $A$.
However, Joyce shows that, for $t$ small enough, one can deform
$\varphi_t$ by exact $3$-forms $d\eta_t$ so that the new
$G_2$-structure $\widetilde{\varphi}=\varphi_t+d\eta_t$ is
torsion-free all over $M^7$. In this way, he constructed families
of closed Riemannian $7$-manifolds with holonomy $G_2$. They are
called {\it Joyce manifolds}. For convenience, we shall call
those from quotients in
Class (1) {\it Joyce manifolds of the first kind}, with
 notation $J(b_1,b_2,c_1,c_3,c_5)$, and those from quotients in
Class (2)
{\it Joyce manifolds of the second kind}, with notation
 $J(u,v,\Lambda)$.

\bigskip

\begin{flushleft}
{\bf Borcea-Voisin threefolds {\rm ([Bo], [G-W], [Ni], [Vo])}.}
\end{flushleft}
Let $X$ be a K3 surface with an involution $\iota$ that acts by
$(-1)$ on the holomorphic $2$-form of $X$. Let $\Sigma$ be the
set of fixed points of $\iota$. Then ([Ni]) $\Sigma$ is a disjoint
union of smooth curves in $X$ as classified below:
\begin{quote}
 \hspace{-1.9em}(1)\hspace{1ex}
 $\Sigma$ is empty;.

 \hspace{-1.9em}(2)\hspace{1ex}
  $\Sigma=C_1\cup C_1^{\prime}$, where $C_1$ and $C_1^{\prime}$
  are both elliptic curves;
   
 \hspace{-1.9em}(3)\hspace{1ex}
  $\Sigma=C_g+E_1+\,\cdots\,+E_k$. where $C_g$ is a curve of
  genus $g$ and $E_i$ are rational curves.
\end{quote}
In all cases, the quotient
$X/\hspace{-.1ex}\mbox{\raisebox{-.4ex}{$\iota$}}$ is a smooth
surface. 

Let $E$ be an elliptic curve, and $j$ be the involution on $E$
acting by negation. The set $S_0$ of fixed-points of $(\iota,j)$
on $X\times E$ consists of four disjoint copies of $\Sigma$.
and the quotient
$(X\times E)/
  \hspace{-.1ex}\mbox{\raisebox{-.4ex}{$(\iota,j)$}}$
has $A_1$-singularities along $S_0$. Simultaneous
blowup along the transverse directions to $S_0$ resovles these
singularities and the result is a smooth Calabi-Yau threefold $Y$.
Such Calabi-Yau manifolds and their mirror symmetry are considered
first in the work of Borcea [Bo] and Voisin [Vo] and we shall call
them {\it Borcea-Voisin threefolds}.
Let $n$ be the number of components of $\Sigma$ and $n^{\prime}$
be the sum of the genus of all these components. Then the Hodge
numbers of a Borcea-Voisin threefold $Y$ are determined by
([Bo] and [Vo])
$$
 h^{1,1}(Y)\;=\;11+5n-n^{\prime}
  \hspace{1em}\mbox{and}\hspace{1em}
 h^{2,1}(Y)\;=\;11+5n^{\prime}-n\,. 
$$

Such Calabi-Yau threefolds could admit K3- and special Lagrangian
${\Bbb T}^3$-fibrations:
\begin{itemize}
 \item
 {\it K3-fibration}$\,$:
 By construction, one has the natural K3-fibration $\pi$ from
 $(X\times E)/\hspace{-.1ex}\mbox{\raisebox{-.4ex}{$(\iota,j)$}}$
 onto $E/\hspace{-.1ex}\mbox{\raisebox{-.4ex}{$j$}}$, which is an
 $S^2(2222)$-orbifold (cf.\ Remark 2.1.4 in Sec.\ 2.1).
 The fixed-point set $S_0$ is contained in the exceptional fibers
 of $\pi$ with one $\Sigma$ in each such fiber. Hence, when
 resloving the singularities of
 $E/\hspace{-.1ex}\mbox{\raisebox{-.4ex}{$j$}}$ along $S_0$ to
 obtain $Y$, the K3-fibration $\pi$ is lifted to a K3-fibration
 $\widetilde{\pi}$ from $Y$ onto the same base
 $E/\hspace{-.1ex}\mbox{\raisebox{-.4ex}{$j$}}$. The exceptional
 fiber over an orbifold point now becomes
 $X/\hspace{-.1ex}\mbox{\raisebox{-.4ex}{$\iota$}}\cup V$,
 where $V=\Sigma\times{\Bbb C}{\rm P}^1$ intersects
 $X/\hspace{-.1ex}\mbox{\raisebox{-.4ex}{$\iota$}}$ transversely
 at the $\Sigma$ from $S_0$.

 \item
 {\it ${\Bbb T}^3$-fibration} ([G-W])$\,$:
 If $X$ has an elliptic fibration invariant under 
 $\iota$ with generic fiber a special Lagrangian submanifold with
 respect to the K\"{a}hler form on $X$, then one can choose a
 ${\Bbb T}^1$-fibration of $E$ and combine these into a special
 Lagrangian ${\Bbb T}^3$-fibration for $Y$. The base of this
 fibration is a $3$-orbifold $Q$ whose underlying topology is $S^3$
 with the singular locus $\Sigma_Q$, which corresponds to the set of
 critical values of the fibration, as indicated in {\sc Figure 1-1}.
 \begin{figure}[htbp]
 \setcaption{{\sc Figure 1-1.}
 \baselineskip 14pt
   The set of critical values of a special Lagrangian
   ${\Bbb T}^3$-fibrations of a Borcea-Voisin threefold
   (cf.\ Figure 3 in [G-W]). The numbers $l$ and $m$ are
   determined by $(X,\iota)$; up to this and a homeomorphism
   of $S^3$, the picture is exact.
 } 
 \centerline{\psfig{figure=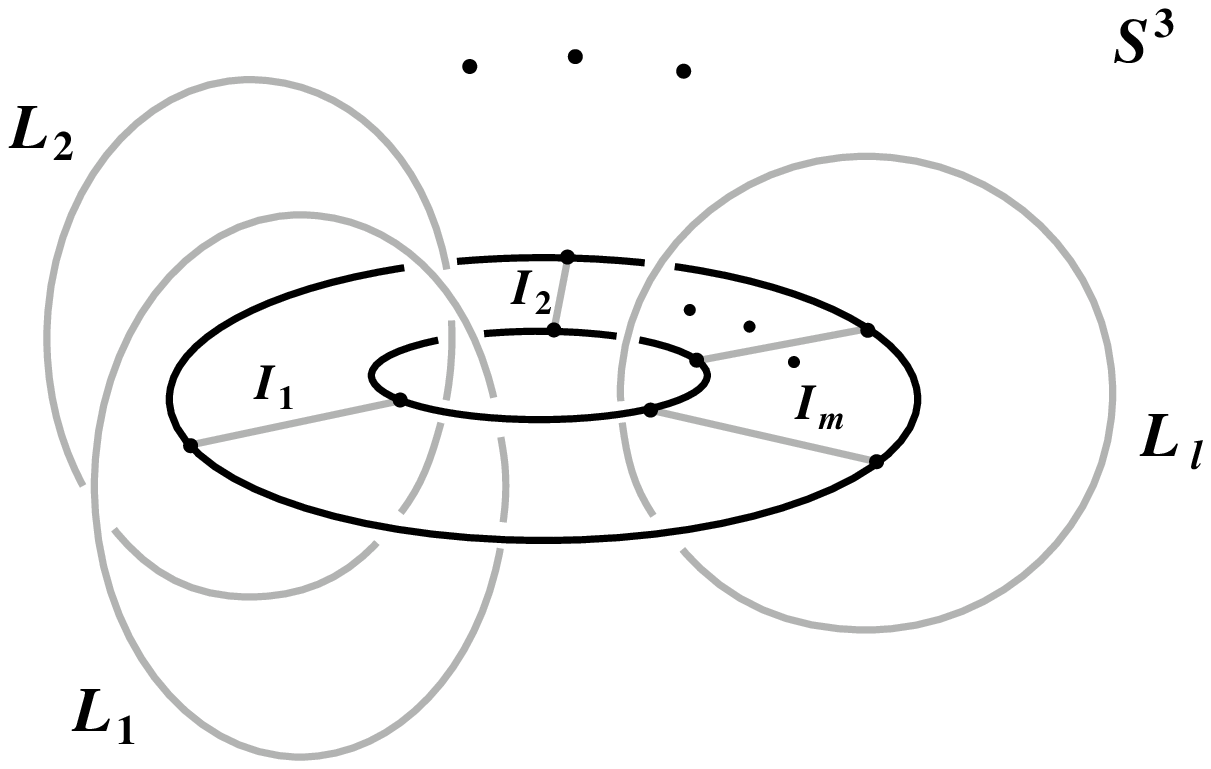,width=13cm,caption=}}
 \end{figure}
 \end{itemize}

\bigskip

With these preparations, let us now turn to the themes of this
paper.

\bigskip

\section{Fibrations of Joyce manifolds from decompositions of
  ${\LargeBbb R}^7$ governed by the $G_2$-structure $\varphi_0$.}

We discuss in Sec.\ 2.1 how some natural fibrations for a
Joyce manifold can arise from $({\Bbb R^7}, \varphi_0)$ and 
discuss how they can be understood in terms of a bundle
$\pi:{\Bbb T}^7/\mbox{\raisebox{-.4ex}{$\Gamma$}}\rightarrow Q$
over an orbifold $Q$.
In Sec.\ 2.2 the effect to $\pi$ after resolving the singular set
$S$ of ${\Bbb T}^7/\mbox{\raisebox{-.4ex}{$\Gamma$}}$ is studied
and, hence, one completes the picture.

\bigskip

\subsection{Fibrations in terms of bundles over orbifolds.}

The following definitions from [H-L] (also [Ha], [Jo1], and [McL])
is the starting point of all the fibrations to be discussed in this
paper.

\bigskip

\noindent
{\bf Definition 2.1.1 [calibration, (co)associative submanifold].} 
 A {\it calibration} on a Riemannian manifold $M$ is a closed
  $k$-form $\omega$ whose restriction to any tangent $k$-plane
  to $M$ is less than or equal to the volume of the $k$-plane.
 A {\it calibrated submanifold} of $(M,\omega)$ is a
  $k$-dimensional submanifold $N$ in $M$ such that $\omega|_N$
  is equal to the volume-form of the induced metric on $N$.
 Let $M^7$ be a $7$-manifold with torsion-free $G_2$-structure
  $\varphi$. Then both $\varphi$ and $\ast\varphi$ are calibrations
  on $M^7$. An {\it associative submanifold} of $M^7$ is a
  calibrated submanifold of $(M^7,\varphi)$ and a
  {\it coassociative} submanifold of $M^7$ is a calibrated
  submanifold of $(M^7,\ast\varphi)$.

\bigskip

\noindent
These definitions generalize automatically to orbifolds and
their sub-orbifolds by considering the complement of their
singular loci.

\bigskip

\begin{flushleft}
{\bf How the fibrations arise.}
\end{flushleft}
Consider ${\Bbb R}^7$ with the standard metric and the
torsion-free $G_2$-structure $\varphi_0$ in Sec.\ 1. From the
explicit expression of $\varphi_0$, one has the following
associative-coassociative product decompositions of
$({\Bbb R}^7,\varphi_0)$:
$$
\begin{array}{llll}
 {\Bbb R}^3_{127}\times{\Bbb R}^4_{3456}\,,
   & {\Bbb R}^3_{136}\times{\Bbb R}^4_{2457}\,,
   & {\Bbb R}^3_{145}\times{\Bbb R}^4_{2367}\,,
   & {\Bbb R}^3_{235}\times{\Bbb R}^4_{1467}\,,  \\[1.5ex]
 {\Bbb R}^3_{246}\times{\Bbb R}^4_{1357}\,,
   & {\Bbb R}^3_{347}\times{\Bbb R}^4_{1256}\,,
   & {\Bbb R}^3_{567}\times{\Bbb R}^4_{1234}\,, &  
\end{array}
$$
where ${\Bbb R}^3_{abc}$ is the standard ${\Bbb R}^3$ with
coordinates $(x_a,x_b,x_c)$ and similarly for ${\Bbb R}^4_{ijkl}$,
and also the tori ${\Bbb T}^3_{abc}$, ${\Bbb T}^4_{ijkl}$ to
appear below.

For Joyce manifolds of the first kind, the associated
decompositions, ${\Bbb T}^3_{abc}\times{\Bbb T}^4_{ijkl}$, of
${\Bbb T}^7$ descending from above are all preserved by $\Gamma$.
Consequently, each of these product decompositions induces a pair
of fibrations on the quotient
${\Bbb T}^7/\mbox{\raisebox{-.4ex}{$\Gamma$}}$ - one with
associative sub-orbifolds as fibers, the other with coassociative
sub-orbifolds as fibers, and fibers from different fibrations in
a same pair intersect each other transversely. Consider any such
fibrations of ${\Bbb T}^7/\mbox{\raisebox{-.4ex}{$\Gamma$}}$. Since
the singular set $S$ of
${\Bbb T}^7/\mbox{\raisebox{-.4ex}{$\Gamma$}}$
arises from the set of fixed ${\Bbb T}^3$ of elements of 
$\Gamma$ on ${\Bbb T}^7$ and any latter ${\Bbb T}^3$ is
also of the form ${\Bbb T}^3_{a^{\prime}b^{\prime}c^{\prime}}$,
which either transverse to all fibers or have empty intersection
with most fibers, resolving $S$ will lead to a fibration for $M^7$.
This fibration in general may not be associative/coassociative.
However, if one lets the tubular neighborhood of $S$ and the
$t$-parameter involved in resolving $S$ (cf.\ Sec.\ 1 and [Jo1])
to be small, then the fibration of $M^7$ thus obtained
will be associative/coassociative except in a region that can be
made arbitrarily small. In the limit, the singular Joyce space 
associative/coassociative fibred 
${\Bbb T}^7/\mbox{\raisebox{-.4ex}{$\Gamma$}}$ is recovered. For
this reason, we shall call such fibration an
{\it asymptotically associative/coassociative fibration} of $M^7$,
abbreviated as {\it a.a./a.c.\ fibration}.

For Joyce manifolds of the second kind, one identifies
${\Bbb C}^3\times{\Bbb R}$ with ${\Bbb R}^7$ by the correspondence
$(z_1,z_2,z_3,x)\mapsto(x_1+x_2i,x_3+x_4i,x_5+x_6i,x_7)$;
then the three decompositions
${\Bbb R}^3_{127}\times{\Bbb R}^4_{3456}$,
${\Bbb R}^3_{347}\times{\Bbb R}^4_{1256}$, and
${\Bbb R}^3_{567}\times{\Bbb R}^4_{1234}$
are preserved by $\Gamma$, each of which descends to a pair
of associative-coassociative foliations for
$({\Bbb C}^3\times{\Bbb R})/
 \hspace{-.1ex}\mbox{\raisebox{-.4ex}{$\langle\Lambda,
                                         \Gamma\rangle$}}$.
In general these may not be realizable as a nice fibration.
However, for most of the concrete examples given by Joyce in [Jo1],
the lattice $\Lambda$ is decomposable into
$\Lambda_{abc}\oplus\Lambda_{ijkl}$, where $\Lambda_{abc}$ is a
lattice in ${\Bbb R}^3_{abc}$, $\Lambda_{ijkl}$ is a lattice in
${\Bbb R}^4_{ijkl}$, and ${\Bbb R}^3_{abc}\oplus{\Bbb R}^4_{ijkl}$
is one of the admissible associative-coassociative product
decomposition above. In this good case, an above decomposition
descends to an $\Gamma$-invariant associative-coassociative
product decomposition ${\Bbb T}^3_{abc}\times{\Bbb T}^4_{ijkl}$
for ${\Bbb T}^7={\Bbb R}^7/\mbox{\raisebox{-.4ex}{$\Lambda$}}$.
This then leads to an a.a./a.c.\ fibration for the Joyce manifold
after resolving the singularity of 
$({\Bbb C}^3\times{\Bbb R})/
 \hspace{-.1ex}\mbox{\raisebox{-.4ex}{$\langle\Lambda,
                                         \Gamma\rangle$}}$
as in the case for Joyce manifolds of the first kind.

\bigskip

\begin{flushleft}
{\bf The associated bundles over orbifolds before resolving $S$.}
\end{flushleft}
Regard ${\Bbb T}^7$, either from
${\Bbb R}^7/\mbox{\raisebox{-.4ex}{${\Bbb Z}^7$}}$ for Joyce
manifolds of the first kind or
$({\Bbb C}^3\times{\Bbb R})/\mbox{\raisebox{-.4ex}{$\Lambda$}}$
for Joyce manifolds of the second kind, as the total space of
the trivial bundle $Y=Z\times X$ associated to an admissible
decomposition of ${\Bbb R}^7$ listed above, with $Z$ being the base
and $X$ the associative/coassociative fiber. Then $\Gamma$ acts
effectively on $Y$ as a finite group of bundle automorphisms. This
induces a $\Gamma$-action on the base $Z$. Let $\Gamma_0$ be the
normal subgroup in $\Gamma$ that consists of all the elements in
$\Gamma$ which acts trivially on $Z$. Then $\Gamma_0$ is contained
in the stabilizer of every fiber of $Y$, after taking the quotient
by $\Gamma$, one obtains a bundle $\pi$ and the commutative
diagram:
$$
\begin{array}{ccll}
 Y & \longrightarrow
            & Y/\mbox{\raisebox{-.4ex}{$\Gamma$}} &\\[.2ex]
 \hspace{3ex}\downarrow &
            & \hspace{1ex}\downarrow\pi & \\[.2ex]
 Z & \stackrel{\pr}{\longrightarrow}
            & Q\;=\;Z/\mbox{\raisebox{-.4ex}{$\Gamma$}} &.
\end{array}
$$
The generic fiber of $\pi$ is
$X/\mbox{\raisebox{-.4ex}{$\Gamma_0$}}$. The following lemma,
which can be justified by tracing definitions, gives the basic
relations among fibers in $Y/\mbox{\raisebox{-.4ex}{$\Gamma$}}$,
$\Gamma_p$ for $p$ in $Q$, and the stabilizer subgroup of fibers
in $Y$:

\bigskip

\noindent
{\bf Lemma 2.1.2.} {\it 
Given a point $p$ in $Q$, let $\widetilde{p}$ be a point in the
preimage of $p$ in $Z$, $F_{\widetilde{p}}$ be the fiber in $Y$
over $\widetilde{p}$, and $\Stab_0(F_{\widetilde{p}})$ be the
smallest subgroup in $\Stab(F_{\widetilde{p}})$ that contains
$\Gamma_0$ and all the elements in $\Gamma$ that act trivially on
$F_{\widetilde{p}}$. Then the group $\Gamma_p$ associated to $p$
is the quotient
$\Stab(F_{\widetilde{p}})/\mbox{\raisebox{-.4ex}{$\Gamma_0$}}$.
The fiber $F_p$ in $Y/\mbox{\raisebox{-.4ex}{$\Gamma$}}$ over
$p$ is the quotient
$F_{\widetilde{p}}\,/\mbox{\raisebox{-.4ex}{$
                                   \Stab(F_{\widetilde{p}})$}}$
with multiplicity equal to the cardinality of
$\Stab(F_{\widetilde{p}})/\mbox{\raisebox{-.4ex}{$
                                   \Stab_0(F_{\widetilde{p}})$}}$.
} 

\bigskip

For Joyce manifolds of the first kind $J(b_1,b_2,c_1,c_3,c_5)$,
$Q$ can be immediately read off by projecting the $\Gamma$-action
on ${\Bbb T}^7$ to some appropriate ${\Bbb T}^3_{abc}$ or
${\Bbb T}^4_{ijkl}$ factor, as listed below:
\begin{itemize}
 \item
 {\it For a.a.\ fibrations}$\,$:
 Let ${\Bbb T}^4={\Bbb R}^4/\mbox{\raisebox{-.4ex}{${\Bbb Z}^4$}}$,
 with coordinates $(x,y,z,w)$,
 $x,y,z,w\in{\Bbb R}/\mbox{\raisebox{-.4ex}{${\Bbb Z}$}}$.
 Then $Q$ is the quotient of ${\Bbb T}^4$ by one of the following
 abelian groups:
 \begin{quote}
  \hspace{-1.9em}
  $\Gamma_{3456}=\langle \overline{\alpha}, \overline{\beta},
     \overline{\gamma} \rangle$, where
  {\small $\overline{\alpha}(x,y,z,w)=(-x,\,-y,\,z,\,w)$,\newline
    $\overline{\beta}(x,y,z,w)=(x,\,y,\,-z,\,-w)$,
    $\overline{\gamma}(x,y,z,w)=(c_3-x,\,y,\,c_5-z,\,w)$; }

  \medskip

  \hspace{-1.9em}
  $\Gamma_{2457}=\langle  \overline{\alpha}, \overline{\beta},
     \overline{\gamma}\rangle$, where
  {\small $\overline{\alpha}(x,y,z,w)=(-x,\,-y,\,z,\,w)$, \newline
    $\overline{\beta}(x,y,z,w)=(b_2-x,\,y,\,-z,\,w)$,
    $\overline{\gamma}(x,y,z,w)=(x,\,y,\,c_5-z,\,-w)$; }

  \medskip

  \hspace{-1.9em}
  $\Gamma_{2367}=\langle \overline{\alpha}, \overline{\beta},
     \overline{\gamma} \rangle$, where
  {\small $\overline{\alpha}(x,y,z,w)=(-x,\,-y,\,z,\,w)$,\newline
    $\overline{\beta}(x,y,z,w)=(b_2-x,\,y,\,-z,\,w)$,
    $\overline{\gamma}(x,y,z,w)=(x,\,c_3-y,\,z,\,-w)$; }

  \medskip

  \hspace{-1.9em}
  $\Gamma_{1467}=\langle \overline{\alpha}, \overline{\beta},
     \overline{\gamma} \rangle$, where
  {\small $\overline{\alpha}(x,y,z,w)=(-x,\,-y,\,z,\,w)$,\newline
    $\overline{\beta}(x,y,z,w)=(b_1-x,\,y,\,-z,\,w)$,
    $\overline{\gamma}(x,y,z,w)=(c_1-x,\,y,\,z,\,-w)$; }

  \medskip

  \hspace{-1.9em}
  $\Gamma_{1357}=\langle \overline{\alpha}, \overline{\beta},
     \overline{\gamma} \rangle$, where
  {\small $\overline{\alpha}(x,y,z,w)=(-x,\,-y,\,z,\,w)$,\newline
    $\overline{\beta}(x,y,z,w)=(b_1-x,\,y,\,-z,\,w)$,
    $\overline{\gamma}(x,y,z,w)=(c_1-x,\,c_3-y,\,c_5-z,\,-w)$;}

  \medskip

  \hspace{-1.9em}
  $\Gamma_{1256}=\langle \overline{\alpha}, \overline{\beta},
     \overline{\gamma} \rangle$, where
  {\small $\overline{\alpha}(x,y,z,w)=(-x,\,-y,\,z,\,w)$,\newline
   $\overline{\beta}(x,y,z,w)=(b_1-x,\,b_2-y,\,-z,\,-w)$,
   $\overline{\gamma}(x,y,z,w)=(c_1-x,\,y,\,c_5-z,\,w)$; }

  \medskip

  \hspace{-1.9em}
  $\Gamma_{1234}=\langle \overline{\alpha}, \overline{\beta},
     \overline{\gamma} \rangle$, where
  {\small $\overline{\alpha}(x,y,z,w)=(-x,\,-y,\,-z,\,-w)$,\newline
    $\overline{\beta}(x,y,z,w)=(b_1-x,\,b_2-y,\,z,\,w)$,
    $\overline{\gamma}(x,y,z,w)=(c_1-x,\,y,\,c_3-z,\,w)$.}
 \end{quote}
\end{itemize}
Such $4$-orbifold $Q^4$ can be described in terms of a flat
${\Bbb T}^2$-, $A^2$-, or $S^2(2222)$-fibrations over a toroidal 
$2$-orbifolds with the exceptional fibers and the monodromy
understood in exactly the same way we try to understand the more
complicated bundle $\pi$. We shall come back to this at the end
of this section.
(Cf.\ Remark 2.1.4; also Examples 3.4 and 3.6 in Sec.\ 3.)

\begin{itemize}
 \item
 {\it For a.c.\ fibrations$\,$}:
 Let ${\Bbb T}^3={\Bbb R}^3/\mbox{\raisebox{-.4ex}{${\Bbb Z}^3$}}$
 with coordinates $(x,y,z)$, where $x,y,z$ are in
 ${\Bbb R}/\mbox{\raisebox{-.4ex}{${\Bbb Z}$}}$. Then $Q$ is the
 quotient of ${\Bbb T}^3$ by one of the following abelian groups:
 \begin{quote}
  \hspace{-1.9em}
  $\Gamma_{127}=\langle\overline{\alpha},\overline{\beta},
                                       \overline{\gamma} \rangle$,
  where
  {\small $\overline{\alpha}(x,y,z)=(-x,\,-y,\,z)$, \newline
    $\overline{\beta}(x,y,z)=(b_1-x,\,b_2-y,\,z)$,
    $\overline{\gamma}(x,y,z)=(c_1-x,\,y,\,-z)$; }

  \medskip

  \hspace{-1.9em}
  $\Gamma_{136}=\langle \overline{\alpha}, \overline{\beta},
     \overline{\gamma} \rangle$, where
  {\small $\overline{\alpha}(x,y,z)=(-x,\,-y,\,z)$, \newline
    $\overline{\beta}(x,y,z)=(b_1-x,\,y,\,-z)$,
    $\overline{\gamma}(x,y,z)=(c_1-x,\,c_3-y,\,z)$; }

  \medskip

  \hspace{-1.9em}
  $\Gamma_{145}=\langle \overline{\alpha}, \overline{\beta},
     \overline{\gamma} \rangle$, where
  {\small $\overline{\alpha}(x,y,z)=(-x,\,-y,\,z)$, \newline
    $\overline{\beta}(x,y,z)=(b_1-x,\,y,\,-z)$,
    $\overline{\gamma}(x,y,z)=(c_1-x,\,y,\,c_5-z)$; }

  \medskip

  \hspace{-1.9em}
  $\Gamma_{235}=\langle \overline{\alpha}, \overline{\beta},
     \overline{\gamma} \rangle$, where
  {\small $\overline{\alpha}(x,y,z)=(-x,\,-y,\,z)$, \newline
    $\overline{\beta}(x,y,z)=(b_2-x,\,y,\,-z)$,
    $\overline{\gamma}(x,y,z)=(x,\,c_3-y,\,c_5-z)$; }

  \medskip

  \hspace{-1.9em}
  $\Gamma_{246}=\langle\overline{\alpha},\overline{\beta}\rangle$,
  where
  {\small $\overline{\alpha}(x,y,z)=(-x,\,-y,\,z)$,
          $\overline{\beta}(x,y,z)=(b_2-x,\,y,\,-z)$; }

  \medskip

  \hspace{-1.9em}
  $\Gamma_{347}=\langle\overline{\alpha},\overline{\gamma}\rangle$,
  where
  {\small $\overline{\alpha}(x,y,z)=(-x,\,-y,\,z)$,
          $\overline{\gamma}(x,y,z)=(c_3-x,\,y,\,-z)$; }

  \medskip

  \hspace{-1.9em}
  $\Gamma_{567}=\langle\overline{\beta},\overline{\gamma}\rangle$,
  where
  {\small $\overline{\beta}(x,y,z)=(-x,\,-y,\,z)$,
          $\overline{\gamma}(x,y,z)=(c_5-x,\,y,\,-z)$.}
 \end{quote}
\end{itemize}
Such $3$-orbifold $Q^3$ can be understood by choosing a
fundamental domain $\Omega$ of the $\Gamma_{abc}$-action on
${\Bbb T}^3$ and examining how  $\Gamma_{abc}$ acts on the
boundary $\partial\Omega$.
(Cf.\ Examples 3.1 and 3.2 in Sec.\ 3.)

For Joyce manifolds of the second kind $J(u,v,\Lambda)$,
since the choice of $\Lambda$ varies with $(u,v)$, one no longer
has general forms as above. However, once $(u,v,\Lambda)$ is
explicitly written down, the base orbifold $Q$ of $\pi$ can still
be understood in the same way as in the case
$J(b_1,b_2,c_1,c_3,c_5)$.
(Cf.\ Examples 3.5 and 3.6 in Sec.\ 3.)

\bigskip

\begin{flushleft}
{\bf The monodromy associated to $\pi$.}
\end{flushleft}
The flat connection on the trivial bundle $Y=Z\times X$ over $Z$
induced from the product structure is $\Gamma$-invariant; hence it
descends to a flat connection on the fibration $\pi$ of
$Y/\mbox{\raisebox{-.4ex}{$\Gamma$}}$ over $Q$. Associated to this
is a monodromy homomorphism
$\rho:\pi_1^{\rm orb}(Q)\rightarrow
   \Aut(X/\mbox{\raisebox{-.4ex}{$\Gamma_0$}})$.
For the Joyce manifolds under consideration,
from the explicit expression of the $\Gamma$-action on $Y$
following Sec.\ 1, any $g$ in $\Gamma$ acting on $Y$ can be
splitted into $(\overline{g},g^f)$, where $\overline{g}$ acts on
$Z$ and $g^f$ acts on $X$; in other words, the $\Gamma$-action on
$Y$ is the diagonal action of $(\overline{\Gamma},\Gamma^f)$,
where $\overline{\Gamma}$ is the projection of the $\Gamma$-action
on $Y$ to $Z$ and $\Gamma^f$ is the projection of the
$\Gamma$-action on $Y$ to $X$. Now recall from Sec.\ 1 that
$\pi_1^{\rm orb}(Q)
  =\langle\overline{\Gamma},\overline{\Lambda}\rangle$,
one thus has
$$
 \rho(\overline{g})\;=\;\overline{g}^f
  \hspace{1em}\mbox{and}\hspace{1em}
 \rho(\overline{\Lambda})\;=\;\{\Id\}\,,
$$
where $\overline{g}^f$ is the automorphism on
$X/\mbox{\raisebox{-.4ex}{$\Gamma_0$}}$ induced from $g$ acting
on $X$.
(Note that in general for this to be well-defined $\Gamma_0$ has
 to lie in the centralizer of $\Gamma$; in our problem $\Gamma_0$
 can be nontrivial only when considering a.a./a.c.\ fibrations for
 Joyce manifolds of the first kind, in which case $\Gamma$ is
 abelian. Thus $\overline{g}^f$ is well-defined.)
Without work, this determines $\rho$ completely.

\bigskip

\noindent
{\it Remark 2.1.3.}
For physicists who feel less familiar with orbifold fundamental
group $\pi_1^{\rm orb}(Q)$, one can also consider directly the
monodromy $\rho^{\prime}$ on the usual $\pi_1(Q-\Sigma_Q)$ as
follows. Let $\pr:Z\rightarrow Q$ be the quotient map. Then
$\pr:Z-\pr^{-1}(\Sigma_Q)\rightarrow Q-\Sigma_Q$ is a regular
covering map ([Sp]) since any two liftings of a closed loop in
$Q-\Sigma_Q$ differ by a transformation via an element in
$\Gamma$ and hence must be either both closed or both open.
This implies that $\pr_{\ast}(\pi_1(Z-\pr^{-1}(\Sigma_Q)))$
is a normal subgroup in $\pi_1(Q-\Sigma_Q)$ and the quotient
$\pi_1(Q-\Sigma_Q)/\mbox{\raisebox{-.4ex}{$\pr_{\ast}
                            (\pi_1(Z-\pr^{-1}(\Sigma_Q)))$}}$
is the group of deck transformations of $\pr$, which is
$\Gamma/\mbox{\raisebox{-.4ex}{$\Gamma_0$}}$ by construction.
Let
$\rho_0:\pi_1(Z-\pr^{-1}(\Sigma_Q))\rightarrow
                  \Aut(X/\mbox{\raisebox{-.4ex}{$\Gamma_0$}})$
be the monodromy homomorphism associated to the flat connection
on $Y$ and
$\rho^{\prime}:\pi_1(Q-\Sigma_Q)\rightarrow
                  \Aut(X/\mbox{\raisebox{-.4ex}{$\Gamma_0$}})$
be the monodromy homomorphism associated to the flat connection
on $Y/\mbox{\raisebox{-.4ex}{$\Gamma$}}$. Then in general we only
know that $\rho^{\prime}$ is an extension of $\rho_0$ in the sense
that $\rho^{\prime}\circ\pr_{\ast}=\rho_0$. In our case, $\rho_0$
is trivial since $Y/\mbox{\raisebox{-.4ex}{$\Gamma_0$}}$ is a
trivial bundle. From the explicit expression of elements in
$\Gamma$, one has a well-defined chain of group homomorphisms:
$$
 \pi_1(Q-\Sigma_Q)\;
 \longrightarrow\; \pi_1(Q-\Sigma_Q)/
   \mbox{\raisebox{-.4ex}{$\pr_{\ast}
                             (\pi_1(Z-\pr^{-1}(\Sigma_Q)))$}}\,
 =\, \Gamma/\mbox{\raisebox{-.4ex}{$\Gamma_0$}}\;
 \stackrel{\jmath}{\longrightarrow}\;
   \Aut(X/\mbox{\raisebox{-.4ex}{$\Gamma_0$}})\,,
$$
where $\jmath$ is the projection of the
$\Gamma/\mbox{\raisebox{-.4ex}{$\Gamma_0$}}$-action on
$Y/\mbox{\raisebox{-.4ex}{$\Gamma_0$}}
 = Z\times(X/\mbox{\raisebox{-.4ex}{$\Gamma_0$}})$ to the
$X/\mbox{\raisebox{-.4ex}{$\Gamma_0$}}$ factor and
$\Aut(X/\mbox{\raisebox{-.4ex}{$\Gamma_0$}})$ is the group of
automorphisms of $X/\mbox{\raisebox{-.4ex}{$\Gamma_0$}}$.
By construction, the homomorphism from $\pi_1(Q-\Sigma_Q)$ to
$\Aut(X/\mbox{\raisebox{-.4ex}{$\Gamma_0$}})$ obtained by the
composition of these homomorphisms coincides with the monodromy
homomorphism $\rho^{\prime}$. Recall from Sec.\ 1 the tautological
homomorphism $\tau:\pi_1(Q-\Sigma_Q)\rightarrow\pi_1^{\rm orb}(Q)$,
then indeed $\rho^{\prime}=\rho\circ\tau$.

For Joyce manifolds of the first kind, $\Gamma$ is abelian;
hence $\rho^{\prime}$ descends to and is determined by a
homomorphism
$$
 \overline{\rho}^{\prime}\;:\;
   H_1(Q-\Sigma_Q;{\Bbb Z})\; \rightarrow\;
             \jmath(\Gamma/\mbox{\raisebox{-.4ex}{$\Gamma_0$}})\,.
$$
When the underlying topology of $Q$ is $S^3$, as in many examples,
$H_1(Q-\Sigma_Q;{\Bbb Z})$ is generated by the set ${\cal C}$ of
meridians associated to real co-dimension $2$ strata of
$\Sigma_Q$ and hence 
$\overline{\rho}^{\prime}(H_1(Q-\Sigma_Q;{\Bbb Z}))
   =\jmath(\Gamma/\mbox{\raisebox{-.4ex}{$\Gamma_0$}})$
is generated by $\overline{\rho}^{\prime}({\cal C})$,
which can be read off also directly from $\Gamma_p$
associated to points $p$ that lie in these strata of $\Sigma_Q$.

\noindent\hspace{14cm} $\Box$

\bigskip

\noindent
{\it Remark 2.1.4 [Conway's notation, toroidal $2$-orbifolds and
       the base of a.a.\ fibrations].}
There are several ways to understand the base $4$-orbifold
$Q^4$ of an a.a.\ fibration $\pi$ above. A most economic one is
to borrow the above discussion for $\pi$ and to regard $Q$ itself
as the total space of a flat bundle over a toroidal $2$-orbifold
$Q^2$. For this and future reference, let us introduce the
Conway's notation to label a $2$-orbifold in terms of its
underlying topology and the combinatorial structure of its
singular locus.
\begin{quote}
 \hspace{-1em}{\it Conway's notation}$\,$: (Cf.\ [Th1].)
 \begin{itemize}
  \item
  The default topology for the underlying space of a
  $2$-orbifold is $S^2$.

  \item
  Each $n_1n_2\cdots$ not preceded by a $^{\ast}$ indicates
  cone-points of order $n_1$, $n_2$, $\cdots$, that lie in the
  interior of the underlying topology.

  \item
  Each $^{\ast}n_1n_2\cdots$ indicates a disk is removed from
  $S^2$ to form a silvered boundary with a string of
  corner-reflectors of order $n_1$, $n_2$, $\cdots$ sitting
  along.

  \item
  Additional topology is indicated by the symbol $^{\circ}$.
  Before a $|\,$, $^{\circ}$ denotes a handle, while after $|$
  it denotes a cross-cap.
 \end{itemize}
\end{quote}
(Cf.\ {\sc Figure 2-1}.)
Though redundant, whenever available we shall combine the above
notation with the symbol for the underlying topology to denote
a $2$-orbifold. For example, $S^2(22)$ is the $2$-orbifold $(22)$
since its underlying topology is $S^2$ and $D^2(^{\ast})$ is the
silvered $2$-disk $(^{\ast})$, etc..

There are 17 toroidal $2$-orbifolds (e.g.\ [Mo] and [Th1]):

\bigskip

\begin{tabular}{|c|}  \hline
 ${\Bbb T^2(\circ|)}$ \hspace{1em} (torus)
    \raisebox{.6ex}{\rule{0em}{1em}}\\[.6ex] \hline
 \begin{tabular} {llll|ll}
  $S^2(2222)$  &  $S^2(236)$  &   $S^2(244)$  &  $S^2(333)$
    &  $A^2(^{\ast\ast})$  & (annulus)
      \raisebox{.6ex}{\rule{0em}{1em}} \\[.6ex] \cline{1-4}
  $D^2(^{\ast}2222)$ & $D^2(^{\ast}236)$
    & $D^2(^{\ast}244)$   & $D^2(^{\ast}333)$
    & $M^2(^{\ast|\circ})$  & (M\"{o}bius strip)
      \raisebox{.6ex}{\rule{0em}{1em}} \\[1ex]
  $D^2(22^{\ast})$   & &  $D^2(4^{\ast}2)$  & $D^2(3^{\ast}3)$
    & $K^2(|^{\circ\circ})$ & (Klein bottle) \\[1ex]
  $D^2(^{\ast}22)$  & & &
    & $P^2(22|^{\circ})$ & (projective plane) \\[.6ex] \hline
 \end{tabular}
\end{tabular}

\bigskip

\noindent
Some $2$-orbifolds $Q^2$ in the list together with a set of
generators for $\pi_1^{\rm orb}(Q^2)$ represented by
based-loops in $Q^2$ are illustrated in {\sc Figure 2-1}. 
\begin{figure}[htbp]
\setcaption{{\sc Figure} 2-1.
\baselineskip 14pt
  Some toroidal $2$-orbifolds $Q^2$ with their Conway notation.
  The boundary that is modelled on
  ${\Bbb C}/
   \mbox{\raisebox{-.4ex}{$\langle z\mapsto\overline{z}\rangle$}}$
  is indicated by a double-line. Such boundary is said to be
   {\it silvered}.
  A generating set of $\pi_1^{\rm orb}(Q^2)$ is represented by
   a collection of based loops.
} 
\centerline{\psfig{figure=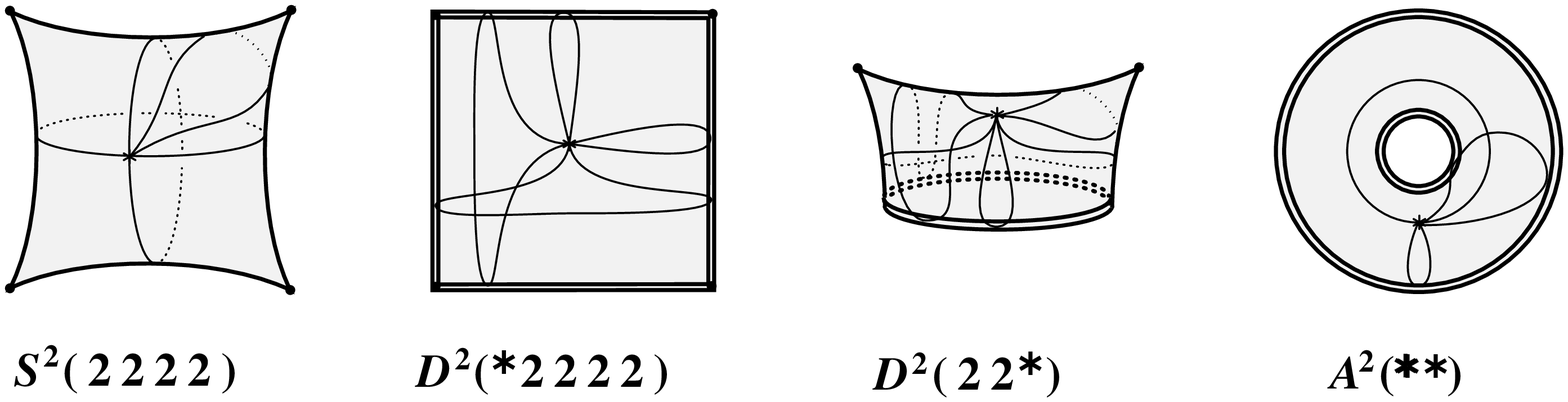,width=13cm,caption=}}
\end{figure}

With these facilities, the base $4$-orbifold $Q^4$ can be easily
visualized. Let us illustrate this by an example.
\begin{quote}
 \hspace{-1em}{\bf Example.}
 $Q^4_{3456}
  ={\Bbb T}^4_{3456}/\mbox{\raisebox{-.4ex}{$\Gamma_{3456}$}}$
 with $(c_3,c_5)=(0,0)$ (cf.\ [Jo1]: II, Example 4). Explicitly,
 ${\Bbb T}^4_{3456}$ is parametrized by $(x_3,x_4,x_5,x_6)$ and
 $\Gamma_{3456}=\langle\overline{\alpha},\overline{\beta},
  \overline{\gamma}\rangle$,
 where
 $$ {\small
 \begin{array}{ll}
  \overline{\alpha}(x_3,x_4,x_5,x_6)
                  =(-x_3,\,-x_4,\,x_5,\,x_6)\,,
    & \overline{\beta}(x_3,x_4,x_5,x_6)
                  =(x_3,\,x_4,\,-x_5,\,-x_6)\,, \\[.4ex]
  \overline{\gamma}(x_3,x_4,x_5,x_6)
                  =(-x_3,\,x_4,\,-x_5,\,x_6)\,. 
 \end{array} } 
 $$
 Regard ${\Bbb T}^4_{3456}$ as the trivial bundle
 ${\Bbb T}^2_{34}\times{\Bbb T}^2_{56}$, with base
 ${\Bbb T}^2_{34}$ and fiber ${\Bbb T}^2_{56}$, and
 $\Gamma_{3456}$ as a group of bundle automorphisms. Then all
 previous discussions for $\pi$ apply and one has a fibration
 $\pi^{3456}_{34}:Q^4_{3456}\rightarrow Q^2_{34}$. By inspection,
 $Q^2_{34}$ is a $D^2(^{\ast}2222)$ and the generic fiber an
 $S^2(2222)$. $\Sigma_p$ for $p\in\Sigma_{Q^2_{34}}$ is
 $\langle\overline{\beta},\overline{\gamma}\rangle=D_2$ and the
 exceptional fiber thereover is a $D^2(^{\ast}2222)$ of
 multiplicity $2$. The global structure of $\pi^{3456}_{34}$ is
 coded in the associated monodromy $\rho^{3456}_{34}$, which can
 be read off directly from the above explicit expression of
 $\overline{\alpha}$, $\overline{\beta}$, and $\overline{\gamma}$:
 With some abuse of notations, let
 $t_3(x_3,x_4)=(x_3+1,x_4)$ and $t_4(x_3,x_4+1)$ be translations
 on ${\Bbb R}^2_{34}$; then $\pi_1^{\rm orb}(Q^2_{34})$ is
 generated by $\overline{\overline{\alpha}}$,
 $\overline{\overline{\gamma}}$, $t_3$, and $t_4$. And
 $\rho^{3456}_{34}$ is determined by 
 $$
 \begin{array}{l}
  \rho^{3456}_{34}(\overline{\overline{\gamma}})\,
   =\,\overline{\gamma}^f\;
   =\;(\,[x_5,x_6]\mapsto[-x_5,x_6]\,)\,,
                        \hspace{2em}\mbox{while}\\[1ex]
  \rho^{3456}_{34}(\overline{\overline{\alpha}})\,
   =\,\rho^{3456}_{34}(t_3)\, =\,\rho^{3456}_{34}(t_4)\,
   =\,\Id\,.
 \end{array}
 $$
 ({\sc Figure 2-2}.)
 \begin{figure}[htbp]
 \setcaption{{\sc Figure 2-2}.
 \baselineskip 14pt
  The monodromy along the loop representing
  $\overline{\overline{\gamma}}$. Notice that the letter ``R"
  on the front face of the $S^2(2222)$-fiber over $\ast$ is
  sent to the back face after the round trip along
  $\overline{\overline{\gamma}}$.
 } 
 \centerline{\psfig{figure=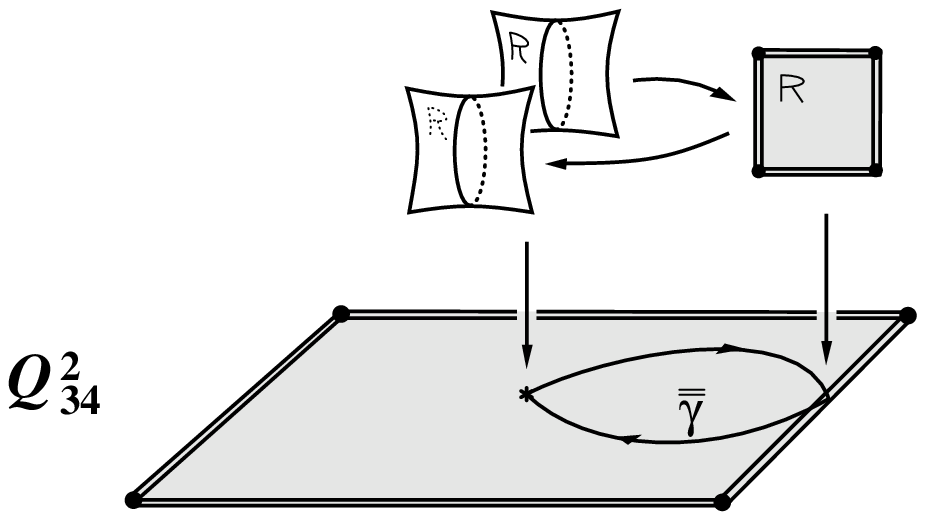,width=13cm,caption=}}
 \end{figure}
\end{quote}

The same method can also be applied to understand an exceptional
fiber of an a.c.\ fibration. This concludes the remark.

\noindent\hspace{14cm} $\Box$

\bigskip

\subsection{Adjustment to the fibration and the monodromy
            after resolving $S$.}

So far the discussion has been centered around the toroidal
orbifold ${\Bbb T}^7/\mbox{\raisebox{-.4ex}{$\Gamma$}}$. To
relate it to the Joyce manifold $M^7$ obtained by resolving
the singular locus $S$ of
${\Bbb T}^7/\mbox{\raisebox{-.4ex}{$\Gamma$}}$, one needs to
understand the effect to $\pi$ when resolving $S$. This can be
accomplished by understanding
\begin{quote}
 \hspace{-1.9em}(1)\hspace{1ex}
  how components of $S$ sit in
  ${\Bbb T}^7/\mbox{\raisebox{-.4ex}{$\Gamma$}}$ over $Q$,
  which gives the information of how components in $S$ intersect
  the fibers of $\pi$, and

 \hspace{-1.9em}(2)\hspace{1ex}
  the restriction of the exceptional locus to a fiber
  when resolving $S$, which gives the information of the new fiber
  after resolving $S$.
\end{quote}
Adjustments to the base $Q$ and the monodromy $\rho$ will then
follow.

\bigskip

\noindent
{\it Remark 2.2.1.}
Recall that a component of $S$ comes from a fixed ${\Bbb T}^3$
of some element in $\Gamma$. For Joyce manifolds of the first
kind, any such fixed ${\Bbb T}^3$ is a coordinate $3$-torus
${\Bbb T}^3_{a^{\prime}b^{\prime}c^{\prime}}$.  Much of the answer
to Item (1) above follows from the label
$(a^{\prime}b^{\prime}c^{\prime})$. This suggests a computer
code to assist the work. (Cf.\ Sec.\ 5, Issue (1).)

\bigskip

Let us now consider the fibrations case by case.

\bigskip

\begin{flushleft}
{\bf a.a.\ fibrations for a Joyce manifold of the first kind.}
\end{flushleft}
For such fibrations, $Y=Z\times X$ with $Z={\Bbb T}^4$,
$X={\Bbb T}^3$, and $\Gamma_0$ is always trivial. Hence the generic
fiber of $\pi$ is always ${\Bbb T}^3$ and the singular set $S$ has
to lie over the singular locus $\Sigma_Q$ of $Q$. Consider now what
happens to the fibers of $\pi$ when resolving $S$. Since a generic
fiber does not intersect $S$, resolving $S$ does not affect it. Let
$F_p$ now be a fiber over a point $p$ in $\Sigma_Q$ and $S_0$ be a
component of $S$. Let $\widetilde{F_p}$, $\widetilde{S_0}$ be a
connected component of their respective preimage in $Y$. Then
$\widetilde{F_p}$ is one of ${\Bbb T}^3_{127}$, ${\Bbb T}^3_{136}$,
${\Bbb T}^3_{145}$, ${\Bbb T}^3_{235}$, ${\Bbb T}^3_{246}$,
${\Bbb T}^3_{347}$, and ${\Bbb T}^3_{567}$. On the other hand,
since $\widetilde{S_0}$ is a fixed ${\Bbb T}^3$ for some element
of $\Gamma$, it can only be one of ${\Bbb T}^3_{567}$,
${\Bbb T}^3_{347}$, ${\Bbb T}^3_{246}$, ${\Bbb T}^3_{127}$,
${\Bbb T}^3_{136}$, ${\Bbb T}^3_{145}$, and ${\Bbb T}^3_{235}$.
Comparing the two lists, one has that
$\widetilde{F_p}\cap\widetilde{S_0}$ can only be empty,
${\Bbb T}^1_{i}$, or ${\Bbb T}^3_{a^{\prime}b^{\prime}c^{\prime}}$,
in the last case $\widetilde{F_p}=\widetilde{S_0}$.
Together with the local models (a) and (b) of $\nu(S_0)$ described
in Sec.\ 1, one concludes that the possible $F_p\cap S_0$ and the
effect to $F_p$ when resolving $S_0$ can only be one of the
following:

\bigskip

\noindent \hspace{1ex}
{\it (a) When} $\;\nu(S_0)={\Bbb T}^3\times({\Bbb C}^2/
               \mbox{\raisebox{-.4ex}{$\langle -1\rangle$}})\,$:
\begin{itemize}
 \item
 {\it $F_p\cap S_0$ empty}$\,$:
 No effect on $F_p$ when resolving $S_0$.
  
 \item
 {\it $F_p\cap S_0\;=$ some disjoint $n$ copies of 
      ${\Bbb T}^1$}$\,$:
 After resolving $S_0$, the new fiber $F^{\prime}_p$ at $p$
 becomes $F_p\cup\,n\,{\Bbb T}^1\times{\Bbb C}{\rm P}^1$ with
 normal crossing singularities along $F_p\cap S_0$.

 \item
 {\it $F_p=S_0$}$\,$:
 Then $p$ is an isolated $A_1$-singularity of $Q$ and resolving
 $S_0$ changes the neighborhood
 $\nu(p)
  \cong{\Bbb C}^2/\mbox{\raisebox{-.4ex}{$\langle -1\rangle$}}$
 of $p$ in $Q$ to
 $\widetilde{\nu(p)}\cong T^{\ast}{\Bbb C}{\rm P}^1$.
 Recall that a fiber of $T^{\ast}{\Bbb C}{\rm P}^1$ here is
 regarded as a subcone
 ${\Bbb C}/\mbox{\raisebox{-.4ex}{$\langle -1\rangle$}}$ in
 ${\Bbb C}^2/\mbox{\raisebox{-.4ex}{$\langle -1\rangle$}}$ and
 the union of their ${\Bbb Z}_2$-cone points form the exceptional
 locus $E$ of the resolution, which is the $0$-section of
 $T^{\ast}{\Bbb C}{\rm P}^1$. Thus, after resolving $S_0$, $Q$ is
 changed to a new orbifold $Q^{\prime}$ with singular locus
 $\Sigma_{Q^{\prime}}=(\Sigma_Q-\{p\})\cup E$, where
 $\Gamma_p={\Bbb Z}_2$ for $p\in E$. The associated new
 fibration inherited from $\pi$ is the product
 $\widetilde{\nu(p)}\times{\Bbb T}^3$ over $\widetilde{\nu(p)}$.
\end{itemize}

\noindent\hspace{1ex}
{\it (b) When} $\;\nu(S_0)=\{{\Bbb T}^3\times({\Bbb C}^2/
                \mbox{\raisebox{-.4ex}{$\langle -1\rangle$}})\}/
                         \mbox{\raisebox{-.4ex}{${\Bbb Z}_2$}}\,$:
\begin{itemize}
 \item
 {\it $F_p\cap S_0$ empty}$\,$:
 No effect on $F_p$ when resolving $S_0$.

 \item
 {\it $F_p\cap S_0\;=$ some disjoint $n$ copies of
      ${\Bbb T}^1$}$\,$:
 Since each component of $F_p\cap S_0$ descends from the
 intersection ${\Bbb T}^1$ of two $3$-tori in $Y$, $F_p\cap S_0$
 is liftable to
 ${\Bbb T}^3\times({\Bbb C}^2/
                 \mbox{\raisebox{-.4ex}{$\langle -1\rangle$}})$
 and the induced ${\Bbb Z}_2$-action on the set of lifted
 ${\Bbb T}^1$ in
 ${\Bbb T}^3\times({\Bbb C}^2/
                   \mbox{\raisebox{-.4ex}{$\langle -1\rangle$}})$
 can only be either free or trivial. In the former case, the new
 fiber $F_p^{\prime}$ at $p$ after resolving $S_0$ becomes
 $F_p\cup\,n\,{\Bbb T}^1\times{\Bbb C}{\rm P}^1$
 with normal crossing singularities along $F_p\cap S_0$, while 
 in the latter case it becomes 
 $F_p\cup\,n\,({\Bbb T}^1\times{\Bbb C}{\rm P}^1)/
                    \mbox{\raisebox{-.4ex}{${\Bbb Z}_2$}}$
 with normal crossing singularities along $F_p\cap S_0$.
 
 \item
 {\it $F_p=S_0$}$\,$:
 In this case $F_p$ is the fiber over $(0,0)$ in the natural
 ${\Bbb T}^3$-fibration of
 $\{{\Bbb T}^3\times({\Bbb C}^2/
    \mbox{\raisebox{-.4ex}{$\langle -1\rangle$}})\}/
       \mbox{\raisebox{-.4ex}{${\Bbb Z}_2$}}$ over
 $\nu(p)= {\Bbb C}^2/
  \mbox{\raisebox{-.4ex}{$(\langle -1\rangle\oplus{\Bbb Z}_2)$}}$,
 where the ${\Bbb Z}_2$ acts on ${\Bbb C}^2$ either
 holomorphically, as generated by $(z_1,z_2)\mapsto(z_1,-z_2)$, or
 antiholomorphically, as generated by 
 $(z_1,z_2)\mapsto(\overline{z_1},\overline{z_2})$, following the
 holomorphicity of the ${\Bbb Z}_2$-action on
 ${\Bbb T}^3\times({\Bbb C}^2/
             \mbox{\raisebox{-.4ex}{$\langle -1\rangle$}})$.
 Let us consider these two cases separately:
 \begin{itemize}
  \item
  {\it ${\Bbb Z}_2$-action holomorphic}$\,$:
  Then $\Sigma_Q\cap\nu(p)$ consists of the image of
  $\{z_1z_2=0\}$ in 
  ${\Bbb C}^2/
   \mbox{\raisebox{-.4ex}{$(\langle -1\rangle\oplus{\Bbb Z}_2)$}}$,
  which is isomorphic to the neighborhood of a double point in
  a complex curve. $\Gamma_p={\Bbb Z}_2$ for $p$ regular point of
  $\Sigma_Q\cap\nu(p)$ and $=\langle -1\rangle\oplus{\Bbb Z}_2$
  for $p$ the double point. The fiber over $\Sigma_Q\cap\nu(p)$ is
  ${\Bbb T^3}/\mbox{\raisebox{-.4ex}{${\Bbb Z}_2$}}$. When $S_0$
  is resolved, the base $Q$ is changed to a new $Q^{\prime}$ with
  $\nu(p)$ resolved to
  $\widetilde{\nu(p)}=T^{\ast}{\Bbb C}{\rm P}^1/
                         \mbox{\raisebox{-.4ex}{${\Bbb Z}_2$}}$.
  The exceptional locus $E$ of the resolution is 
  ${\Bbb C}{\rm P}^1/\mbox{\raisebox{-.4ex}{${\Bbb Z}_2$}}$, where
  ${\Bbb C}{\rm P}^1={\Bbb C}\cup\{\infty\}$ is the $0$-section of
  $T^{\ast}{\Bbb C}{\rm P}^1$ parametrized by $z$ and the
  ${\Bbb Z}_2$-action is generated by $z\mapsto -z$.
  Since the locus $\{z_1z_2=0\}$ selects two complex lines in
  $T_{(0,0)}{\Bbb C}^2$ that are inequivalent under the
  $\langle -1\rangle\oplus{\Bbb Z}_2$-action, {\it the new singular
  locus $\Sigma_{Q^{\prime}}$ is the union of the resolution of
  $\Sigma_Q$ at the double point by normalization and the new
  component $E$ as in} {\sc Figure 2-3} (i).
  $\Gamma_p=\langle -1\rangle$ for $p\in E-\{p_0,p_{\infty}\}$ and 
  $=\langle -1\rangle\oplus{\Bbb Z}_2$ for $p=p_0$ or $p_{\infty}$.
  (Intrinsically $E$ is an $S^2(22)$-orbifold.)

  \item
  {\it ${\Bbb Z}_2$-action antiholomorphic}$\,$:
  Then $\Sigma_Q\cap\nu(p)$ consists of the image of
  $$
   C\;=\; \{\,(z_1,z_2)\,|\,
      \mbox{$z_1$, $z_2$ both real or both purely imaginary}\,\}
  $$
  in
  ${\Bbb C}^2/
   \mbox{\raisebox{-.4ex}{$(\langle -1\rangle\oplus{\Bbb Z}_2)$}}$,
  whose topology is still the same as a neighborhood of a double
  point in a complex curve. $\Gamma_p={\Bbb Z}_2$ for $p$ regular
  point of $\Sigma_Q\cap\nu(p)$ and
  $=\langle -1\rangle\oplus{\Bbb Z}_2$ for $p$ the double point.
  The fiber over $\Sigma_Q\cap\nu(p)$ is
  ${\Bbb T^3}/\mbox{\raisebox{-.4ex}{${\Bbb Z}_2$}}$.
  When $S_0$ is resolved, the base $Q$ is changed to
  $Q^{\prime}$ with $\nu(p)$ resolved to
  $\widetilde{\nu(p)}=T^{\ast}{\Bbb C}{\rm P}^1/
                      \mbox{\raisebox{-.4ex}{${\Bbb Z}_2$}}$.
  The exceptional locus $E$ of the resolution is
  ${\Bbb C}{\rm P}^1/\mbox{\raisebox{-.4ex}{${\Bbb Z}_2$}}$, where
  ${\Bbb C}{\rm P}^1={\Bbb C}\cup\{\infty\}$ is the $0$-section of
  $T^{\ast}{\Bbb C}{\rm P}^1$ parametrized by $z$ and the
  ${\Bbb Z}_2$-action is generated by $z\mapsto\overline{z}$.
  The locus $C$ now selects an $S_1$-family of complex lines in
  $T_{(0,0)}{\Bbb C}^2$ that are inequivalent under the
  $\langle -1\rangle\oplus{\Bbb Z}_2$-action; hence, {\it the new
  singular locus $\Sigma_{Q^{\prime}}$ is the union of the
  resolution of $\Sigma_Q$ at the double point by tubing and the
  new component $E$ in} {\sc Figure 2-3} (ii).
  $\Gamma_p={\Bbb Z}_2$ for $p$ in the interior of $E$ and
  $=\langle -1\rangle\oplus{\Bbb Z}_2$ for $p$ in its boundary.
  (Intrinsically $E$ is a silvered $2$-disk $D^2(^{\ast})$.)
 \end{itemize}
 \begin{figure}[htbp]
 \setcaption{{\sc Figure 2-3}
 \baselineskip 14pt
  The two local models for $\nu(S_0)$ in Case (b) lead to different
  resolutions $\widetilde{\nu(p)}$ of $\nu(p)$. The shaded part is
  the exceptional locus $E$, which becomes an additional component
  in $\Sigma_{Q^{\prime}}$. In (i), $E$ is an $S^2(22)$-orbifold,
  while in (ii) $E$ is a silvered $2$-disk. The singular locus
  $\Sigma_Q\cap\nu(p)$ and what it is resolved to are also
  indicated.
 } 
 \centerline{\psfig{figure=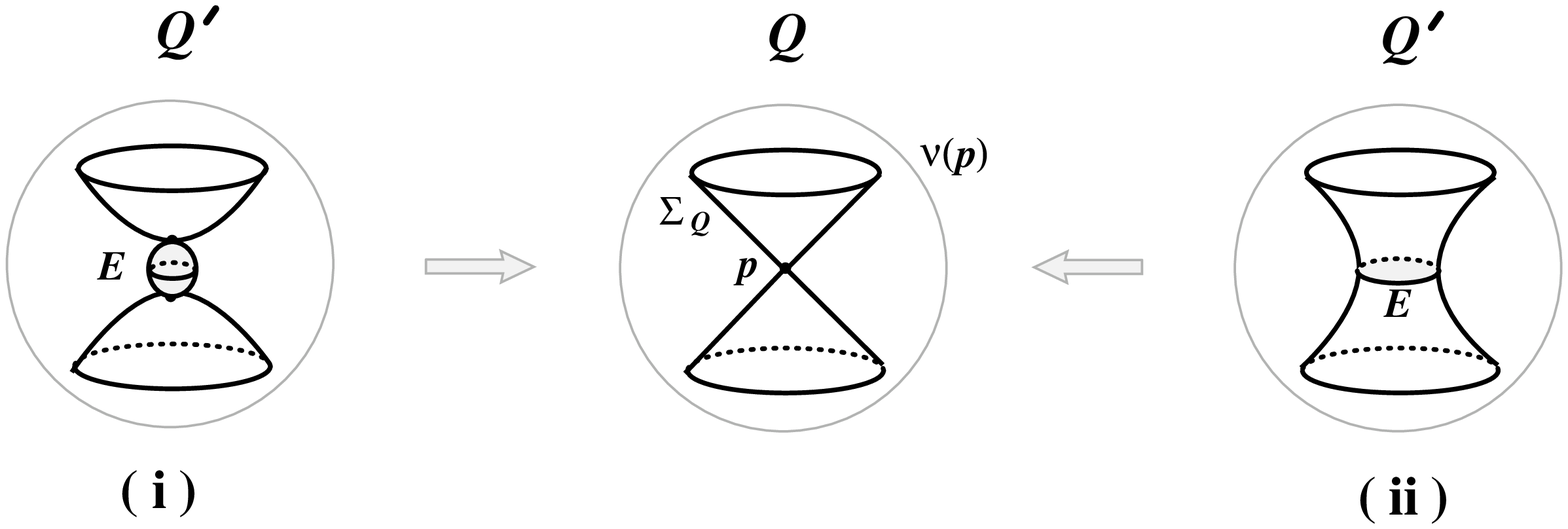,width=13cm,caption=}}
 \end{figure}
\end{itemize}

\bigskip

\begin{flushleft}
{\bf a.c.\ fibrations for a Joyce manifold of the first kind.}
\end{flushleft}
For such fibrations, $Y=Z\times X$ with $Z={\Bbb T}^3$ and
$X={\Bbb T}^4$. One can check that there are two cases:
\begin{quote}
 \hspace{-1.9em}(1)\hspace{1ex} {\it $\Gamma_0=0\,$}:
 In this case $Y/\mbox{\raisebox{-.4ex}{$\Gamma$}}$ is a
 ${\Bbb T}^4$-fibration over $Q$ and $S$ lies over $\Sigma_Q$.

 \hspace{-1.9em}(2)\hspace{1ex}
 {\it $\Gamma_0=\langle g\rangle={\Bbb Z}_2\,$ for some $g$ in
      $\Gamma$}$\,$:
 In this case $g$ acts on $X={\Bbb T}^4$ by negation and the
 generic fiber of the fibration
 $\pi:Y/\mbox{\raisebox{-.4ex}{$\Gamma$}}
    \rightarrow Q=Z/\mbox{\raisebox{-.4ex}{$\Gamma$}}$
 is the singular K3 surface
 $X/\mbox{\raisebox{-.4ex}{$\langle g\rangle$}}$.
 The components of $S$ that are associated to the fixed $3$-tori
 of $g$ now appear as multi-sections of $\pi$ while all other
 components of $S$ must sit over $\Sigma_Q$.
\end{quote}
Let us now consider what happens to a fiber of $\pi$ when
resolving $S$.

Consider first the special components of $S$ that come from the
fixed $3$-tori of $g$ in Case (2) above. Since their preimages
in $Y$ are transverse to $X$, they are transverse to all fibers in
$Y/\mbox{\raisebox{-.4ex}{$\Gamma$}}$. Conversely, if a componnet
${\Bbb T}^3$ of $S$ is transverse to some fiber, then its preimage
in $Y$ must be transverse to $X$ and hence must come from a fixed
${\Bbb T}^3$ of $g$. Thus they accounts for all the isolated
singularities in a fiber - generic or exceptional alike - of
$\pi$. Resolving such components of $S$ will resolve
simultaneously all the isolated singularities in fibers and,
hence, leads to a {\it K3-fibration} for the Joyce
manifold $M^7$.
                                                                                                                                             
For all other components of $S$, let $F_p$ be a fiber of $\pi$
over $p$ and $S_0$ be a component of $S$ that lies over
$\Sigma_Q$. Let $\widetilde{F_p}$, $\widetilde{S_0}$ be a
connected component of their respective preimage in $Y$. A similar
consideration as in the discussion for a.a.\ fibrations implies
that $\widetilde{F_p}$ is a coordinate ${\Bbb T}^4$,
$\widetilde{S_0}$ is a coordinate ${\Bbb T}^3$, and their
intersection can only be either empty or a ${\Bbb T}^2_{ij}$.
Together with the local model (a) and (b) of $\nu(S_0)$ described
in Sec.\ 1, one concludes that the possible connected component of
$F_p\cap S_0$ and the effect to $F_p$ when resolving $S_0$ can
only be one of the following:

\bigskip

\noindent\hspace{1ex}
{\it (a) When} $\;\nu(S_0)={\Bbb T}^3\times({\Bbb C}^2/
               \mbox{\raisebox{-.4ex}{$\langle -1\rangle$}})\,$:
\begin{itemize}
 \item
 {\it $F_p\cap S_0$ empty}$\,$:
 No effect on $F_p$ when resolving $S_0$.

 \item
 {\it $F_p\cap S_0\;=$
      some disjoint $n$ copies of ${\Bbb T}^2$}$\,$:
 After resolving $S_0$, the new fiber $F^{\prime}_p$ at $p$
 becomes $F_p\cup\,n\,{\Bbb T}^2\times{\Bbb C}{\rm P}^1$ with
 normal crossing singularities along $F_p\cap S_0$.
\end{itemize}

\noindent\hspace{1ex}
{\it (b) When} $\;\nu(S_0)=\{{\Bbb T}^3\times({\Bbb C}^2/
            \mbox{\raisebox{-.4ex}{$\langle -1\rangle$}})\}/
                        \mbox{\raisebox{-.4ex}{${\Bbb Z}_2$}}\,$:
\begin{itemize}
 \item
 {\it $F_p\cap S_0$ empty}$\,$:
 No effect on $F_p$ when resolving $S_0$.

 \item
 {\it $F_p\cap S_0\;=$
      some disjoint $n$ copies of ${\Bbb T}^2$}$\,$:
 Afain $F_p\cap S_0$ is liftable to
  ${\Bbb T}^3\times({\Bbb C}^2/
             \mbox{\raisebox{-.4ex}{$\langle -1\rangle$}})$
 and the induced ${\Bbb Z}_2$-action on the set of lifted
 ${\Bbb T}^2$ in
 ${\Bbb T}^3\times({\Bbb C}^2/
             \mbox{\raisebox{-.4ex}{$\langle -1\rangle$}})$
 is either free or trivial. In the former case, the new
 fiber $F_p^{\prime}$ at $p$ after resolving $S_0$ becomes
 $F_p\cup\,n\,{\Bbb T}^2\times{\Bbb C}{\rm P}^1$
 with normal crossing singularities along $F_p\cap S_0$, while
 in the latter case it becomes
 $F_p\cup\,n\,({\Bbb T}^2\times{\Bbb C}{\rm P}^1)/
                    \mbox{\raisebox{-.4ex}{${\Bbb Z}_2$}}$
 with normal crossing singularities along $F_p\cap S_0$.
\end{itemize}

\bigskip

\begin{flushleft}
{\bf a.a./a.c.\ fibrations for Joyce manifolds of the second kind.}
\end{flushleft}
For such fibrations, $\Gamma_0$ is always trivial. The singular
set $S$ of $Y/\mbox{\raisebox{-.4ex}{$\Gamma$}}$ sits over
$\Sigma_Q$. From Sec.\ 1, the tubular neighborhood $\nu(S_0)$
of a component $S_0$ of $S$ is always homeomorphic to
${\Bbb T}^3\times({\Bbb C}^2/
             \mbox{\raisebox{-.4ex}{$\langle -1\rangle$}})$;
thus, if a fiber $F_p$ happens to be some $S_0$ in an a.a.\
fibration $\pi$, resolving $S_0$ changes $Q$ to $Q^{\prime}$
obtained by blowing up $Q$ at $p$. Other possible intersections
of $F_p$ and $S_0$ and the effect to $F_p$ when resolving $S_0$
are similar to the discussion for Joyce manifolds of the first
kind. For such Joyce manifold, an a.a.\ fibration is a
${\Bbb T}^3$-fibration and an a.c.\ fibration is a
${\Bbb T}^4$-fibration.

\bigskip

\begin{flushleft}
{\bf Adjustment to the monodromy.}
\end{flushleft}
Let $\widetilde{\pi}:M^7\rightarrow\widetilde{Q}$ be the induced
fibration from
$\pi:Y/\mbox{\raisebox{-.4ex}{$\Gamma$}}\rightarrow Q$
after resolving $S$. Then the flat connection on $\pi$ can be
lifted to a flat connection on $\widetilde{\pi}$.
(For $\widetilde{\pi}$ a K3-fibration, see more details below.)
Let $\widetilde{\rho}$ be the associated monodromy of
$\widetilde{\pi}$.

\bigskip

\noindent
{\it (i) When $\pi$ is an a.a.\ fibration}$\,$:

\medskip

Both $\pi$ and $\widetilde{\pi}$ are ${\Bbb T}^3$-fibrations.
When $\pi$ contains no component of $S$ as a fiber, then
$\widetilde{Q}=Q$ and the restriction of $\widetilde{\pi}$ to
$\widetilde{Q}-\Sigma_{\widetilde{Q}}$ coincides
with the restriction of $\pi$ to $Q-\Sigma_Q$. Thus
$\widetilde{\rho}=\rho$.

On the other hand, if $\pi$ contains some components of $S$ as
fibers, then one has a resolution $r:\widetilde{Q}\rightarrow Q$
as discussed earlier. Since the restriction of $\widetilde{\pi}$
to $\widetilde{Q}-\Sigma_{\widetilde{Q}}$ still coincides with
the restriction of $\pi$ to $Q-\Sigma_Q$, 
$\widetilde{\rho}=\rho\circ r_{\ast}$, where
$r_{\ast}:\pi_{1}^{\rm orb}(\widetilde{Q})
                                \rightarrow\pi_1^{\rm orb}(Q)$
is the induced homomorphism of $r$.

\bigskip

\noindent
{\it Remark 2.2.1.}
 From the local model of $\nu(S)$, the monodromy along a meridian
 $C$ associated to an exceptional locus $E$ in
 $\Sigma_{\widetilde{Q}}$ of the resolution $r$ above is always
 trivial since $C$ is homologous to the unit circle in a fiber of
 $T^{\ast}{\Bbb C}{\rm P}^1$ and the fibration is trivial when
 restricted thereover.

\bigskip

\noindent
{\it (ii) When $\pi$ is an a.c.\ fibration}$\,$:

\medskip

For such fibration, $\widetilde{Q}$ and $Q$ are always the same.
If $\Gamma_0=\{0\}$, then both $\pi$ and $\widetilde{\pi}$ are
${\Bbb T}^4$-fibrations and the restriction of $\widetilde{\pi}$
to $\widetilde{Q}-\Sigma_{\widetilde{Q}}$ coincides with the
restriction of $\pi$ to $Q-\Sigma_Q$. Thus
$\widetilde{\rho}=\rho$.

On the other hand, if $\Gamma_0={\Bbb Z}_2$, then resolving
$S$ - though does not change the base $Q$ - alters the generic
fiber from a toroidal $4$-orbifold to a smooth K3 surface via
the resolution of $A_1$-singularities
$f:\widetilde{X/\mbox{\raisebox{-.4ex}{$\langle -1\rangle$}}}
   \rightarrow X/\mbox{\raisebox{-.4ex}{$\langle -1\rangle$}}$.
However, from Sec.\ 1, as bundle automorphisms of $Y$ over $Z$,
elements in $\Gamma$ restricted to fibers are either holomorphic
or anti-holomorphic. Thus $\Gamma$ can be lifted to a group of
bundle automorphisms of $Y^{\prime}=Z\times\widetilde{X}$ over
$Z$, where $\widetilde{X}$ is the blow up of $X$ at the fixed
points of $\Gamma_0$. The flat connection on $Y^{\prime}$ induces
then a flat connection on
$\widetilde{\pi}$ which is compatible with the flat connection
on $Y/\mbox{\raisebox{-.4ex}{$\Gamma$}}$ under the resolution
$M^7\rightarrow Y/\mbox{\raisebox{-.4ex}{$\Gamma$}}$ that resolves
$S$; hence $\widetilde{\rho}$ is a lifting of $\rho$. Indeed,
if one recalls Remark 2.1.3, then since any map in
$\jmath(\Gamma/\mbox{\raisebox{-.4ex}{$\Gamma_0$}})$
is either holomorphic or anti-holomorphic and the resolution
$\widetilde{X/\mbox{\raisebox{-.4ex}{$\langle -1\rangle$}}}
  \rightarrow X/\mbox{\raisebox{-.4ex}{$\langle -1\rangle$}}$
is through blowups, 
$\jmath(\Gamma/\mbox{\raisebox{-.4ex}{$\Gamma_0$}})$ is liftable
in a unique way into
$\Aut(\widetilde{X/\mbox{\raisebox{-.4ex}{$\langle -1\rangle$}}})$,
which gives the monodromy $\widetilde{\rho}$ of 
$\widetilde{\pi}$.

\bigskip

This concludes our general discussions for the a.a./a.c.\
fibrations of a Joyce manifold.

\bigskip

\section{Examples in the 5-step-routine.}
 
The discussion in Sec.\ 2 suggests a {\it 5-step-routine} to
understand the a.a./a.c.\ fibrations of any given Joyce manifold.
In this section, we choose  
$J(0,\frac{1}{2},\frac{1}{2},\frac{1}{2},0)$ in [Jo1] (I) and
$J(e^{\pi i/3},e^{2\pi i/3},\Lambda)$ in [Jo1] (II) to illuminate
this procedure.

\bigskip

\noindent
{\bf Example 3.1 [$J(0,\frac{1}{2},\frac{1}{2},\frac{1}{2},0)\,$:
     a.c.\ K3-fibration].} (Cf.\ [Jo1]: I; also II, Example 3.)
\newline
$\Gamma$ is generated by
$$
\begin{array}{rcl}
 \alpha(x_1,\cdots,x_7) & =
     & (-x_1,-x_2,-x_3,-x_4,x_5,x_6,x_7)\,, \\[.2ex]
 \beta(x_1,\cdots,x_7) & =
     & (-x_1,\frac{1}{2}-x_2,x_3,x_4,-x_5,-x_6,x_7)\,, \\[.2ex]
 \gamma(x_1,\cdots,x_7) & =
     & (\frac{1}{2}-x_1,x_2,\frac{1}{2}-x_3,x_4,-x_5,x_6,-x_7)\,.
\end{array}
$$
From [Jo1] the quotient
${\Bbb T}^7/\mbox{\raisebox{-.4ex}{$\Gamma$}}$ has a singular
set $S$ of $A_1$-singularities that consists of $12$ disjoint
${\Bbb T}^3$ arising from the fixed ${\Bbb T}^3$ of $\alpha$,
$\beta$, and $\gamma$. The tubular neighborhood of each component
of $S$ is modelled on
${\Bbb T}^3\times(\,{\Bbb C}^2/
                 \mbox{\raisebox{-.4ex}{$\langle -1\rangle$}})$.
After resolving $S$, one obtains a Joyce manifold.

Consider the associative-coassociative decomposition
$Y={\Bbb T}^7={\Bbb T}^3_{567}\times{\Bbb T}^4_{1234}$
with $Z={\Bbb T}^3_{567}$ and $X={\Bbb T}^4_{1234}$.
One then has the fibration
$\pi_{567}:{\Bbb T}^7/\mbox{\raisebox{-.4ex}{$\Gamma$}}
                                           \rightarrow Q^3_{567}$,
where $Q^3_{567}$ is the orbifold
${\Bbb T}^3_{567}/\mbox{\raisebox{-.4ex}{
     $\langle\overline{\beta},\overline{\gamma}\rangle$}}$ with 
$\overline{\beta}(x_5,x_6,x_7)=(-x_5,-x_6,x_7)$ and
$\overline{\gamma}(x_5,x_6,x_7)=(-x_5,x_6,-x_7)$.

\bigskip

\noindent
{\it (i) The base orbifold $Q^3_{567}\,$:}
By choosing a fundamental domain $\Omega$ of the
$\langle\overline{\beta},\overline{\gamma}\rangle$-action on
${\Bbb T}^3_{567}$ to be, e.g.,
$[0,1]\times[0,\frac{1}{2}]\times[0,\frac{1}{2}]$ and examining
how $\langle\overline{\beta},\overline{\gamma}\rangle$ acts on
the boundary $\partial\Omega$, one concludes that $Q^3_{567}$ is
a Euclidean $3$-orbifold whose underlying topology is $S^3$ and
whose singular locus $\Sigma_Q$ is the $1$-skeleton of a $3$-cube.
The group $\Gamma_p$ associated to an interior point $p$ of each
edge of $\Sigma_Q$ is ${\Bbb Z}_2$ - generated by
$\overline{\beta}\overline{\gamma}$, $\overline{\gamma}$, or 
$\overline{\beta}$, depending on whether the edge is parallel to
$x_5$-, $x_6$-, or $x_7$-axis respectively -
and to each vertex
${\Bbb Z}_2\oplus{\Bbb Z}_2
   =\langle\overline{\beta},\overline{\gamma}\rangle$.
({\sc Figure 3-1-1}.)

\begin{figure}[htbp]
 \setcaption{{\sc Figure 3-1-1}.
 \baselineskip 14pt
  After the identification of $\partial\Omega$ governed by
  $\langle\overline{\beta},\overline{\gamma}\rangle$,
  one obtains the Euclidean orbifold $Q^3_{567}$ with underlying
  topology $S^3$ and singular locus $\Sigma_Q$ the $1$-skeleton
  of a $3$-cube.
 } 
 \centerline{\psfig{figure=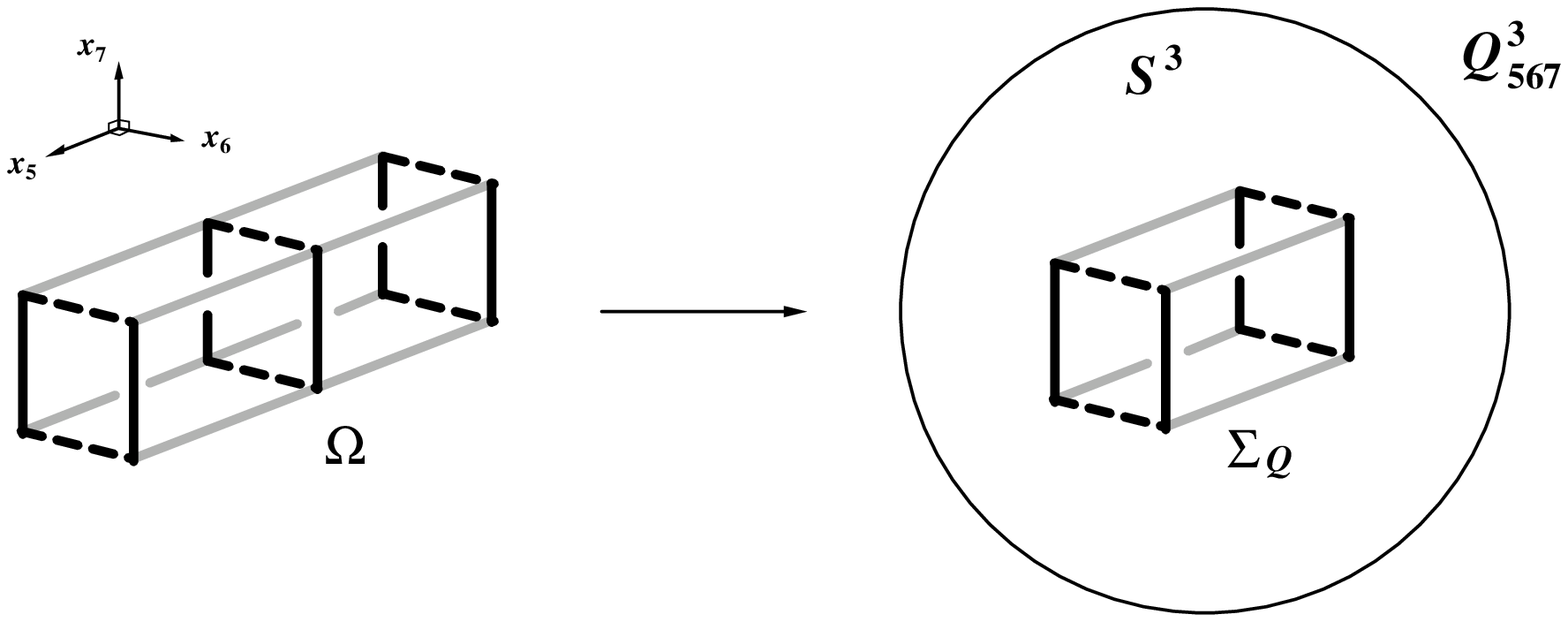,width=13cm,caption=}}
\end{figure}
 
\bigskip

\noindent
{\it (ii) The fibers}$\,$:
$\Gamma_0={\Bbb Z}_2=\langle\alpha\rangle$ in this example. 
The fiber $F_p$ of the fibration $\pi_{567}$ over a point $p$ in
$Q^3_{567}$ is listed below:
\begin{itemize}
 \item
 For $p\in Q^3_{567}-\Sigma_Q$,
 $$
  F_p\; =\; X_0\; =\;{\Bbb T}^7_{1234}/
    \mbox{\raisebox{-.4ex}{$\langle\alpha\rangle$}}\;
  =\; {\Bbb T}^4_{1234}/\mbox{\raisebox{-.4ex}{\small$\langle
     (x_1,x_2,x_3,x_4)\mapsto(-x_1,-x_2,-x_3,-x_4)\rangle$}}\,,
 $$
 which is a singular K3 surface with $16$ $A_1$-singularities.

 \item
 For $p$ an interior point of an edge of $\Sigma_Q$,
 $\Gamma_p$ is
 $\langle\overline{\alpha},
                   \overline{\beta}\overline{\gamma}\rangle$,
 $\langle\overline{\alpha},\overline{\gamma}\rangle$, or 
 $\langle\overline{\alpha},\overline{\beta}\rangle$, depending
 on whether the edge is parallel to $x_5$-, $x_6$-, or $x_7$-axis
 respectively. Thus $F_p=X_e$ is the (isomorphic)
 ${\Bbb Z}_2$-quotient
 $X_0/\mbox{\raisebox{-.4ex}{$\langle\beta\gamma\rangle$}}$,
 $X_0/\mbox{\raisebox{-.4ex}{$\langle\gamma\rangle$}}$, or
 $X_0/\mbox{\raisebox{-.4ex}{$\langle\beta\rangle$}}$ accordingly,
 where it is understood that, e.g.\
 \begin{eqnarray*}
  \lefteqn{
   X_0/\mbox{\raisebox{-.4ex}{$\langle\beta\gamma\rangle$}}\;
   =\; {\Bbb T}^7_{1234}/
    \mbox{\raisebox{-.4ex}{$\langle\alpha,\beta\gamma\rangle$}} }\\
   & =\;{\Bbb T}^4_{1234}/\mbox{\raisebox{-.4ex}{\footnotesize
    $\langle(x_1,x_2,x_3,x_4)\mapsto(-x_1,-x_2,-x_3,-x_4),
      (x_1,x_2,x_3,x_4)\mapsto(\frac{1}{2}+x_1,\frac{1}{2}-x_2,-x_3,x_4)
       \rangle$}}\,. 
 \end{eqnarray*}
 Such $F_p$ has multiplicity $2$.

 \item
 For $p$ a vertex of $\Sigma_Q$,
 $\Gamma_p=\overline{\Gamma}$. Thus
 $F_p=X_v
   =X_0/\mbox{\raisebox{-.4ex}{$\langle\beta,\gamma\rangle$}}$,
 which has multiplicity $4$.
\end{itemize}

These exceptional fibers have various realizations as a
${\Bbb T}^2$-bundle over toroidal $2$-orbifolds as demonstrated
by the example in Remark 2.1.4.

\bigskip

\noindent
{\it (iii) How $S$ sits over $Q^3_{567}$}$\,$:
By inspection, how the $12$ components of $S$ sit in
${\Bbb T}^7/\mbox{\raisebox{-.4ex}{$\Gamma$}}$ with respect to
$\pi_{567}$ is listed below:
\begin{itemize}
 \item
 Since $\Gamma_0=\langle\alpha\rangle$, the $4$ ${\Bbb T}^3$ of $S$
 that descend from the $16$ fixed ${\Bbb T}^3$ of $\alpha$ become
 $4$ disjoint multi-sections $\sigma_i$, $i=1,\ldots,4$, of
 $\pi_{567}$. The union $\cup_i\,\sigma_i$ contains exactly the
 isolated singularities of fibers $F_p$.

 \item
 The $4$ ${\Bbb T}^3$ from the fixed $3$-tori of $\beta$ are
 mapped under $\pi_{567}$ onto the $4$ edges of $\Sigma_Q$ that are
 parallel to the $x_7$-axis with one ${\Bbb T}^3$ to one edge; and
 similarly for the other $4$ ${\Bbb T}^3$ from the fixed $3$-tori
 of $\gamma$ to the edges of $\Sigma_Q$ that are parallel to
 $x_6$-axis. Up to permutations of coordinates, the restriction of
 $\pi_{567}$ to each of these ${\Bbb T}^3$ is of the form
 $$
  \begin{array}{ccc}
   {\Bbb T}^3 & \rightarrow & [0,\mbox{$\frac{1}{2}$}]\\[1ex]
   (x,y,z)    & \mapsto     & \overline{z}
  \end{array}, \hspace{2ex}
  \mbox{where $x,y,z\in
        {\Bbb R}/\hspace{-.1ex}\mbox{\raisebox{-.4ex}{${\Bbb Z}$}}$
    and} \hspace{2ex}
  \begin{array}{l}
       \overline{z}=\left\{\begin{array}{cl}
                     z & \mbox{if $z\in [0,\frac{1}{2}]$}\\[1ex]
                     1-z & \mbox{if $z\in[\frac{1}{2},1]$}
                     \end{array}\right. \\[3ex]
       \mbox{(i.e.\ folding of ${\Bbb T}^1$)}
  \end{array},
 $$
 \begin{figure}[htbp]
 \setcaption{{\sc Figure 3-1-2.}
 \baselineskip 14pt
   How the components of $S$ sit over $\Sigma_Q$ is indicated.
 } 
 \centerline{\psfig{figure=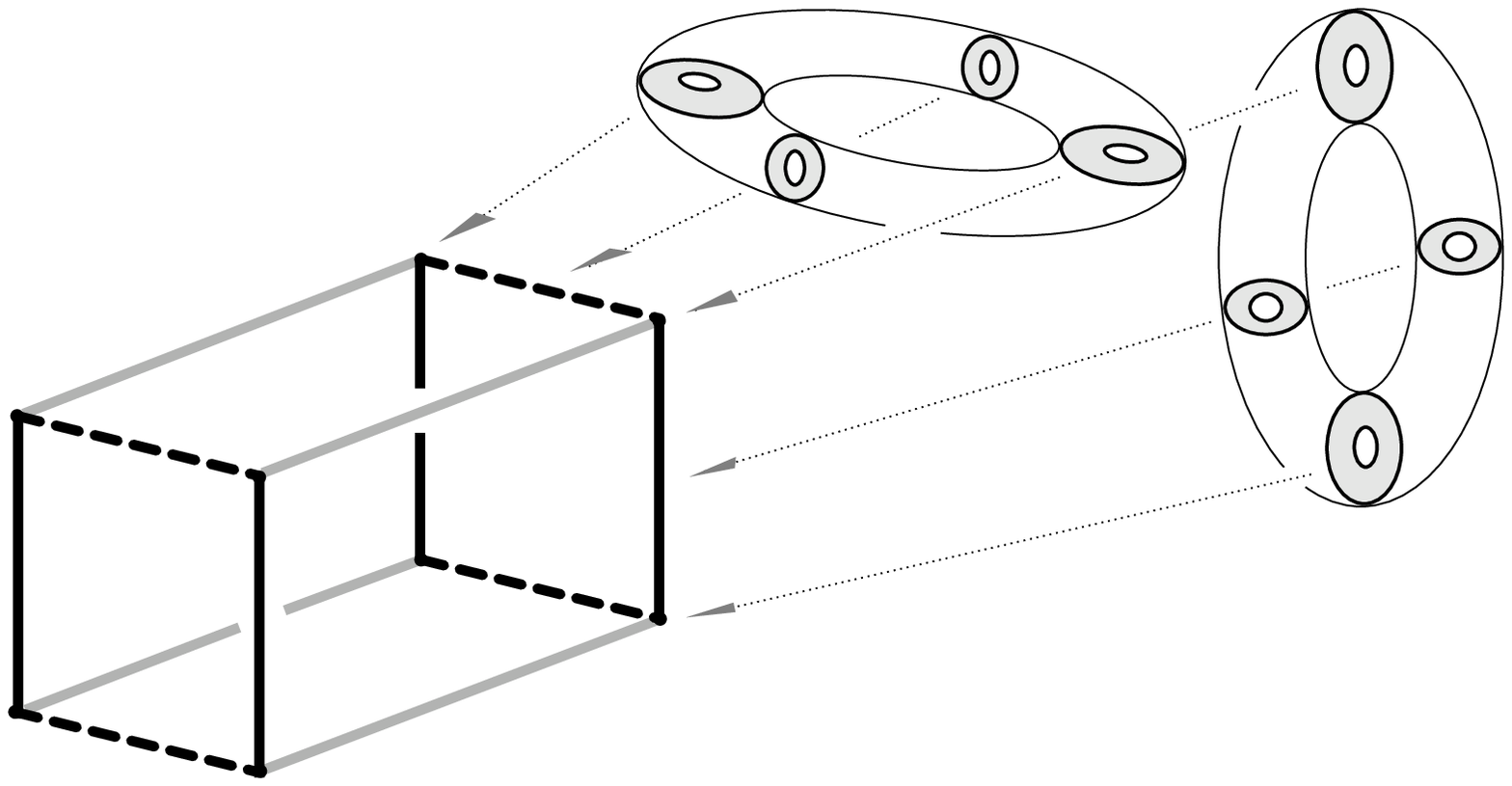,width=13cm,caption=}}
 \end{figure}
\end{itemize}

\bigskip

\noindent
{\it (iv) Adjustment after resolving $S$}$\,$:
Following Sec.\ 2.2, after resolving $S$, one has a fibration
$\widetilde{\pi}_{567}$ of $M^7$ over $Q^3_{567}$ with generic
fiber $\widetilde{X_0}$, which is a smooth K3 surface, and
exceptional fibers:
\begin{itemize}
 \item
 $\widetilde{X_e}$ over the interior of the edges of
 $\Sigma_Q$ parallel to $x_5$, which is a smooth
 ${\Bbb Z}_2$-quotient of $\widetilde{X_0}$;

 \item
 $\widetilde{X_e}\cup\,2\,{\Bbb T}^2\times{\Bbb C}{\rm P}^1$
 over the interior of other edges of $\Sigma_Q$;

 \item
 $\widetilde{X_v}\cup\,2\,{\Bbb T}^2\times{\Bbb C}{\rm P}^1$
 over a vertex of $\Sigma_Q$.
\end{itemize}

\bigskip

\noindent
{\it (v) Monodromy}$\,$:
From Sec.\ 2, the monodromy of $\widetilde{\pi}_{567}$ is
determined by a representation $\overline{\rho}$ from
$H_1(Q^3_{567}-\Sigma_Q;{\Bbb Z})$ to $\Aut(X_0)$. Since
$Q^3_{567}-\Sigma_Q$ is homeomorphic to the complement of the
bouquet $\vee_5\,S^1$ of five circles in $S^3$,
$H_1(Q^3_{567}-\Sigma_Q;{\Bbb Z})={\Bbb Z}^5$, whose generators
may be chosen to be the meridians $C_{5,1}$, $C_{5,2}$, $C_{6,1}$,
$C_{6,2}$, and $C_{6,3}$ as indicated in {\sc Figure 3-1-3}.
\begin{figure}[htbp]
 \setcaption{{\sc Figure} 3-1-3.
 \baselineskip 14pt
  The set of generating cycles
  $\{\,C_{5,1},\,C_{5,2},\,C_{6,1},\,C_{6,2},\,C_{6,3}\}$
  for $H_1(Q^3_{567}-\Sigma_Q;{\Bbb Z})$ are indicated by
  thick loops.
 } 
 \centerline{\psfig{figure=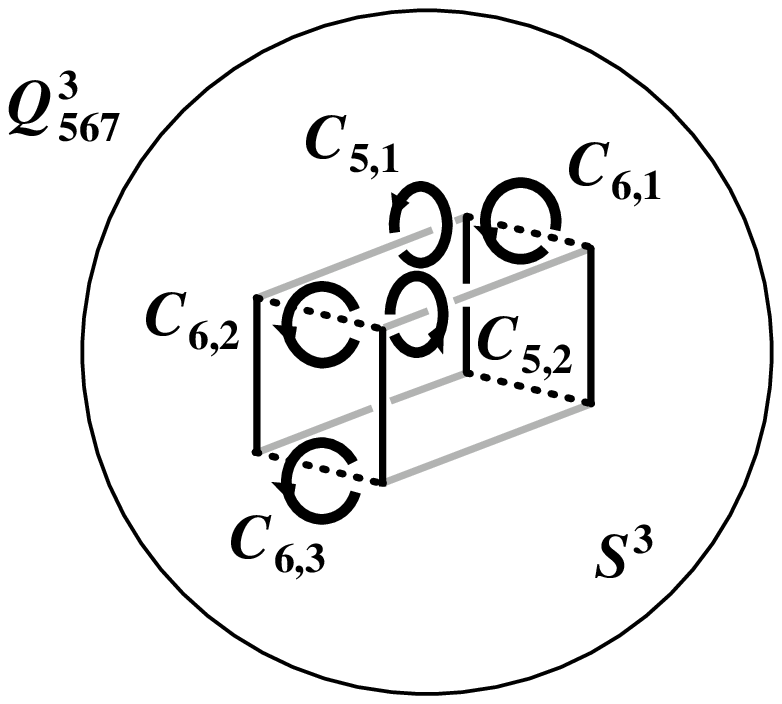,width=13cm,caption=}}
\end{figure}
One then has
$\overline{\rho}(C_{5,i})\;=\;\jmath(\beta\gamma)$ for $i=1,2$, and 
$\overline{\rho}(C_{6,i})\;=\;\jmath(\gamma)$ for $i=1,2,3$.

This concludes the example.

\noindent\hspace{14cm} $\Box$

\bigskip

\noindent
{\bf Example 3.2 [$J(0,\frac{1}{2},\frac{1}{2},\frac{1}{2},0)\,$:
   a.c.\ K3-fibration].}
In Example 3.1, consider instead the decomposition
$Y={\Bbb T}^7={\Bbb T}^3_{347}\times{\Bbb T}_{1256}$ with
$Z={\Bbb T}^3_{347}$ and $X={\Bbb T}^4_{1256}$.
One then has the fibration
$\pi_{347}:{\Bbb T}^7/\mbox{\raisebox{-.4ex}{$\Gamma$}}
                                           \rightarrow Q^3_{347}$,
where $Q^3_{347}$ is the orbifold
${\Bbb T}^3_{347}/\mbox{\raisebox{-.4ex}{
     $\langle\overline{\alpha},\overline{\gamma}\rangle$}}$ with
$\overline{\alpha}(x_3,x_4,x_7)=(-x_3,-x_4,x_7)$ and
$\overline{\gamma}(x_3,x_4,x_7)
                      =(\mbox{$\frac{1}{2}$}-x_3,x_4,-x_7)$.

\bigskip

\noindent
{\it (i) The base orbifold $Q^3_{347}\,$:}
By choosing a fundamental domain $\Omega$ in ${\Bbb T}^3_{347}$ to
be $[0, 1]\times[0,\frac{1}{2}]\times[0,\frac{1}{2}]$ and examining
how $\langle\overline{\alpha},\overline{\gamma}\rangle$ acts on
$\partial\Omega$, one concludes that $Q^3_{347}$ is an Euclidean
$3$-orbifold whose underlying topology is $S^3$ and whose singular
locus $\Sigma_Q$ is a doubled Hopf link, which has $4$ components:
$S^1_{4,1}$ and $S^1_{4,2}$ parallel to the $x_4$-axis, and
$S^1_{7,1}$ and $S^1_{7,2}$ parallel to the $x_7$-axis.
For $p$ in $S^1_{4,i}$, $\Sigma_p=\langle\overline{\gamma}\rangle$;
and, for $p$ in $S^1_{7,i}$,
$\Sigma_p=\langle\overline{\alpha}\rangle$. ({\sc Figure 3-2-1}.)
\begin{figure}[htbp]
 \setcaption{{\sc Figure 3-2-1.}
 \baselineskip 14pt
  After the identification of $\partial\Omega$ governed by
  $\langle\overline{\alpha},\overline{\gamma}\rangle$, one obtains
  the Euclidean orbifold $Q^3_{347}$ with underlying topology
  $S^3$ and singular locus $\Sigma_Q$ a doubled Hopf link.
 }  
\centerline{\psfig{figure=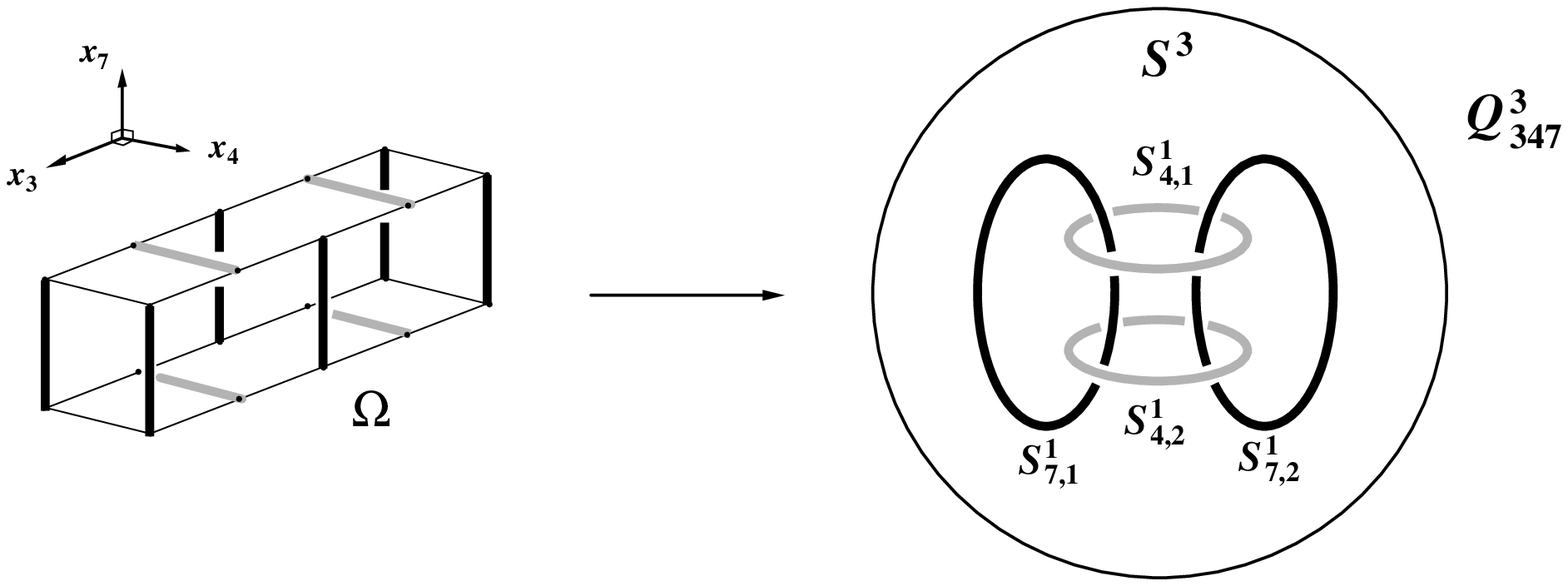,width=13cm,caption=}}
\end{figure}

\bigskip

\noindent
{\it (ii) The fibers}$\,$:
In this example, $\Gamma_0={\Bbb Z}_2=\langle\beta\rangle$.
The fiber $F_p$ of the fibration $\pi_{347}$
over a point $p$ in $Q^3_{347}$ is listed below:
\begin{itemize}
 \item
 For $p$ in $Q^3_{347}-\Sigma_Q$,
 $$
  F_p\; =\; X_0\; =\;{\Bbb T}^7_{1256}/
    \mbox{\raisebox{-.4ex}{$\langle\beta\rangle$}}\;
  =\; {\Bbb T}^4_{1256}/\mbox{\raisebox{-.4ex}{\small$\langle
     (x_1,x_2,x_5,x_6)
             \mapsto(-x_1,\frac{1}{2}-x_2,-x_5,-x_6)\rangle$}}\,,
 $$
 which is a singular K3 surface with $16$ $A_1$-singularity.

 \item
 For $p$ in $S^1_{4,i}$ (resp.\ $S^1_{7,i}$), $\Gamma_p$ is 
 $\langle\overline{\beta},\overline{\gamma}\rangle$
 (resp.\ $\langle\overline{\alpha},\overline{\beta}\rangle$).
 Thus $F_p=X_e$ is the (isomorphic) ${\Bbb Z}_2$-quotient
 $X_0/\mbox{\raisebox{-.4ex}{$\langle\gamma\rangle$}}$ or 
 $X_0/\mbox{\raisebox{-.4ex}{$\langle\alpha\rangle$}}$
 accordingly,
\end{itemize}

\bigskip

\noindent
{\it (iii) How $S$ sits over $Q^3_{347}$}$\,$:
\begin{itemize}
 \item
 The $4$ ${\Bbb T}^3$ of $S$ that descend from the $16$ fixed
 ${\Bbb T}^3$ of $\beta$ become $4$ disjoint multi-sections
 $\sigma_i$, $i=1,\ldots,4$, of $\pi_{347}$. The union
 $\cup_i\,\sigma_i$ contains exactly the isolated singularities of
 fibers $F_p$.

 \item
 Two of the $4$ ${\Bbb T}^3$ in $S$ from the fixed $3$-tori of
 $\alpha$ are mapped onto $S^1_{7,1}$ under $\pi_{347}$ and the
 other two onto $S^1_{7,2}$. Similarly for the $4$ ${\Bbb T}^3$
 in $S$ from the fixed $3$-tori of $\gamma$: two onto $S^1_{4,1}$
 and two onto $S^1_{4,2}$. The restriction of $\pi_{347}$ to each
 of these ${\Bbb T}^3$ is simply a standard projection of
 ${\Bbb T}^3$ to ${\Bbb T}^1$.
\end{itemize}

\bigskip

\noindent
{\it (iv) Adjustment after resolving $S$}$\,$:
After resolving $S$, one obtains a different K3-fibration for the
same $M^7$ in Example 3.1:
$\widetilde{\pi}_{347}:M^7\rightarrow Q^3_{347}$. Its set of
critical values is $\Sigma_Q$ and its degenerate fiber is
$\widetilde{X_e}\cup\,2\,{\Bbb T}^2\times{\Bbb C}{\rm P}^1$.

\bigskip

\noindent
{\it (v) Monodromy}$\,$:
For monodromy
$\overline{\rho}:
          H_1(Q^3_{347}-\Sigma_Q;{\Bbb Z})\rightarrow\Aut(X_0)$,
$H_1(Q^3_{347}-\Sigma_Q;{\Bbb Z})={\Bbb Z}^4$ is generated by the
meridians $C_{4,1}$, $C_{4,2}$, $C_{7,1}$, and $C_{7,2}$
associated to $S^1_{4,1}$, $S^1_{4,2}$, $S^1_{7,1}$, and
$S^1_{7,2}$ respectively, as indicated in {\sc Figure 3-2-2}.
\begin{figure}[htbp]
 \setcaption{{\sc Figure 3-2-2.}
 \baselineskip 14pt
  The set of generating cycles
  $\{\,C_{4,1},\,C_{4,2},\,C_{7,1},\,C_{7,2}\,\}$
  for $H_1(Q^3_{347}-\Sigma_Q;{\Bbb Z})$ are indicated by
      thick loops.
 } 
 \centerline{\psfig{figure=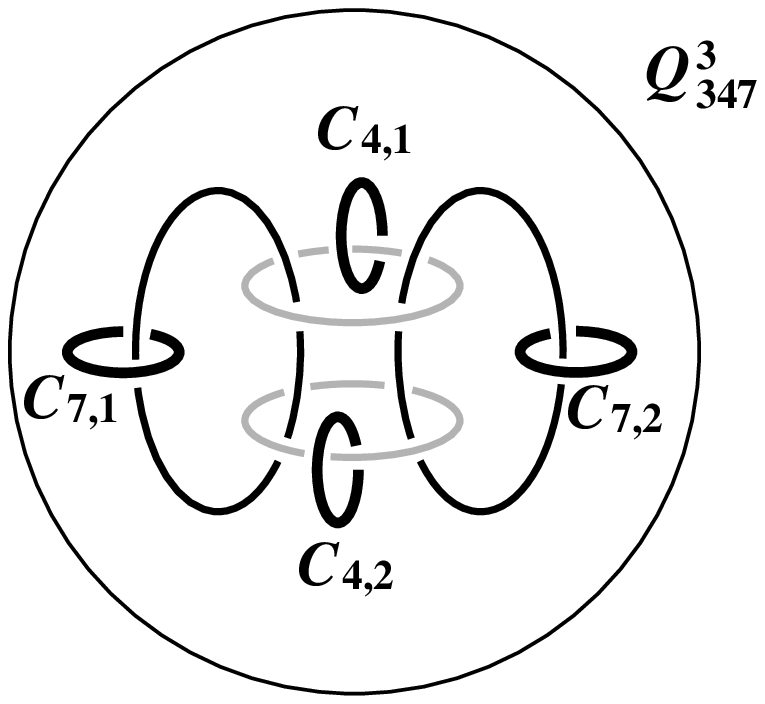,width=13cm,caption=}}
\end{figure}
One thus has
$\overline{\rho}(C_{4,i})\;=\;\jmath(\gamma)$ and 
$\overline{\rho}(C_{7,i})\;=\;\jmath(\alpha)$, $i=1,2$.

\bigskip

This concludes the example.

\noindent\hspace{14cm} $\Box$

\bigskip

\noindent
{\it Remark 3.2.1.}
 Examples 3.1 and 3.2 together show that a Joyce manifold can have
 inequivalent a.a.\ K3-fibrations.

\bigskip

\noindent
{\bf Example 3.3 [$J(0,\frac{1}{2},\frac{1}{2},\frac{1}{2},0)\,$:
   a.c.\ ${\Bbb T}^4$-fibration].}
In Example 3.1, consider instead the decomposition
$Y={\Bbb T}^7={\Bbb T}^3_{127}\times{\Bbb T}_{3456}$ with
$Z={\Bbb T}^3_{127}$ and $X={\Bbb T}^4_{3456}$.
One then has the fibration
$\pi_{127}:{\Bbb T}^7/\mbox{\raisebox{-.4ex}{$\Gamma$}}
                                           \rightarrow Q^3_{127}$,
where $Q^3_{127}$ is the orbifold
${\Bbb T}^3_{127}/\mbox{\raisebox{-.4ex}{$\langle
   \overline{\alpha}, \overline{\beta},\overline{\gamma}\rangle$}}$
with
$\overline{\alpha}(x_1,x_2,x_7)=(-x_1,-x_2,x_7)$,
$\overline{\beta}(x_1,x_2,x_7)=(-x_1,\frac{1}{2}-x_2,x_7)$,
 and
$\overline{\gamma}(x_1,x_2,x_7)
                      =(\mbox{$\frac{1}{2}$}-x_1,x_2,-x_7)$.

\bigskip

\noindent
{\it (i) The base orbifold $Q^3_{127}$}$\,$:
By choosing a fundamental domain $\Omega$ in ${\Bbb T}^3_{127}$ to
be $[0,\frac{1}{2}]\times[0,\frac{1}{2}]\times[0,\frac{1}{2}]$ and
examining how
$\langle\overline{\alpha},\overline{\beta},\overline{\gamma}\rangle$
acts on $\partial\Omega$, one concludes that $Q^3_{127}$ is an
Euclidean $3$-orbifold whose underlying topology is $S^3$ and whose
singular locus $\Sigma_Q$ is a doubled Hopf link, which has $4$
components:
$S^1_{2,1}$ and $S^1_{2,2}$ parallel to the $x_2$-axis, and
$S^1_{7,\alpha}$ and $S^1_{7,\beta}$ parallel to the $x_7$-axis.
For $p$ in $S^1_{2,i}$, $\Sigma_p=\langle\overline{\gamma}\rangle$;
for $p$ in $S^1_{7,\alpha}$,
$\Sigma_p=\langle\overline{\alpha}\rangle$; and, for $p$ in
$S^1_{7,\beta}$, $\Sigma_p=\langle\overline{\beta}\rangle$.
({\sc Figure 3-3-1}.)
\begin{figure}[htbp]
 \setcaption{{\sc Figure 3-3-1.}
 \baselineskip 14pt
  After the identification of $\partial\Omega$ governed by
  $\langle\overline{\alpha},
      \overline{\beta},\overline{\gamma}\rangle$,
  one obtains the Euclidean orbifold $Q^3_{127}$ with underlying
  topology $S^3$ and singular locus $\Sigma_Q$ a doubled Hopf link.
 }  
\centerline{\psfig{figure=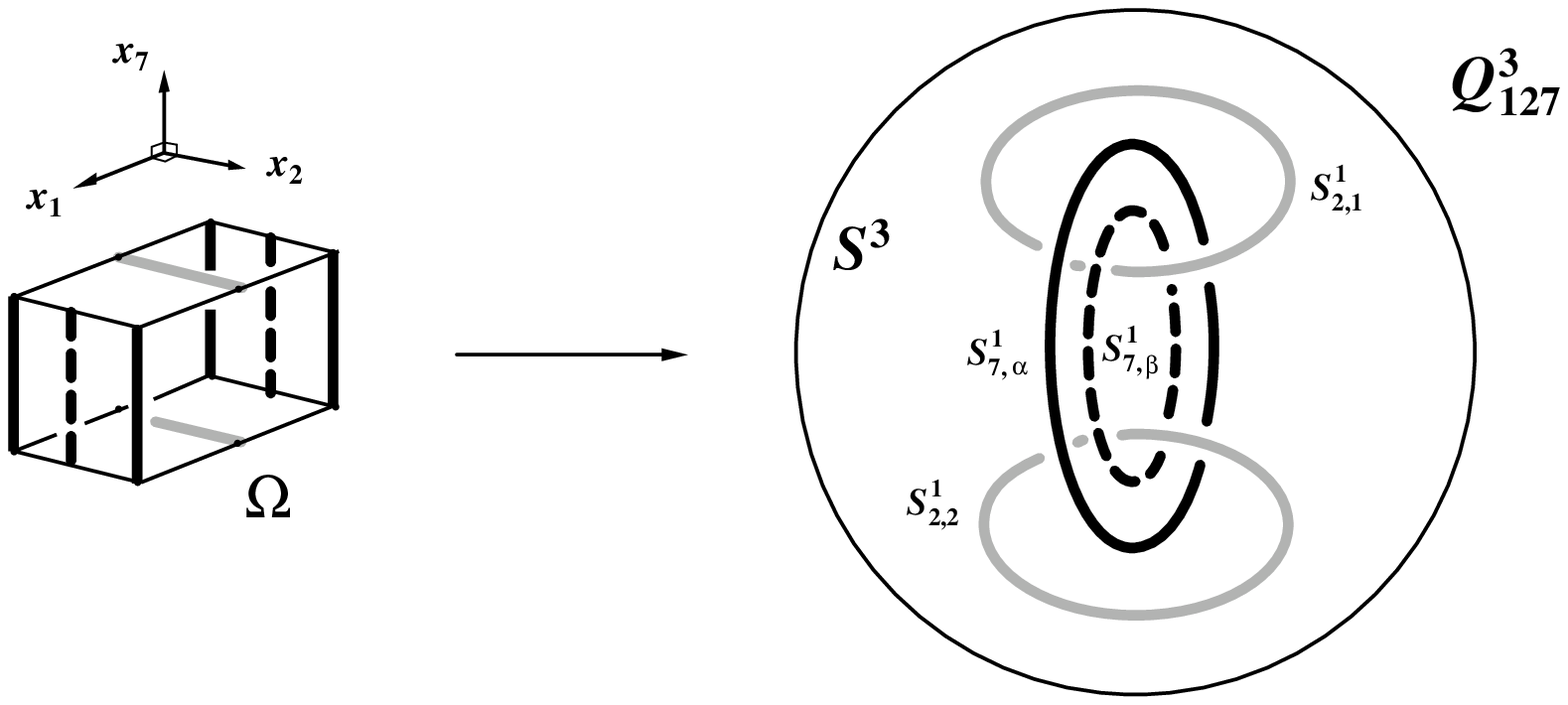,width=13cm,caption=}}
\end{figure}

\bigskip

\noindent
{\it (ii) The fibers}$\,$:
In this example, $\Gamma_0=\{0\}$. The fiber $F_p$ of the fibration
$\pi_{127}$ over a point $p$ in $Q^3_{127}$ is listed below:
\begin{itemize}
 \item
 For $p$ in $Q^3_{347}-\Sigma_Q$, $F_p=X={\Bbb T}^4_{3456}$.

 \item
 For $p$ in $S^1_{2,i}$ (resp.\ $S^1_{7,\alpha}$, $S^1_{7,\beta}$),
 $\Gamma_p$ is $\langle\overline{\gamma}\rangle$
 (resp.\ $\langle\overline{\alpha}\rangle$,
  $\langle\overline{\beta}\rangle$).
 Thus $F_p$ is the ${\Bbb Z}_2$-quotient
 $X/\mbox{\raisebox{-.4ex}{$\langle\gamma\rangle$}}$ (resp.\  
 $X/\mbox{\raisebox{-.4ex}{$\langle\alpha\rangle$}}$,
 $X/\mbox{\raisebox{-.4ex}{$\langle\beta\rangle$}}$).
 They are all isomorphic to $X_e=S^2(2222)\times{\Bbb T}^2$.
\end{itemize}

\bigskip

\noindent
{\it (iii) How $S$ sits over $\Sigma_Q$}$\,$:
The $4$ ${\Bbb T}^3$ in $S$ from the fixed $3$-tori
${\Bbb T}^3_{567}$ of $\alpha$ are mapped onto $S^1_{7,\alpha}$
by $\pi_{347}$ as a standard projection.
Similarly for the $4$ ${\Bbb T}^3$ in $S$ from the fixed $3$-tori
${\Bbb T}^3_{347}$ of $\beta$ onto $S^1_{7,\beta}$.
Finally, $2$ of the $4$ ${\Bbb T}^3$ in $S$ from the fixed
$3$-tori ${\Bbb T}^3_{246}$ of $\gamma$ are mapped onto $S^1_{2,1}$
by $\pi_{347}$ as a composition of a standard projection and a
double covering map from $S^1$ to $S^1$. Similarly for the
remaining $2$ ${\Bbb T}^3$ of $S$ onto $S^1_{2,2}$.

\bigskip

\noindent
{\it (iv) Adjustment after resolving $S$}$\,$:
After resolving $S$, one obtains ${\Bbb T}^4$-fibration for the
same $M^7$ in Example 3.1:
$\widetilde{\pi}_{127}:M^7\rightarrow Q^3_{127}$.
Its set of critical values is $\Sigma_Q$ and its degenerate
fiber is $X_e\cup\,4\,{\Bbb T}^2\times{\Bbb C}{\rm P}^1$.

\bigskip

\noindent
{\it (v) Monodromy}$\,$:
For monodromy
$\overline{\rho}:
          H_1(Q^3_{347}-\Sigma_Q;{\Bbb Z})\rightarrow\Aut(X_0)$,
$H_1(Q^3_{347}-\Sigma_Q;{\Bbb Z})={\Bbb Z}^4$ is generated by the
meridians $C_{2,1}$, $C_{2,2}$, $C_{7,\alpha}$, and $C_{7,\beta}$
associated to $S^1_{2,1}$, $S^1_{2,2}$, $S^1_{7,\alpha}$, and
$S^1_{7,\beta}$ respectively (cf.\ {\sc Figures} 3-2-2 and 3-3-1).
One thus has
$\overline{\rho}(C_{2,i})\;=\;\jmath(\gamma)$, $i=1,2$,
$\overline{\rho}(C_{7,\alpha})\;=\;\jmath(\alpha)$, and
$\overline{\rho}(C_{7,\beta})\;=\;\jmath(\beta)$.

\bigskip

This concludes the example.

\noindent\hspace{14cm} $\Box$

\bigskip

\noindent
{\bf Example 3.4 [$J(0,\frac{1}{2},\frac{1}{2},\frac{1}{2},0)\,$:
   a.a.\ ${\Bbb T}^3$-fibration].}
In Example 3.1, consider instead the decomposition
$Y={\Bbb T}^7={\Bbb T}^4_{1234}\times{\Bbb T}^3_{567}$ with
$Z={\Bbb T}^4_{1234}$ and $X={\Bbb T}^3_{567}$.
One then has the fibration
$\pi_{1234}:{\Bbb T}^7/\mbox{\raisebox{-.4ex}{$\Gamma$}}
                                        \rightarrow Q^4_{1234}$,
where $Q^4_{1234}$ is the orbifold
${\Bbb T}^4_{1234}/\mbox{\raisebox{-.4ex}{$\langle
   \overline{\alpha},\overline{\beta},\overline{\gamma}\rangle$}}$
with
$$
\begin{array}{l}
 \overline{\alpha}(x_1,x_2,x_3,x_4)\;
                  =\;(-x_1,-x_2,-x_3,-x_4)\,,  \\[1ex]
 \overline{\beta}(x_1,x_2,x_3,x_4)\;
                  =\;(-x_1,\frac{1}{2}-x_2,x_3,x_4)\,,\\ [1ex]
 \overline{\gamma}(x_1,x_2,x_3,x_4)\;
                  =\;(\frac{1}{2}-x_1,x_2,\frac{1}{2}-x_3,x_4)\,.
\end{array}
$$

\bigskip

\noindent
{\it (i) The base orbifold $Q^4_{1234}$}$\,$:
Similar to the discussions in [Jo1], one can check that 
the only fixed-points of the $\overline{\Gamma}$-action on
${\Bbb T}^4_{1234}$ are
the $16$ fixed-points of $\overline{\alpha}$,
the $4$ fixed ${\Bbb T}^2_{34}$ of $\overline{\beta}$,
and the $4$ fixed ${\Bbb T}^2_{24}$ of $\gamma$. 
These fixed points or $2$-tori are all disjoint and the action
of $\langle\overline{\beta},\overline{\gamma}\rangle$
on $16$ fixed-points of $\overline{\alpha}$, 
$\langle\overline{\alpha},\overline{\gamma}\rangle$ on the
set of $4$ fixed ${\Bbb T}^2_{34}$ of $\overline{\beta}$, and
$\langle\overline{\alpha},\overline{\beta}\rangle$ on the set of
$4$ fixed ${\Bbb T}^2_{24}$ of $\overline{\gamma}$ are all free.
Thus the singuar set $\Sigma_Q$ of the orbifold $Q^4_{1234}$
consists of $4$ isolated points $\{q_1,q_2,q_3,q_4\}$
and two disjoint copies ${\Bbb T}^2$, denoted by
${\Bbb T}^2_{34,\beta}$ and ${\Bbb T}^2_{24,\gamma}$ respectively.

The decomposition
${\Bbb T}^4_{1234}={\Bbb T}^2_{12}\times{\Bbb T}^2_{34}$
induces a ${\Bbb T}^2_{34}$-fibration
$\pi^{1234}_{12}:Q^4_{1234}\rightarrow Q^2_{12}=D^2(22^{\ast})$.
By inspection, one may write
$\Sigma_{Q^2_{12}}=S^1_{\gamma}\cup\{p_{\alpha},p_{\beta}\}$
with
$\Gamma_{p_{\alpha}}=\langle\overline{\overline{\alpha}}\rangle$,
$\Gamma_{p_{\beta}}=\langle\overline{\overline{\beta}}\rangle$, and
$\Gamma_p=\langle\overline{\overline{\gamma}}\rangle$ for
$p\in S^1_{\gamma}$. The exceptional fiber is an $S^2(2222)$ over
$p_{\alpha}$ and an $A^2(^{\ast\ast})$ over $p\in S^1_{\gamma}$.
(The fiber over $p_{\beta}$ ia a regular ${\Bbb T}^2_{34}$.)
The monodromy
$\rho^{1234}_{12}:\pi_1^{\rm orb}(Q^2_{12})
   \rightarrow\Aut({\Bbb T}^2_{34})$
is determined by
$$
\begin{array}{l}
 \rho^{1234}_{12}(\overline{\overline{\alpha}})\,
   =\,\overline{\alpha}^f\;
   =\;(\,(x_3,x_4)\mapsto (-x_3,x_4)\,)\,,   \\[1ex]
 \rho^{1234}_{12}(\overline{\overline{\gamma}})\,
  =\,\overline{\gamma}^f\;
  =\;(\,(x_3,x_4)\mapsto (\mbox{$\frac{1}{2}$}-x_3,x_4)\,)\,,
                         \hspace{1em}\mbox{and}\\[1ex]
 \rho^{1234}_{12}(\overline{\overline{\beta}})\,
  =\,\rho^{1234}_{12}(t_1)\, =\,\rho^{1234}_{12}(t_2)\,
  =\,\Id\,.
\end{array}
$$
In terms of $\pi^{1234}_{12}$, the $\Sigma_{Q}$ sits in
$Q^4_{1234}$ as illustrated in {\sc Figure 3-4-1}.
\begin{figure}[htbp]
 \setcaption{{\sc Figure 3-4-1.} 
 \baselineskip 14pt
  The $4$-orbifold $Q^4_{1234}$ and its singular locus
  $\Sigma_Q=\{q_1,\cdots,q_4\}
    \cup{\Bbb T}^2_{34,\beta}\cup{\Bbb T}^2_{24,\gamma}$.
  The component ${\Bbb T}^2_{24,\gamma}$ consists
  of the silvered boundary of the $A^2(^{\ast\ast})$-fibers
  along the silvered boundary of the base $Q^2_{12}$, it winds
  over $\partial Q^2_{12}$ twice (only sketched). 
 } 
 \centerline{\psfig{figure=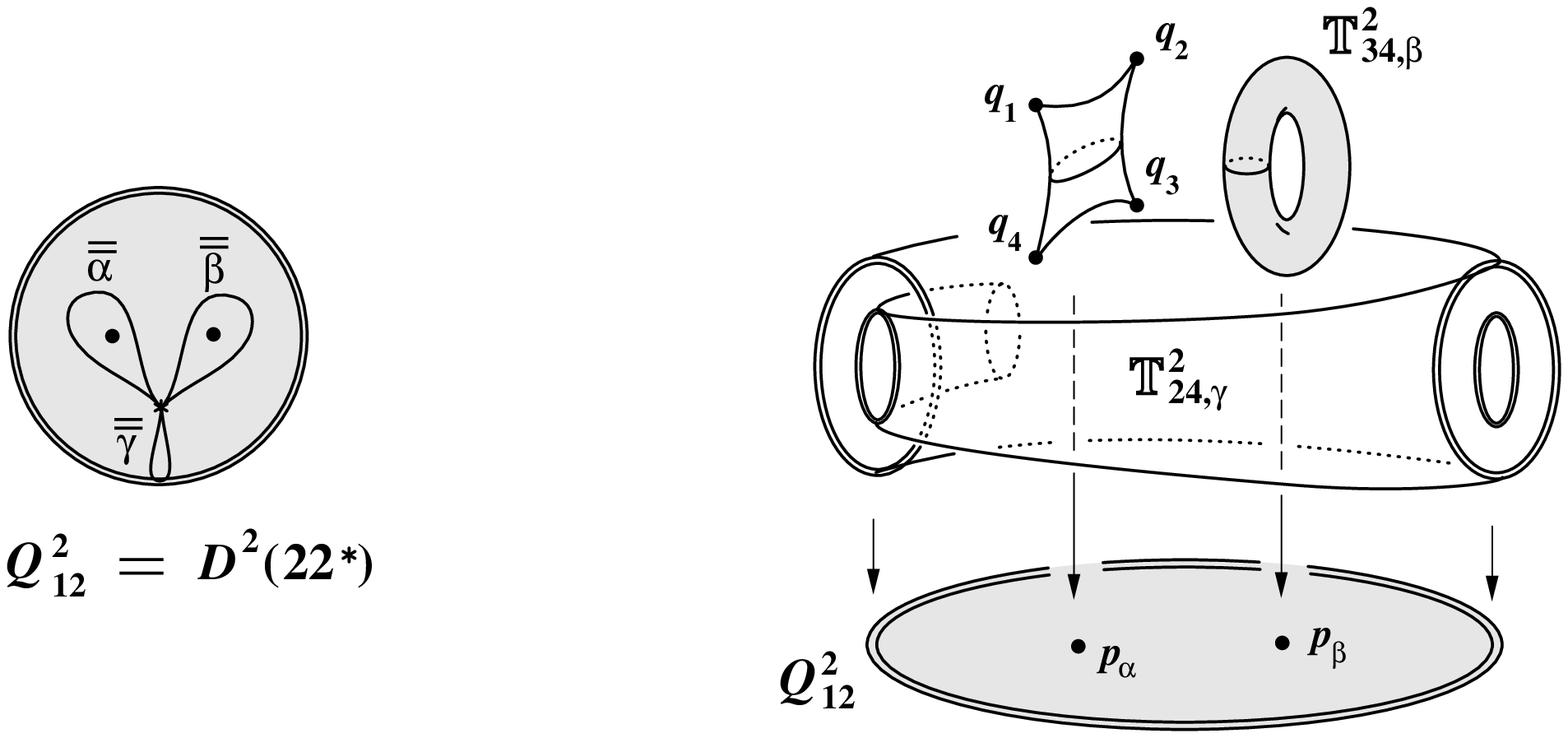,width=13cm,caption=}}
\end{figure}

\noindent
{\it (ii) The fibers}$\,$:
\begin{itemize}
 \item
 For $p$ in $Q^4_{1234}-\Sigma_Q$, $F_p=X={\Bbb T}^3_{567}$.

 \item
 For $p\in\{q_1,q_2,q_3,q_4\}$,
 $\Gamma_p=\langle\overline{\alpha}\rangle$; thus
 $F_p={\Bbb T}^3_{567}/
         \mbox{\raisebox{-.4ex}{$\langle\alpha\rangle$}}
     ={\Bbb T}^3_{567}$, a regular fiber.

 \item
 For $p\in{\Bbb T}^2_{34,\beta}$,
 $\Gamma_p=\langle\overline{\beta}\rangle$; thus
 $F_p={\Bbb T}^3_{567}/
         \mbox{\raisebox{-.4ex}{$\langle\beta\rangle$}}$.
 Likewise, for $p\in{\Bbb T}^2_{24,\gamma}$,
 $\Gamma_p=\langle\overline{\gamma}\rangle$; thus
  $F_p={\Bbb T}^3_{567}/
          \mbox{\raisebox{-.4ex}{$\langle\gamma\rangle$}}$.
 They are all isomorphic to $X_e=S^2(2222)\times S^1$.
\end{itemize}

\bigskip

\noindent
{\it (iii) How $S$ sits}$\,$:
The $4$ ${\Bbb T}^3$ in $S$ from the fixed $3$-tori
 ${\Bbb T}^3_{567}$ of $\alpha$ are mapped onto
 $\{q_1,q_2,q_3,q_4\}$ by $\pi_{1234}$.
Each of the $4$ ${\Bbb T}^3$ in $S$ from the fixed $3$-tori
 ${\Bbb T}^3_{347}$ of $\beta$ onto ${\Bbb T}^2_{34,\beta}$
 as standard projection.
Similarly, each of the $4$ ${\Bbb T}^3$ in $S$ from the fixed
 $3$-tori ${\Bbb T}^3_{246}$ of $\gamma$ are mapped onto
 ${\Bbb T}^2_{24,\gamma}$ as standard projection.

\bigskip

\noindent
{\it (iv) Adjustment after resolving $S$}$\,$:
After resolving $S$, one obtains a ${\Bbb T}^3$-fibration for the
same $M^7$ in Example 3.1:
$\widetilde{\pi}_{1234}:M^7\rightarrow\widetilde{Q}^4_{1234}$.
where $r:\widetilde{Q}^4_{1234}$ is the resolution of $Q^4_{1234}$.
$\Sigma_{\widetilde{Q}}$ is isomorphic to
$\Sigma_Q\cup\,4\,{\Bbb C}{\rm P}^1$.
The set of critical values of $\widetilde{\pi}_{1234}$ is the
$\Sigma_Q$-part of $\Sigma_{\widetilde{Q}}$ and the exceptional
fiber is $X_e\cup\,4\,{\Bbb T}^1\times{\Bbb C}{\rm P}^1$ with
normal crossing singularities.

\bigskip

\noindent
{\it (v) Monodromy}$\,$:
Let $t_i$ be the translation of $x_i$ in ${\Bbb R}$ by $1$; then
$\pi_1^{\rm orb}(Q^4_{1234})
  =\langle\,\overline{\alpha},\overline{\beta},\overline{\gamma};
               t_1,t_2,t_3,t_4\,\rangle$
and 
$\widetilde{\rho}=\rho\circ r_{\ast}:
 \pi_1^{\rm orb}(\widetilde{Q}_{1234})
  \rightarrow\Aut({\Bbb T}^3_{567})$
is determined by 
$$
\begin{array}{l}
 \rho(\overline{\beta})\,
  =\,\beta^f\;
  =\;(\,(x_5,x_6,x_7)\mapsto(-x_5,-x_6,x_7)\,)\,,\\[1ex]
 \rho(\overline{\gamma})\,
  =\,\gamma^f\;
  =\;(\,(x_5,x_6,x_7)\mapsto(-x_5,x_6,-x_7)\,)\,,
         \hspace{1em}\mbox{and}\\[1ex]
 \rho(\overline{\alpha})\, =\,\rho(t_i)\, =\,\Id\,,\; i=1,2,3,4.
\end{array}
$$

\bigskip

This concludes the example.

\noindent\hspace{14cm} $\Box$

\bigskip

\noindent
{\bf Example 3.5 [$J(e^{\pi i/3}, e^{2\pi i/3},\Lambda)\,$:
    a.c.\ ${\Bbb T}^4$-fibration].} (Cf.\ [Jo1]: II, Example 11.)
\newline
$\Gamma$ is generated by
$$
\begin{array}{rcl}
 \alpha(z_1,z_2,z_3,x) & =
  & (e^{\pi i/3}z_1, e^{2\pi i/3}z_2, -z_3, x+\frac{1}{6}) \\[.2ex]
 \beta(z_1,z_2,z_3,x)  & =
  & (-\overline{z_1}, -\overline{z_2}, -\overline{z_3}, -x)\,,
\end{array}
$$
where $z_1=x_1+ix_2$, $z_2=x_3+ix_4$, $z_3=x_5+ix_6$, $x=x_7$; and
$$
 \Lambda\;=\;({\Bbb Z}+e^{2\pi i/3}{\Bbb Z})\,
  \oplus\, ({\Bbb Z}+e^{2\pi i/3}{\Bbb Z})\,
  \oplus\,({\Bbb Z}+i{\Bbb Z})\,\oplus\,{\Bbb Z}\,.
$$
From [Jo1] (II), the quotient
${\Bbb T}^7/\mbox{\raisebox{-.4ex}{$\Gamma$}}$ has a singular
set $S$ of $A_1$-singularities that consists of $4$ disjoint
copies ${\Bbb T}^3$: $2$ copies arising from the fixed
${\Bbb T}^3$ of any element in $\Gamma$ conjugate to $\beta$ and
another $2$ copies arising from the fixed ${\Bbb T}^3$ of any
element in $\Gamma$ conjugate to $\beta\alpha$. The tubular
neighborhood of each component of $S$ is modelled on
${\Bbb T}^3\times(\,{\Bbb C}^2/
                 \mbox{\raisebox{-.4ex}{$\langle -1\rangle$}})$.
After resolving $S$, one obtains a Joyce manifold $M^7$.

Consider the associative-coassociative decomposition
$Y={\Bbb T}^7={\Bbb T}^3_{127}\times{\Bbb T}^4_{3456}$,
where
$$
 Z\;=\;{\Bbb T}^3_{127}\;
 =\;\{{\Bbb C}/\mbox{\raisebox{-.4ex}{$({\Bbb Z}
                          \oplus e^{2\pi i/3}{\Bbb Z})$}}\}\,
  \times\,{\Bbb R}/\mbox{\raisebox{-.4ex}{${\Bbb Z}$}}
$$
is parametrized by $(z_1,x)$ and
$$
 X\;=\;{\Bbb T}^4_{3456}\;
 =\; \{ {\Bbb C}/\mbox{\raisebox{-.4ex}{$({\Bbb Z}
                             \oplus e^{2\pi i/3}{\Bbb Z})$}} \}\,
 \times\,\{ {\Bbb C}/\mbox{\raisebox{-.4ex}{$({\Bbb Z}
                             \oplus i{\Bbb Z})$}} \}
$$
is parametrized by $(z_2,z_3)$.
One then has the fibration
$\pi_{127}:{\Bbb T}^7/\mbox{\raisebox{-.4ex}{$\Gamma$}}
                                        \rightarrow Q^3_{127}$,
where $Q^3_{127}$ is the orbifold
${\Bbb T}^3_{127}/\mbox{\raisebox{-.4ex}{$\langle
                   \overline{\alpha},\overline{\beta}\rangle$}}$
with
$\overline{\alpha}(z_1,x)=(e^{\pi i/3}z_1,x+\frac{1}{6})$ and
$\overline{\beta}(z_1,x)=(-\overline{z_1},-x)$.

\bigskip

\noindent
{\it (i) The base orbifold $Q^3_{127}$}$\,$:
There are two ways to understand $Q^3_{127}$. Consider first the
general method by studying a fundamental domain $\Omega$ of the
$\Gamma$-action on ${\Bbb T}^1_{127}$. One can choose $\Omega$ to
be ${\Bbb T}^2_{12}\times [0,\frac{1}{12}]$, then
$\partial\Omega={\Bbb T}^2_{12}\times\{0,\frac{1}{12}\}$. Since
$\Stab({\Bbb T}^2_{12}\times\{0\})=\langle\overline{\beta}\rangle$
and
$\Stab({\Bbb T}^2_{12}\times\{\frac{1}{12}\})
                  =\langle\overline{\alpha}\overline{\beta}\rangle$,
$Q^3_{127}$ is obtained from $\Omega$ by identifying $p$ with
$\overline{\beta}(p)$ for $p\in{\Bbb T}^2_{12}\times\{0\}$ and
$p^{\prime}$ with $\overline{\alpha}\overline{\beta}(p^{\prime})$
for $p^{\prime}\in{\Bbb T}^2_{12}\times\{\frac{1}{12}\}$.
These orientation-reversing maps on $\partial\Omega$ and their
fixed locus are indicated in {\sc Figure 3-5-1}. The quotient of
$\partial\Omega$ by them are two M\"{o}bius strips.
From this, one concludes that the underlying topology of
$Q^3_{127}$ is a closed orientable $3$-manifold and the singular
locus $\Sigma_Q$ consists of two copies of $S^1$: one from the
fixed locus of $\overline{\beta}$ in ${\Bbb T}^2_{12}\times\{0\}$
and the other from the fixed locus of
$\overline{\alpha}\overline{\beta}$ in
${\Bbb T}^2_{12}\times\{\frac{1}{12}\}$. We shall denote the former
by $S^1_{\beta}$ and the latter by $S^1_{\alpha\beta}$.

\begin{figure}[htbp]
 \setcaption{{\sc Figure 3-5-1.}
 \baselineskip 14pt
  The fundamental domain $\Omega$ of the $\Gamma$-action on
  ${\Bbb T}^3_{127}$. The fixed locus $S^1_{\beta}$ of
  $\overline{\beta}$ on ${\Bbb T^2_{12}}\times\{0\}$ and
  the fixed locus ${S^1_{\alpha\beta}}$ of
  $\overline{\alpha}\overline{\beta}$ on
  ${\Bbb T}^2_{12}\times\{\frac{1}{12}\}$ are indicated.
  Faces of the polyhedron are identified appropriately to get
  $\Omega={\Bbb T}^2_{12}\times[0,\frac{1}{12}]$.
} 
\centerline{\psfig{figure=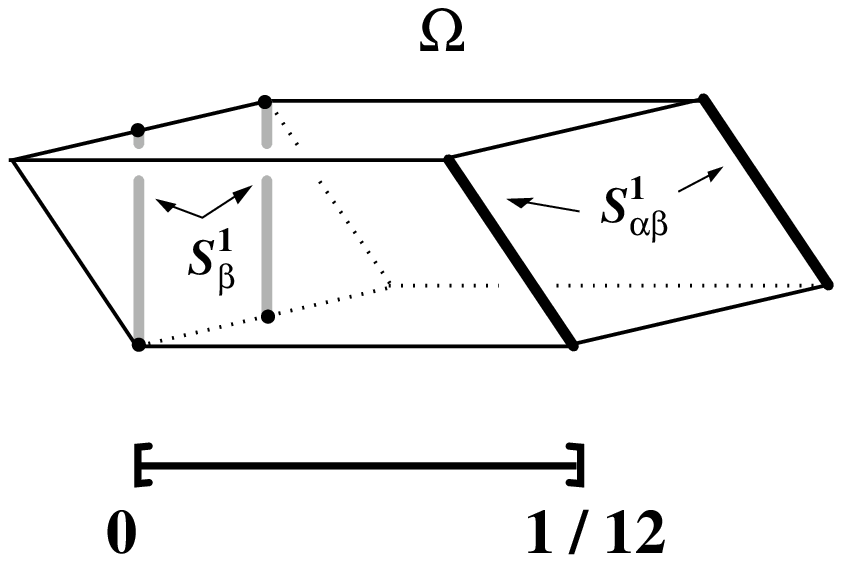,width=13cm,caption=}}
\end{figure}

Denote the underlying $3$-manifold of $Q^3_{127}$ by $M^3$. Then
one can show that $\pi_1(M^3)$ is trivial as follows: Regard $M^3$
as pasted from the two halves, $M^3_-$ from
${\Bbb T}^2_{12}\times[0,\frac{1}{24}]$ and $M^3_+$ from
${\Bbb T}^2_{12}\times[\frac{1}{24},\frac{1}{12}]$, along the
shared boundary ${\Bbb T}^2_{12}\times\{\frac{1}{24}\}$.
Then $\pi_1(M^3_-)={\Bbb Z}$ is generated by the core loop
$\gamma_-$ of the M\"{o}bius strip
$({\Bbb T}^2_{12}\times\{0\})/
    \mbox{\raisebox{-.4ex}{$\langle\overline{\beta}\rangle$}}$
since $M^3_-$ is strong deformation retractable to $\gamma_-$.
Similarly $\pi_1(M^3_+)={\Bbb Z}$ is generated by the core loop
$\gamma_+$ of the M\"{o}bius strip
$({\Bbb T}^2_{12}\times\{\frac{1}{12}\})/
    \mbox{\raisebox{-.4ex}{$\langle
            \overline{\alpha}\overline{\beta}\rangle$}}$.
By choosing a generating set for
$\pi_1({\Bbb T}^2_{12}\times\{\frac{1}{24}\})$ to be
$\{u_1,u_2\}$, where $u_1$ corresponds to the path in ${\Bbb C}$
from $0$ to $1$ and $u_2$ from $0$ to $e^{2\pi i/3}$,
and using the Van Kampen's theorem [Sp], one obtains that
$\pi_1(M^3)$ has a presentation
$\langle [\gamma_-],[\gamma_+]\,|\,
               [\gamma_+]=1,[\gamma_-]=[\gamma_+]^2 \rangle$,
which is $\{1\}$. Thus, if one assumes that the Poincare
conjecture is correct, then $M^3$ is an $S^3$ and $\Sigma_Q$ is a
two-component link in $S^3$. Though one may try to push on to gain
more information on $Q^3_{127}$, we shall now turn to the second
method.

The second method is due to the observation that, for the current
example, if one identifies $(z_1,x)$ in ${\Bbb T}^3_{127}$ with
$(z_1, 2\pi x)$ in the unit tangent bundle $T_1({\Bbb T}^2_{12})$
of ${\Bbb T}^2_{12}$, where $(z_1,\theta=0)$ corresponds to a global
unit tangent vector field $\xi$ on ${\Bbb T}^2_{12}$ that is
parallel to the fixed direction of $\overline{\beta}$ on
${\Bbb T}^2_{12}$, then the $\Gamma$-action on ${\Bbb T}^3_{127}$
is indeed the induced action of the $\Gamma$-action on
${\Bbb T}^2_{12}$, defined by
$\overline{\alpha}(z_1)=e^{\pi i/3}z_1$ and
$\overline{\beta}(z_1)=-\overline{z_1}$. Consequently, $Q^3_{127}$
is the unit tangent bundle of the $2$-orbifold
${\Bbb T}^2_{12}/\mbox{\raisebox{-.4ex}{$\langle
               \overline{\alpha},\overline{\beta}\rangle$}}$,
which is the $D^2(^{\ast}236)$-orbifold. Recall now the following
fact from [Th1]: 

\bigskip

\noindent
{\bf Fact 3.5.1.} ([Th1]: Sec.\ 13.4.) {\it
 Let $Q^2$ be a $2$-orbifold whose combinatorial type is a
 polygon. Then its unit tangent bundle $T_1Q^2$ is a $3$-orbifold
 whose underlying topology is an $S^3$ and the singular locus
 is a link, constructed as indicated in {\sc Figure 3-5-2}.
 } 
\begin{figure}[htbp]
 \setcaption{{\sc Figure 3-5-2.} 
 \baselineskip 14pt
  The unit tangent bundle $Q^3=T_1Q^2$ of a $2$-orbifold
  $Q^2=D^2(m_1\,\cdots\,m_l\,^{\ast}\,n_1\,\cdots\,n_k)$ and its
  singular locus $\Sigma_Q$. $\Sigma_p={\Bbb Z}_2$ for 
  $p\in\Sigma_Q$. Notice that, though the cone-points
  of $Q^2$ influence the bundle structure of $Q^3$, they play
  no role in determining $\Sigma_Q$ as a link in $S^3$.
 } 
 \centerline{\psfig{figure=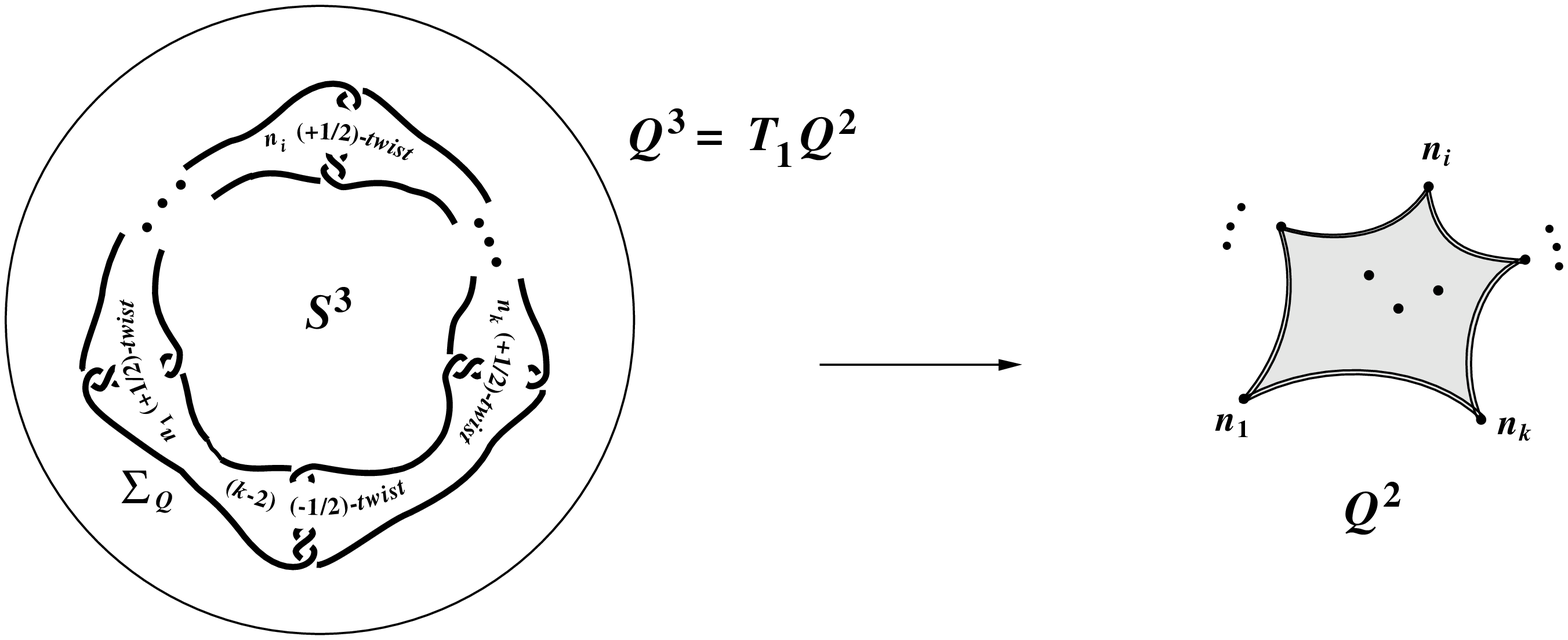,width=13cm,caption=}}
\end{figure}

\bigskip

\noindent
Thus $Q^3_{127}=T_1D^2(^{\ast}236)$ has the underlying topology
$S^3$ and $\Sigma_Q$ as in {\sc Figure 3-5-3}, which is a
two-component link. To see which component is $S^1_{\beta}$ and
which component is $S^1_{\alpha\beta}$, recall from Sec.\ 1 that,
up to an overall conjugation by $\Gamma$, one has subgroups
$\Gamma_{(2)}$, $\Gamma_{(3)}$, and $\Gamma_{(6)}$ in $\Gamma$
associated to the corner reflectors labelled in {\sc Figure 3-5-3}
by $(2)$, $(3)$, and $(6)$ resplectively and subgroups
$\Gamma_{(23)}$, $\Sigma_{(26)}$, and $\Sigma_{(36)}$ in $\Gamma$
associated to points in the silvered edge of $D^2(^{\ast}236)$
that connects corner reflectors $\{(2), (3)\}$, $\{(3),(6)\}$,
and $\{(2),(6)\}$ respectively. One can check directly that
$\Gamma_{(23)}$ and $\Gamma_{(36)}$ conjugate in $\Gamma_{(3)}$
and both are conjugate to $\langle\overline{\beta}\rangle$ in
$\Gamma$ while $\Gamma_{(26)}$ is conjugate to
$\langle\overline{\alpha}\overline{\beta}\rangle$. Consequently,
in {\sc Figure 3-5-3}, $K_1$ corresponds to $S^1_{\beta}$ while
$K_2$ corresponds to $S^1_{\alpha\beta}$.

\begin{figure}[htbp]
 \setcaption{{\sc Figure 3-5-3.} 
 \baselineskip 14pt
  $Q^3_{127}$ realized as the unit tangent bundle
  $T_1D^2(^{\ast}236)$. Its singular locus $\Sigma_Q$ is a
  $2$-component link.
 } 
 \centerline{\psfig{figure=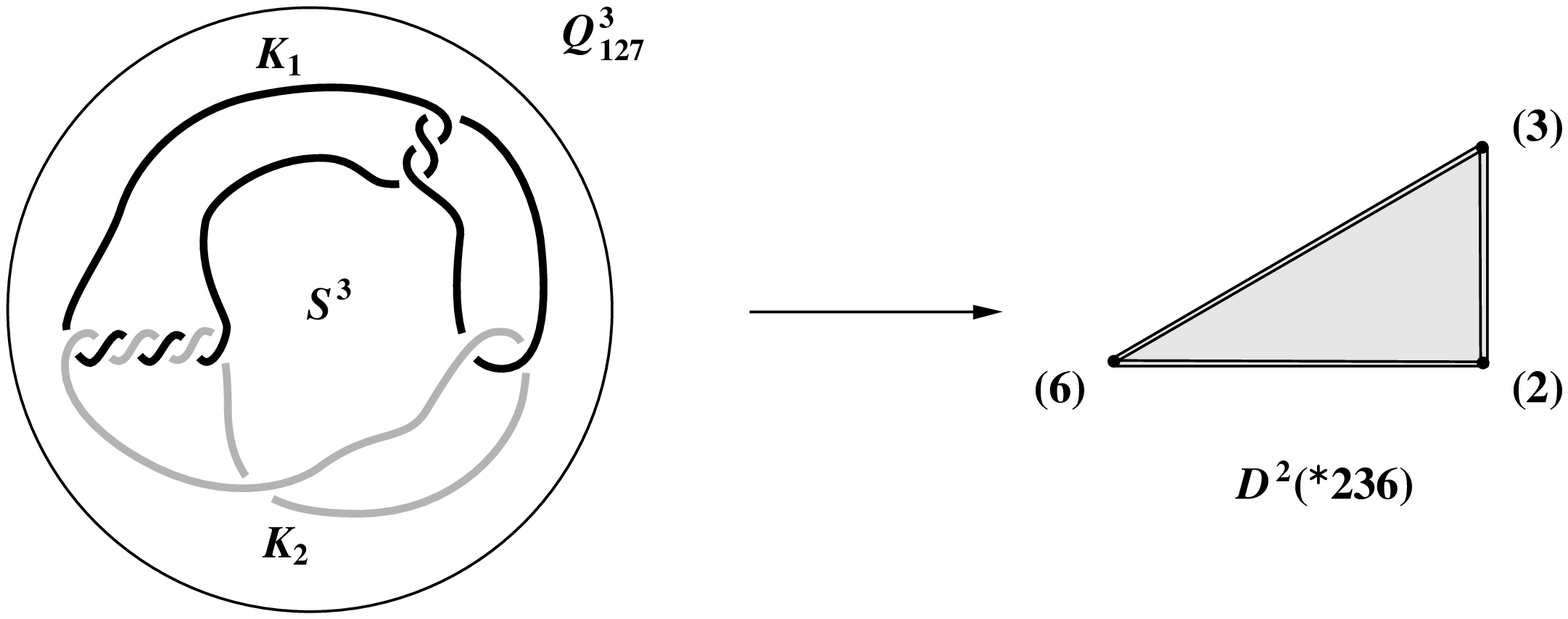,width=13cm,caption=}}
\end{figure}

\bigskip

\noindent
{\it (ii) The fibers}$\,$:
\begin{itemize}
 \item
 For $p$ in $Q^3_{127}-\Sigma_Q$, $F_p=X={\Bbb T}^4_{3456}$.

 \item
 For $p$ in $S^1_{\beta}$, $\Sigma_p=\langle\overline{\beta}\rangle$;
 hence,
 $F_p={\Bbb T}^4_{3456}/
       \mbox{\raisebox{-.4ex}{$\langle\beta\rangle$}}$.
 For $p$ in $S^1_{\alpha\beta}$,
 $\Sigma_p=\langle\overline{\alpha}\overline{\beta}\rangle$; hence,
 $F_p={\Bbb T}^4_{3456}/
       \mbox{\raisebox{-.4ex}{$\langle\alpha\beta\rangle$}}$.
 These fibers are isomorphic and will be denotes by $F_e$.
 They have multiplicity $2$ and can be realized as a
 ${\Bbb T}^2$-bundle over a toroidal $2$-orbifold.
\end{itemize}

\bigskip

\noindent
{\it (iii) How $S$ sits}$\,$:
Under $\pi_{127}$, the $2$ copies of ${\Bbb T}^3$ in $S$
associated to $\beta$ are mapped onto $S^1_{\beta}$, while the
other $2$ copies of ${\Bbb T}^3$ in $S$ are mapped onto
$S^1_{\alpha\beta}$. All these maps are standard projections.

\bigskip

\noindent
{\it (iv) Adjustment after resolving $S$}$\,$:
After resolving $S$, one obtains a ${\Bbb T}^4$-fibration:
$\widetilde{\pi}_{127}:M^7\rightarrow Q^3_{127}$. Its set of
critical values is $\Sigma_Q$ and its exceptional fiber is
$X_e\cup\,2\,{\Bbb T}^2\times{\Bbb C}{\rm P}^1$.

\bigskip

\noindent
{\it (v) Monodromy}$\,$:
Let $u_1(z_1,x)=(z_1+1,x)$, $u_2(z_1,x)=(z_1+e^{2\pi i/3},x)$, and
$t_7(z_1,x)=(z_1,x+1)$;
then
$$
 \pi_1^{\rm orb}(\widetilde{Q}^3_{127})\;
 =\;\pi_1^{\rm orb}(Q^3_{127})\;
 =\; \langle\,\overline{\alpha},\overline{\beta};
                  u_1,u_2,t_7\,\rangle
$$
and
$\widetilde{\rho}=\rho:\pi_1^{\rm orb}(Q^3_{1234})
  \rightarrow\Aut({\Bbb T}^4_{3456})$
is determined by
$$
\begin{array}{l}
 \rho(\overline{\alpha})\,=\,\alpha^f\;
  =\;(\,(z_2,z_3)\mapsto(e^{2\pi i/3}z_2,-z_3 )\,)\,,\\[1ex]
 \rho(\overline{\beta})\,=\,\beta^f\;
  =\;(\,(z_2,z_3)\mapsto(-\overline{z_2},-\overline{z_3})\,)\,,
         \hspace{1em}\mbox{and}\\[1ex]
 \rho(u_1)\, =\, \rho(u_2)\, =\, \rho(t_7)\, =\,\Id\,.
\end{array}
$$

\bigskip

This concludes the example.

\noindent\hspace{14cm} $\Box$

\bigskip

\noindent
{\bf Example 3.6 [$J(e^{\pi i/3}, e^{2\pi i/3},\Lambda)\,$:
    a.a.\ ${\Bbb T}^3$-fibration].} 
In Example 3.5, consider the same 
$Y={\Bbb T}^7={\Bbb T}^3_{127}\times{\Bbb T}^4_{3456}$ but now
with $Z={\Bbb T}^4_{3456}$ and $X={\Bbb T}^3_{127}$. One then has
the fibration
$\pi_{3456}:{\Bbb T}^7/\mbox{\raisebox{-.4ex}{$\Gamma$}}
                                         \rightarrow Q^4_{3456}$,
where $Q^4_{3456}$ is the orbifold
${\Bbb T}^3_{3456}/\mbox{\raisebox{-.4ex}{
     $\langle\overline{\alpha},\overline{\beta}\rangle$}}$ with
$\overline{\alpha}(z_2,z_3)=(e^{2\pi i/3}z_2,-z_3)$ and
$\overline{\beta}(z_2,z_3)=(-\overline{z_2},-\overline{z_3})$.

\bigskip

\noindent
{\it (i) The base orbifold $Q^4_{3456}$}$\,$:
Denote 
${\Bbb T}^4_{3456}\,
 =\,\{ {\Bbb C}/\mbox{\raisebox{-.4ex}{$({\Bbb Z}
 \oplus e^{2\pi i/3}{\Bbb Z})$}} \}\,
 \times\,\{ {\Bbb C}/\mbox{\raisebox{-.4ex}{$({\Bbb Z}
 \oplus i{\Bbb Z})$}} \}$
by ${\Bbb T}^2_{34}(e^{2\pi i/3})\times{\Bbb T}^2_{56}(i)$
and regard it as a trivial bundle with base
${\Bbb T}^2_{34}(e^{2\pi i/3})$, fiber ${\Bbb T}^2_{56}(i)$,
and $\Gamma$ as a group of bundle automorphisms. Since this action
has the subgroup $\langle\overline{\alpha}^3\rangle$ that acts
trivially on the base, while acting as negation on the fiber,
the induced fibration
$\pi^{3456}_{34}:{\Bbb T}^4_{3456}/
  \mbox{\raisebox{-.4ex}{$\Gamma$}} \rightarrow
 Q^2_{34}={\Bbb T}^2_{34}(e^{2\pi i/3})/
  \mbox{\raisebox{-.4ex}{$\Gamma$}}$
is an $S^2(2222)$-fibration over a $D^2(^{\ast}333)$. All the
exceptional fibers are isomorphic to $D^2(^{\ast}2222)$ and have 
multiplicity $2$.

Let us now locate the singular locus $\Sigma_Q$ of $Q^4_{3456}$
in $\pi^{3456}_{34}$. As bundle automorphisms, $\overline{\alpha}$
has three invariant ${\Bbb T}^2_{56}$-fibers (at $z_2=0$,
$\frac{2+\omega}{3}$, $\frac{2i}{3}$), on each of which there are
$4$ fixed points; they correspond to the $4$ corner-reflectors of
the exceptional $D^2(^{\ast}2222)$-fiber over the $3$
corner-reflectors of $Q^2_{34}$. $\overline{\alpha}^2$ has the
above three $D^2(^{\ast}2222)$-fibers as the fixed-point set.
The $4$ fixed ${\Bbb T}^2_{34}$ of $\overline{\alpha}^3$
correspond to the $4$ sections of $\pi^{3456}_{34}$ whose union
consists of the $4$ cone-points of a generic fiber $S^2(2222)$
and the corner-reflectors of an exceptional fiber
$D^2(^{\ast}2222)$. This takes care of the rotation part of
$\Gamma$. For the reflection part, since they form two conjugacy
classes,
$\{\overline{\beta},\overline{\alpha}^2\overline{\beta},
 \overline{\alpha}^4\overline{\beta}\}$
and
$\{\overline{\alpha}\overline{\beta},
 \overline{\alpha}^3\overline{\beta},
 \overline{\alpha}^5\overline{\beta}\}$,
one only needs to consider the fixed locus of $\overline{\beta}$
and the fixed locus of $\overline{\alpha}\overline{\beta}$ on
${\Bbb T}^4_{3456}$. By inspection, both has $2$ copies of
${\Bbb T}^2$ as fixed locus. Each descends to a silvered annulus
$A^2(^{\ast\ast})$ in $Q^4_{3456}$ mapped onto $\Sigma_{Q^2_{34}}$.
When appropriately parametrized, the map is modelled by
$(y_{34},y_{56})\mapsto(y_{34}, -y_{56})$, where
$y_{34}, y_{56}\in S^1$. Together these $4$ annuli form a
$2$-torus $\Sigma_{Q^2_{34}}\times\Sigma_{D^2(^{\ast}2222)}$.
Thus one concludes the following decomposition
$$
 \Sigma_Q\;
 =\; 2\,A^2_{\beta}(^{\ast\ast})\,
   \cup\,2\,A^2_{\alpha\beta}(^{\ast\ast})\,
      \cup\, 3\,D^2_{\alpha^2}(^{\ast}2222)\,
         \cup\, 4\,D^2_{\alpha^3}(^{\ast}333)\,.
$$
From the discussion, $\Gamma_p$ for $p\in Q^4_{3456}$ can also be
obtained. ({\sc Figure 3-6-1}.)
\begin{figure}[htbp]
 \setcaption{{\sc Figure 3-6-1.} 
 \baselineskip 14pt
  The decomposition of $\Sigma_Q$. Topologically, $\Sigma_Q$ is a
  $2$-complex obtained by a
  ${\Bbb T}^2$ (i.e.\ $2A^2_{\beta}\cup 2A^2_{\alpha\beta}$)
  attached by $4$ $2$-disks (i.e.\ $D^2_{\alpha^3}$) along disjoint
  $(1,0)$-loops and $3$ $2$-disks (i.e.\ $D^2_{\alpha^2}$) along
  disjoint $(0,1)$-loops. (The ${\Bbb T}^2$ sits above
  $\partial Q^2_{34}$ and is transparent.)
 } 
 \centerline{\psfig{figure=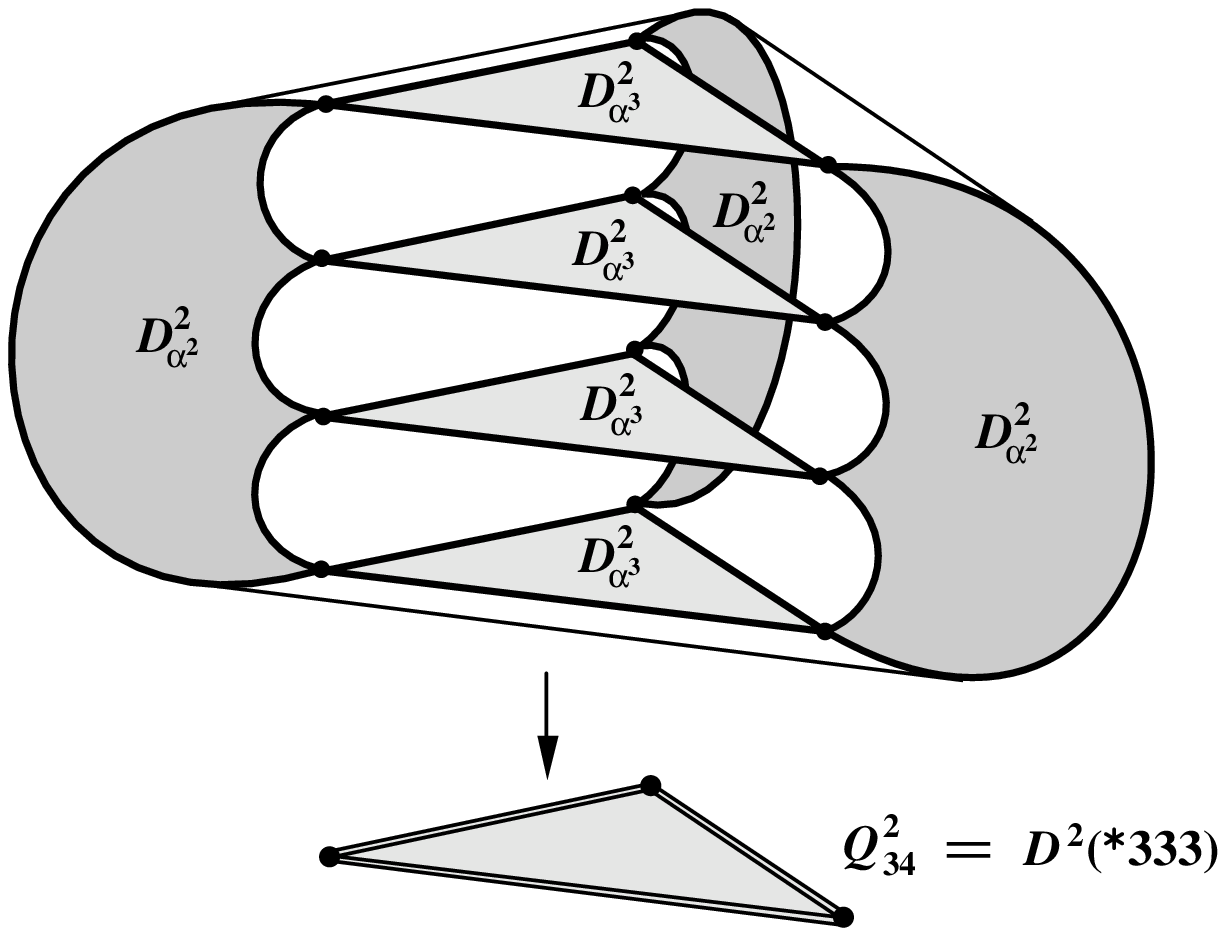,width=13cm,caption=}}
\end{figure}

\bigskip

\noindent
{\it (ii) The fibers}$\,$:
The generic fiber of $\pi_{3456}$ is ${\Bbb T}^3_{127}$. Its
exceptional fibers follow from {\sc Figure 3-6-1}:
(Note that in the following table
$p|Q^2$ means $p$ is a regular point of $Q^2$ and $m$ is the
multiplicity of the fiber $F_p$ over $p$.)
$$
\begin{array}{c}
 \mbox{
  \begin{tabular}{|c|ccccc|} \hline
   $p$
     &  $D^2_{\alpha^3}(^{\ast}333)$
     &  $D^2_{\alpha^2}(^{\ast}2222)$
     &  $A^2_{\beta}(^{\ast\ast})$
     &  $A^2_{\alpha\beta}(^{\ast\ast})$
     &  $D^2_{\alpha^3}\cap A^2_{\beta}\cap A^2_{\alpha\beta}$
           \raisebox{.6ex}{\rule{0em}{1em}}  \\[.6ex]  \hline
   $F_p$
     &  ${\Bbb T}^3_{127}/
         \mbox{\raisebox{-.4ex}{$\langle\alpha^3\rangle$}}$
     &  ${\Bbb T}^3_{127}/
         \mbox{\raisebox{-.4ex}{$\langle\alpha^2\rangle$}}$
     &  ${\Bbb T}^3_{127}/
         \mbox{\raisebox{-.4ex}{$\langle\beta\rangle$}}$
     &  ${\Bbb T}^3_{127}/
         \mbox{\raisebox{-.4ex}{$\langle\alpha\beta\rangle$}}$
     &  ${\Bbb T}^3_{127}/
         \mbox{\raisebox{-.4ex}{$\langle\alpha,\beta\rangle$}}$
           \raisebox{.6ex}{\rule{0em}{1em}} \\[.6ex] \hline
   $m$  & $2$  & $3$  & $2$  & $2$  & $12$
           \raisebox{.6ex}{\rule{0em}{1em}}  \\[.6ex] \hline
  \end{tabular} }
 \\[2.5em]
 \mbox{
  \begin{tabular}{|ccc|}\hline
   $D^2_{\alpha^2}\cap A^2_{\beta}$
      &  $D^2_{\alpha^2}\cap A^2_{\alpha\beta}$
      &  $D^2_{\alpha^2}\cap D^2_{\alpha^3}
          \cap A^2_{\beta}\cap A^2_{\alpha\beta}$
         \raisebox{.6ex}{\rule{0em}{1em}}  \\[.6ex]  \hline
   ${\Bbb T}^3_{127}/
          \mbox{\raisebox{-.4ex}{$\langle\alpha^2,\beta\rangle$}}$
      &  ${\Bbb T}^3_{127}/\mbox{\raisebox{-.4ex}{$\langle
         \alpha^2,\alpha\beta\rangle$}}$
      &  ${\Bbb T}^3_{127}/\mbox{\raisebox{-.4ex}{$\langle
         \alpha,\beta\rangle$}}$
            \raisebox{.6ex}{\rule{0em}{1em}}  \\[.6ex] \hline
   $6$  & $6$  &  $12$
        \raisebox{.6ex}{\rule{0em}{1em}} \\  \hline
  \end{tabular}  }
\end{array}
$$

All these exceptional fibers are realizable as Seifert
$3$-orbifolds ([B-S]).

\bigskip

\noindent
{\it (iii) How $S$ sits}$\,$:
Recall the 4 ${\Bbb T}^3$-components of $S$ from Example 3.5.
The standard projection ${\Bbb T}^7\rightarrow{\Bbb T}^4_{3456}$
takes a fixed ${\Bbb T}^3$ of $\beta$ (resp.\ $\alpha\beta$) on
${\Bbb T}^7$ to a fixed ${\Bbb T}^2$ of $\overline{\beta}$
(resp.\ $\overline{\alpha}\overline{\beta}$) on
${\Bbb T}^4_{3456}$, modelled by
$(y_{12}, y_{34}, y_{56})\rightarrow(y_{34}, y_{56})$, where
$y_{12}, y_{34},y_{56}\in S^1$. Composed further by the quotient
map ${\Bbb T}^4_{3456}\rightarrow Q^4_{3456}$, one concludes that 
$S$ sits over
$2\,A^2_{\beta}(^{\ast\ast})\,
  \cup\,2\,A^2(\alpha\beta)(^{\ast\ast})$,
with one ${\Bbb T}^3$ to one $A^2$.
The restriction of $\pi_{3456}$ to a such
${\Bbb T}^3\rightarrow A^2$ is modelled by the folding
$(y_{12}, y_{34}, y_{56})\mapsto(y_{34},\overline{y}_{56})$,
where $y_{12}, y_{34}, y_{56}\in
{\Bbb R}/\hspace{-.1ex}\mbox{\raisebox{-.4ex}{${\Bbb Z}$}}$ and
$\overline{y}_{56}=y_{56}$ if $y_{56}\in [0,\frac{1}{2}]$,
$ =1-y_{56}$ if $z\in[\frac{1}{2},1]$.

\bigskip

\noindent
{\it (iv) Adjustment after resolving $S$}$\,$:
After resolving $S$, one obtains a ${\Bbb T}^3$-fibration
$\widetilde{\pi}_{3456}$ of $M^7$ over
$\widetilde{Q}_{3456}=Q^4_{3456}$. From Step (iii) above, only
the fibers of $\pi_{3456}$ that sit over
$2\,A^2_{\beta}\,\cup\,2\,A^2_{\alpha\beta}$ is modified,
by $F_p\rightarrow F_p\cup\,2\,{\Bbb T}^1\times{\Bbb C}{\rm P}^1$;
all other fibers of $\pi_{3456}$ go to the fibers of
$\widetilde{\pi}_{3456}$ unchanged.

\bigskip

\noindent
{\it (v) Monodromy}$\,$:
Let $u_3(z_2,z_3)=(z_2+1,z_3)$,
$u_4(z_2,z_3)=(z_2+e^{2\pi i/3},z_3)$, $u_5(z_2,z_3)=(z_2,z_3+1)$,
and $u_6(z_2,z_3)=(z_2,z_3+i)$; then 
$$
 \pi_1^{\rm orb}(\widetilde{Q}^4_{3456})\;
 =\;\pi_1^{\rm orb}(Q^4_{3456})\;
 =\; \langle\,\overline{\alpha},\overline{\beta};
                           u_3,u_4,u_5,u_6\,\rangle
$$
and the monodromy 
$\widetilde{\rho}=\rho:\pi_1^{\rm orb}(Q^4_{3456})
  \rightarrow\Aut({\Bbb T}^3_{127})$
is determined by
$$
\begin{array}{l}
 \rho(\overline{\alpha})\,=\,\alpha^f\;
  =\;(\,(z_1,x)\mapsto(e^{\pi i/3}z_1,x+\frac{1}{6})\,)\,,\\[1ex]
 \rho(\overline{\beta})\,=\,\beta^f\;
  =\;(\,(z_1,x)\mapsto(-\overline{z_1},-x)\,)\,,
         \hspace{1em}\mbox{and}\\[1ex]
 \rho(u_3)\, =\,\rho(u_4)\, =\,\rho(u_5)\, =\,\rho(u_6)\,=\,\Id\,.
\end{array}
$$

\bigskip

This concludes the example.

\noindent\hspace{14cm} $\Box$

\bigskip

We hope these examples are enough to demonstrate the method
(cf.\ Sec.\ 5, Issue (1)).

\bigskip

\section{Fibrations of Joyce manifolds by Borcea-Voisin
         threefolds.}

In addition to the a.a/a.c.\ fibrations discussed so far, a Joyce
manifold of the first kind admits also natural fibrations by
Borcea-Voisin threefolds, as we shall now discuss.

\bigskip

\begin{flushleft}
{\bf Fibrations by Borcea-Voisin threefolds.}
\end{flushleft}
Recall from Sec.\ 1 the group
$\Gamma=\langle\alpha,\beta,\gamma\rangle$ that acts on
$({\Bbb T}^7,\varphi_0)$ by
$$
\begin{array}{rcl}
 \alpha(x_1,\cdots,x_7) & =
      & (-x_1,-x_2,-x_3,-x_4,x_5,x_6,x_7)\,, \\[.2ex]
 \beta(x_1,\cdots,x_7) & =
      & (b_1-x_1,b_2-x_2,x_3,x_4,-x_5,-x_6,x_7)\,, \\[.2ex]
 \gamma(x_1,\cdots,x_7) & =
      & (c_1-x_1,x_2,c_3-x_3,x_4,c_5-x_5,x_6,-x_7)
\end{array}
$$
with $b_1$, $b_2$, $c_1$, $c_3$, $c_5$ some appropriate constants
in $\{0,\frac{1}{2}\}$ and how this leads to Joyce manifolds
$M^7$ in $J(b_1,b_2,c_1,c_3,c_5)$.

The $7$-orbifold
${\Bbb T}^7/\mbox{\raisebox{-.4ex}{$\Gamma$}}$ can be realized
as the quotient of
${\Bbb T}^7/\mbox{\raisebox{-.4ex}{$\langle\alpha\rangle$}}$ by
$\langle\beta,\gamma\rangle$. If one chooses the complex coordinates
on ${\Bbb T}^6_{123456}$ by $z_1=x_1+ix_2$, $z_2=x_3+ix_4$,
$z_3=x_5+ix_6$, then $\beta$ acts on
${\Bbb T}^7/\mbox{\raisebox{-.4ex}{$\langle\alpha\rangle$}}$
holomorphically while $\gamma$ acts on
${\Bbb T}^7/\mbox{\raisebox{-.4ex}{$\langle\alpha\rangle$}}$
antiholomorphically. Both lift to an automorphism of the
resolution $\mbox{\rm K3}\,\times{\Bbb T}^2_{56}\times{\Bbb T}^1_7$
of ${\Bbb T}^7/\mbox{\raisebox{-.4ex}{$\langle\alpha\rangle$}}$
with $\beta$ holomorphic and $\gamma$ antiholomorphic on the
$(\mbox{\rm K3}\,\times{\Bbb T}^2_{56})$-component.
Furthermore, when the lifted $\beta$ restricts to the
$(\mbox{\rm K3}\,\times{\Bbb T}^2_{56})$-component, it serves as
the automorphism $(\iota,j)$ in the construction of Borcea-Voisin
threefolds since the $\beta$-action on the K3-component of
$\mbox{\rm K3}\,\times{\Bbb T}^2_{56}$ is a
holomorphic automorphism that acts by $(-1)$ on its holomorphic
$2$-form lifted from $dz_1\wedge dz_2$ on ${\Bbb T}^4_{1234}$ and
on the ${\Bbb T}^2_{56}$ by negation. The set of fixed-points of
the $\iota$ in this case is the union of two copies of
${\Bbb T}^2_{34}$. Let $Y$ be the Borcea-Voisin threefold obtained
by resolving the singularities of
$(\mbox{\rm K3}\times{\Bbb T}^2_{56})/
          \mbox{\raisebox{-.4ex}{$\langle\beta\rangle$}}$;
then the corresponding resolution $W^7_{\alpha,\beta}$ of
${\Bbb T}^7/\mbox{\raisebox{-.4ex}{$\langle\alpha,\beta\rangle$}}$
is a trivial bundle ${\Bbb T}^1_7\times Y$ over ${\Bbb T}^1_7$.
Now $\gamma$ acts on $W^7_{\alpha,\beta}$ as a bundle automorphism
with two invariant fibers: one over $x_7=0$ and the other over
$x_7=\frac{1}{2}$. Hence one obtains a fibration
$\pi_7:
  W^7_{\alpha,\beta}/\mbox{\raisebox{-.4ex}{$\langle\gamma\rangle$}}
 \longrightarrow [0,\frac{1}{2}]=
 {\Bbb T}^1_7/\mbox{\raisebox{-.4ex}{$\langle -1\rangle$}}$.

Let $S_{\gamma}$ be the singular set of
$W^7_{\alpha,\beta}/
    \mbox{\raisebox{-.4ex}{$\langle\gamma\rangle$}}$;
then it consists of the $3$-tori or the free ${\Bbb Z}_2$-quotient
of $3$-tori descending from the fixed ${\Bbb T}^3$'s of
$\gamma$, $\alpha\gamma$, $\beta\gamma$, or $\alpha\beta\gamma$
on ${\Bbb T}^7$ and is contained in the two exceptional fibers
$F_e=Y/\mbox{\raisebox{-.4ex}{$\langle\gamma\rangle$}}$ of
$\pi_7$. On the other hand, since all the singularities involved
arise from the fixed ${\Bbb T}^3$ of elements in $\Gamma$ and the
resolution is by transverse blowups,
$W^7_{\alpha,\beta}/
   \mbox{\raisebox{-.4ex}{$\langle\gamma\rangle$}}$
is the same orbifold as obtained from
${\Bbb T}^7/\mbox{\raisebox{-.4ex}{$\Gamma$}}$
by resolving the ${\Bbb T}^3$-components in the singular
set $S$ that arise from the fixed tori associated to
$\langle\alpha,\beta\rangle$ on ${\Bbb T}^7$.
Thus, after resolving $S_{\gamma}$, one recovers the Joyce
manifold $M^7$, which now fibers over $[0,\frac{1}{2}]$ with
generic fiber the Borcea-Voisin threefold $Y$ and exceptional
fibers over $x_7=0$ and $x_7=\frac{1}{2}$. These two exceptional
fibers are isomorphic since the construction is invariant under
the map
$(x_1,\cdots,x_6,x_7)\mapsto(x_1,\cdots,x_6,\frac{1}{2}-x_7)$,
which flips $x_7=0$ and $x_7=\frac{1}{2}$ to each other.
Both fibers have multiplicity $2$.

\bigskip

\noindent
{\it Remark 4.1 [topology of $Y$].}
 From Sec.\ 1, the Borcea-Voisin threefold $Y$ has Hodge numbers
 $h^{1,1}(Y)=h^{2,1}(Y)=19$.
 
\bigskip

\noindent
{\it Remark 4.2 [exceptional fiber].}
 The exceptional fiber $\widetilde{F_e}$ in $M^7$ is a connected
 $6$-space that is a $6$-manifold-with-boundary except at some
 singularities on the boundary. To see this, first notice that,
 over $x_7\in\{0,\frac{1}{2}\}$ the fixed-point set of the
 $\gamma$-action on $Y$ coincides with $S_{\gamma}$; hence,
 $F_e=Y/\mbox{\raisebox{-.4ex}{$\langle\gamma\rangle$}}$ is a
 manifold except at $S_{\gamma}$. Now let $S_0$ be a component of
 $S_{\gamma}$ and consider the effect of resolving $S_0$ to $F_e$
 case by case, following the local models in Sec.\ 1 for the
 tubular neighborhood $\nu(S_0)$ of $S_0$:
 \begin{itemize}
  \item
  {\it Case (a)} \hspace{1ex}
  $\nu(S_0)={\Bbb T}^3_{246}\times({\Bbb R}^4_{1357}/
    \mbox{\raisebox{-.4ex}{$\langle -1\rangle$}})\,$:
  Identify ${\Bbb R}^4_{1357}$ with ${\Bbb C}^2$, say, via the
  decomposition ${\Bbb R}^2_{13}\times{\Bbb R}^2_{57}$ and recall
  from Sec.\ 1 the resolution
  $\psi:T^{\ast}{\Bbb C}{\rm P}^1\rightarrow
      {\Bbb C}^2/\mbox{\raisebox{-.4ex}{$\langle -1\rangle$}}$,
  whose exceptional locus $E$ is the $0$-section of
  $T^{\ast}{\Bbb C}{\rm P}^1$. The transverse directions to $S_0$
  in $F_e$ is given by
  ${\Bbb R}^3_{135}/\mbox{\raisebox{-.4ex}{$\langle -1\rangle$}}
   =({\Bbb C}\times{\Bbb R}_5)
       /\mbox{\raisebox{-.4ex}{$\langle -1\rangle$}}$ in
  ${\Bbb C}^2/\mbox{\raisebox{-.4ex}{$\langle -1\rangle$}}$.
  Thus resolving $S_0$ changes $\nu(S_0)$ to
  ${\Bbb T}^3_{246}\times
     \{\psi^{-1}(({\Bbb C}\times{\Bbb R}_5)
        /\mbox{\raisebox{-.4ex}{$\langle -1\rangle$}})\}$.
  To understand
  $\psi^{-1}(({\Bbb C}\times{\Bbb R}_5)/
           \mbox{\raisebox{-.4ex}{$\langle -1\rangle$}})$,
  recall the foliation ${\cal F}$ of
  $({\Bbb C}^2-\{(0,0)\})/
             \mbox{\raisebox{-.4ex}{$\langle -1\rangle$}}$
  by the complex cone-lines
  ${\Bbb C}^{\ast}_{\lambda}/
      \mbox{\raisebox{-.4ex}{$\langle -1\rangle$}}
   ={\Bbb C}^{\ast}\cdot(1,\lambda)\,/
      \mbox{\raisebox{-.4ex}{$\langle -1\rangle$}}$,
  where ${\Bbb C}^{\ast}={\Bbb C}-\{0\}$ and
  $\lambda\in{\Bbb C}{\rm P}^1={\Bbb C}\cup\{\infty\}$,
  and the fact that $\psi$ is obtained by blowing up the
  singularity $(0,0)$ in
  ${\Bbb C}^2/\mbox{\raisebox{-.4ex}{$\langle -1\rangle$}}$ to
  the leaf space ${\Bbb C}{\rm P}^1$ of ${\cal F}$, which is
  identified with $E$. ${\cal F}$ now induces an equivalence
  relation $\sim$ on
  $({\Bbb C}\times{\Bbb R}_5-\{(0,0)\})/
               \mbox{\raisebox{-.4ex}{$\langle -1\rangle$}}$
  by setting $(z,x_5)\sim(z^{\prime}, x_5^{\prime})$ if both lie
  in a same
  ${\Bbb C}^{\ast}_{\lambda}/
        \mbox{\raisebox{-.4ex}{$\langle -1\rangle$}}$.
  Since every
  ${\Bbb C}^{\ast}_{\lambda}/
      \mbox{\raisebox{-.4ex}{$\langle -1\rangle$}}$
  intersects
  $({\Bbb C}\times{\Bbb R}_5-\{(0,0)\})/
               \mbox{\raisebox{-.4ex}{$\langle -1\rangle$}}$,
  $E$ is also the exceptional locus for the restriction of
  $\psi$ on
  $\psi^{-1}(({\Bbb C}\times{\Bbb R}_5)/
               \mbox{\raisebox{-.4ex}{$\langle -1\rangle$}})$.
  When restricted to the unit projective space
  ${\Bbb R}{\rm P}^2
    =S^2/\mbox{\raisebox{-.4ex}{$\langle -1\rangle$}}$
  in
  $({\Bbb C}\times{\Bbb R}_5)/
               \mbox{\raisebox{-.4ex}{$\langle -1\rangle$}}$,
  each $\sim$-equivalence class is simply a point in
  ${\Bbb R}{\rm P}^2$ except the class associated to
  ${\Bbb C}^{\ast}_0/
     \mbox{\raisebox{-.4ex}{$\langle -1\rangle$}}$,
  which is a generating circle $C$ for
  $\pi_1({\Bbb R}{\rm P}^2)$. Since $\nu(C)$ in
  ${\Bbb R}{\rm P}^2$ is a M\"{o}bius strip, one has that
  $\psi^{-1}(({\Bbb C}\times{\Bbb R}_5)/
               \mbox{\raisebox{-.4ex}{$\langle -1\rangle$}})$
  is a manifold with boundary
  $E=S^2={\Bbb R}{\rm P}^2/\mbox{\raisebox{-.4ex}{$\sim$}}$, 
  on which there is an isolated singularity $\ast$ whose
  link $\partial\nu(\ast)$ is a M\"{o}bius strip instead of a
  $2$-disk for all other points on the boundary. This implies
  that resolving such $S_0$ contributes to $\widetilde{F_e}$
  a boundary component ${\Bbb T}^3\times S^2$ with a singular
  locus ${\Bbb T}^3\times\{\ast\}$.

  \item
  {\it Case (b)} \hspace{1ex}
  $\nu(S_0)=\{{\Bbb T}^3\times({\Bbb C}^2/
         \mbox{\raisebox{-.4ex}{$\langle -1\rangle$}})\}/
                     \mbox{\raisebox{-.4ex}{${\Bbb Z}_2$}}\,$:
  Since the free ${\Bbb Z}_2$-action on
  ${\Bbb T}^3\times({\Bbb C}^2/
         \mbox{\raisebox{-.4ex}{$\langle -1\rangle$}})$
  arises from a subgroup in $\langle\alpha,\beta\rangle$ and its
  restriction to the
  ${\Bbb C}^2/
       \mbox{\raisebox{-.4ex}{$\langle -1\rangle$}}$-component
  is either holomorphic or anti-holomorphic, it leaves
  $({\Bbb C}\times{\Bbb R})/
            \mbox{\raisebox{-.4ex}{$\langle -1\rangle$}}$
  in Case (a) invariant and is compatibe with the equivalence
  relation $\sim$. Together with (a), this implies that
  resolving such $S_0$ contributes to $\widetilde{F_e}$
  a boundary component, which is the free quotient
  $({\Bbb T}^3\times S^2)/\mbox{\raisebox{-.4ex}{${\Bbb Z}_2$}}$
  with a singular locus
  $({\Bbb T}^3/\mbox{\raisebox{-.4ex}{${\Bbb Z}_2$}})
                                               \times\{\ast\}$.
 \end{itemize}
 This concludes the remark.

\bigskip

\noindent 
{\it Remark 4.3 [non-uniqueness].}
In general, the fibration of a Joyce manifold of the first kind
by Borcea-Voisin threefolds is not unique. One can identify the
complex-real coordiates $(z_1,z_2,z_3,x)$ with $(x_1,\cdots,x_7)$
in several ways. Each may give different factorization
$\mbox{\rm K3}\times{\Bbb T}^2\times{\Bbb T}^1$ in the
intermediate step, which then leads to different fibrations of
$M^7$ by Borcea-Voisin threefolds $Y$. However the Borcea-Voisin
manifolds $Y$ that appear in these different fibrations are
homeomorphic with Hodge numbers as in Remark 4.1.

\bigskip

\noindent
{\it Remark 4.4 [$7$-space from rolling Calabi-Yau].}
The existence of fibrations of a Joyce manifold $M^7$ of the first
kind by Borcea-Voisin threefolds renders such $M^7$ similar to the
$7$-spaces obtained by rolling Calabi-Yau threefolds, as discussed
in [Li].

\bigskip

\begin{flushleft}
{\bf The harmony with a.a./a.c.\ fibrations.}
\end{flushleft}
The Borcea-Voisin threefold $Y$ constructed above is naturally
fibred over the $2$-orbifold $Q^2(2222)$ with generic fiber the 
K3 surface $X$ obtained by resolving
${\Bbb R}^4_{1234}/\mbox{\raisebox{-.4ex}{$\langle -1\rangle$}}$.
This in turn induces a fibration of $M^7$ by $X$ over a
$3$-orbifold. On the other hand, since the fibration of $Y$ by $X$
can be regarded as descending from the decomposition
${\Bbb T}^6_{123456}={\Bbb T}^4_{1234}\times{\Bbb T}^2_{56}$,
which is compactible with the associative-coassociative product
decomposition ${\Bbb T}^7={\Bbb T}^3_{567}\times{\Bbb T}^4_{1234}$,
the induced fibration of $M^7$ is simply an a.c.\ K3-fibration
discussed in Sec.\ 2.

Likewise, with respect to the complex structure on $X$ induced
from the above identification of ${\Bbb R}^4_{1234}$ with
${\Bbb C}^2$, the special Lagrangian ${\Bbb T}^3$-fibrations for
$Y$ as constructed in [G-W] come from either of the decompositions
${\Bbb T}^6_{123456}={\Bbb T}^3_{135}\times{\Bbb T}^3_{246}$ and
${\Bbb T}^6_{123456}={\Bbb T}^3_{136}\times{\Bbb T}^3_{245}$,
which are compatible with the associative-coassociative product
decompositions ${\Bbb T}^7={\Bbb T}^3_{246}\times{\Bbb T}^4_{1357}$
and ${\Bbb T}^7={\Bbb T}^3_{136}\times{\Bbb T}^3_{2457}$
respectively. Hence the induced ${\Bbb T}^3$-fibrations for $M^7$
are a.a.\ ${\Bbb T}^3$-fibrations discussed in Sec.\ 2.
In addition to these, one also has ${\Bbb T}^3$-fibrations of $Y$
induced from the decompositions
${\Bbb T}^3_{145}\times{\Bbb T}^3_{236}$
 (from ${\Bbb T}^3_{145}\times{\Bbb T}^3_{2367}$) and
${\Bbb T}^3_{235}\times{\Bbb T}^3_{146}$
 (from ${\Bbb T}^3_{235}\times{\Bbb T}^3_{1467}$) of
${\Bbb T}^6_{123456}$. These induce also a.a.\
${\Bbb T}^3$-fibrations for $M^7$ discussed in Sec.\ 2.

\bigskip

\noindent
{\bf Example 4.5 [$J(0,\frac{1}{2},\frac{1}{2},0,0)$].}
([Jo1]: II, Example 4.)
Following [Jo1] and previous discussions, the fixed $3$-tori
${\Bbb T}^3_{246}$ of $\gamma$ on ${\Bbb T}^7$ descends to
$S_{\gamma}$, which consists of $8$ copies of
${\Bbb T}^3/\mbox{\raisebox{-.4ex}{${\Bbb Z}_2$}}$ in
$W^7_{\alpha,\beta}/\mbox{\raisebox{-.4ex}{$\langle\gamma\rangle$}}$
with $4$ copies over $x_0=0$ and $4$ copies over $x_7=\frac{1}{2}$.
After resolving $S_{\gamma}$, $M^7$ has a fibration over
$[0,\frac{1}{2}]$ with generic fiber $Y$.
From Remark 4.2, the exceptional fiber $\widetilde{F_e}$ is a
connected $6$-space that is a manifold with $4$ boundary components
$\{{\Bbb T}^3\times S^2\}/\mbox{\raisebox{-.4ex}{${\Bbb Z}_2$}}$
except for a singular locus
${\Bbb T^3}/\mbox{\raisebox{-.4ex}{${\Bbb Z}_2$}}$
on each boundary component. The K3-fibration of $Y$ induces the
a.c.\ K3-fibration of $M^7$. The standard projection
from ${\Bbb T}^3_{567}$ to ${\Bbb T}^1_7$ induces a fibration of
$Q^3_{567}$ by $Q^2_{56}(2222)$. Likewise, the decomposition
${\Bbb T}^6={\Bbb T}^3_{135}\times{\Bbb T}^3_{246}$ induces a
${\Bbb T}^3$-fibration of $Y$ over a $3$-orbifold $Q^3_{135}$,
which then induces an a.a.\ ${\Bbb T}^3$-fibration of $M^7$. The
prejection of ${\Bbb T}^4_{1357}$ to ${\Bbb R}_7$ induces a
fibration of $Q^4_{1357}$ by a $3$-orbifold $Q^3_{135}$.
Diagramatically, one has
$$
\begin{array}{ccccc}
 \mbox{\rm K3} & \longrightarrow & Y
                  & \longrightarrow & M^7  \\[1ex]
 & & \downarrow & & \downarrow   \\[1ex]
 & & Q^2_{56}(2222) & \longrightarrow & Q^3_{567}  \\[1ex]
 & & & & \downarrow  \\[1ex]
 & & & & [0,\frac{1}{2}] 
\end{array}
  \hspace{2em}\mbox{and}\hspace{2em}
\begin{array}{cccccl}
 {\Bbb T}^3_{246} & \longrightarrow & Y
                  & \longrightarrow & M^7 & \\[1ex]
 & & \downarrow & & \downarrow & \\[1ex]
 & & Q^3_{135} & \longrightarrow & Q^4_{1357} & \\[1ex]
 & & & & \downarrow & \\[1ex]
 & & & & [0,\frac{1}{2}] &,
\end{array}
$$
where ``$\rightarrow$" stands for ``the generic fiber of"
and ``$\downarrow$" stands for ``fibred over".
The discussion for all other cases are similar. 

\noindent\hspace{14cm} $\Box$

\bigskip

This concludes our discussions on the fibrations of Joyce manifolds.

\bigskip

\section{Remarks on further questions/works.} 

As readers may have noticed, the various fibrations of Joyce
manifolds discussed and illustrated by examples in this paper come
freely and naturally from the work of Joyce; we strongly feel that
there must be some place for them in string/M-theory. Indeed this
has been pursued vigorously by Acharya ([Ar1] and [Ar2]). On the
other hand, the issue of dualities in string/M-theory requires us
to go beyond this present work. Among many further questions, let
us list three issues$^2$ with some comments for future
pursuit/cooperations:
\footnotetext[2]{
 I would like to thank Prof.\ Jacques Distler who kindly gave me
 a long discussion/lecture on physics related to fibrations of
 Joyce manifolds, which I have yet to digest.
} 
\begin{itemize}
 \item
 {\it (1) Complete table}$\,$:
 Give a complete table of the a.a./a.c.\ fibrations of all the
 Joyce manifolds of the first and the second kind, up to fibration
 isomorphism (cf.\ Table 2 in [Jo1]:II).
 Design some elegant notations to code the combinatorial data of
 the fibrations. Give more informations of the exceptional fibers,
 e.g.\ their various numerical topological invariants.
 This work seems endurable, though experience told us that it could
 be very tedious or even brain-racking. Since many steps are
 straightforward, likely a computer code should be able to
 realize quite a part of the work.

 \item
 {\it (2) How far from associative/coassociative fibrations}$\,$:
 The associative/coassociative fibrations for a $7$-manifold of
 holonomy $G_2$ is the analogue of the fibrations of a Calabi-Yau
 threefold by supersymmetric cycles (i.e.\ special Lagrangian
 submanifolds). Thus, one likes to design some quantity of
 measuring how far the fibrations given in the table in Issue (1)
 are from being a true associative/coassociative fibration and
 to understand the physical meaning of this quantity.

 \item
 {\it (3) The role of global structure}$\,$:
 Before a comprehensive table in Issue (1) is completed,
 following the $5$-step-routine in Sec.\ 3, one can still
 understand the global structure of an a.a./a.c.\ fibrations of
 a Joyce manifold. For examples of Joyce manifolds that has
 already appeared in physics literature (e.g.\ [Ar1] and [Ar2]),
 one of the the next major questions is then:
 \begin{quote}
  \hspace{-1.8em}{\bf Q.} {\it
  How could its global structure play roles in string/M-theory
  dualities?}
 \end{quote}
 In other words, in a string/M-theory duality involving fibrations
 of Joyce manifolds, where is the global structure of the
 fibration hidden?
\end{itemize}

With these open question/work/project in mind, let us conclude
this paper.


\enddocument

%% file: psfig-97.tex
\catcode`\@=11\relax
\newwrite\@unused
\def\typeout#1{{\let\protect\string\immediate\write\@unused{#1}}}

\def\psglobal#1{
\immediate\special{ps: plotfile #1 }}
\def\psfiginit{\typeout{psfiginit}
\immediate\psglobal{figtex.pro}%
\special{ps:: /TeXMagnification {\the\mag} def}
}

\def\@nnil{\@nil}
\def\@empty{}
\def\@psdonoop#1\@@#2#3{}
\def\@psdo#1:=#2\do#3{\edef\@psdotmp{#2}\ifx\@psdotmp\@empty \else
    \expandafter\@psdoloop#2,\@nil,\@nil\@@#1{#3}\fi}
\def\@psdoloop#1,#2,#3\@@#4#5{\def#4{#1}\ifx #4\@nnil \else
       #5\def#4{#2}\ifx #4\@nnil \else#5\@ipsdoloop #3\@@#4{#5}\fi\fi}
\def\@ipsdoloop#1,#2\@@#3#4{\def#3{#1}\ifx #3\@nnil
       \let\@nextwhile=\@psdonoop \else
      #4\relax\let\@nextwhile=\@ipsdoloop\fi\@nextwhile#2\@@#3{#4}}
\def\@tpsdo#1:=#2\do#3{\xdef\@psdotmp{#2}\ifx\@psdotmp\@empty \else
    \@tpsdoloop#2\@nil\@nil\@@#1{#3}\fi}
\def\@tpsdoloop#1#2\@@#3#4{\def#3{#1}\ifx #3\@nnil
       \let\@nextwhile=\@psdonoop \else
      #4\relax\let\@nextwhile=\@tpsdoloop\fi\@nextwhile#2\@@#3{#4}}
\def\psdraft{
	\def\@psdraft{0}
	\def\@psdraftspecial{100}
}
\def\psdraftspecial{
	\def\@psdraft{0}
	\def\@psdraftspecial{0}
}
\def\psfull{
	\def\@psdraft{100}
}
\psfull

\newif\if@prologfile
\newif\if@postlogfile
\newif\if@bbllx
\newif\if@bblly
\newif\if@bburx
\newif\if@bbury
\newif\if@height
\newif\if@width
\newif\if@rheight
\newif\if@rwidth
\newif\if@clip
\newif\if@right
\newif\if@left
\newif\if@toplines
\newif\if@box
\newif\if@caption
\newif\if@surround
\newif\if@captionwidth
\newif\if@captionwrite
\newif\if@captionopen
\def\@p@@sclip#1{\@cliptrue}
\def\@p@@sfile#1{
		\def\@p@sfile{#1}
}
\def\@p@@sfigure#1{
		\def\@p@sfile{#1}
}
\def\@p@sfake{\hbox to 0pt{\hss Whatever\hss}}
\def\@p@@sbbllx#1{
		\@bbllxtrue
		\@d@mscratch=#1
		\edef\@p@sbbllx{\number\@d@mscratch}
}
\def\@p@@sbblly#1{
		\@bbllytrue
		\@d@mscratch=#1
		\edef\@p@sbblly{\number\@d@mscratch}
}
\def\@p@@sbburx#1{
		\@bburxtrue
		\@d@mscratch=#1
		\edef\@p@sbburx{\number\@d@mscratch}
}
\def\@p@@sbbury#1{
		\@bburytrue
		\@d@mscratch=#1
		\edef\@p@sbbury{\number\@d@mscratch}
}
\def\@p@@sheight#1{
		\@heighttrue
		\@d@mscratch=#1
   		\edef\@p@sheight{\number\@d@mscratch}
}
\def\@p@@swidth#1{
		\@widthtrue
		\@d@mscratch=#1
		\edef\@p@swidth{\number\@d@mscratch}
}
\def\@p@@srheight#1{
		\@rheighttrue
		\@d@mscratch=#1
		\edef\@p@srheight{\number\@d@mscratch}
}
\def\@p@@srwidth#1{
		\@rwidthtrue
		\@d@mscratch=#1
		\edef\@p@srwidth{\number\@d@mscratch}
}
\def\@p@@sright#1{\@righttrue \@surroundtrue}
\def\@p@@sleft#1{\@lefttrue \@surroundtrue}
\def\@p@@sextraheight#1{\@d@mextraheight=#1}
\def\@p@@sbox#1{\@boxtrue}
\def\@p@@scaption#1{\@captiontrue}
\def\@p@@stoplines#1{
		\@toplinestrue
		\@c@ttoplines=#1
}
\def\@p@@scaptionwidth#1{
		\@captionwidthtrue
	  	\@d@mcaptionwidth=#1
}
\def\@p@@scaptionwrite#1{
		\global\@captionwritetrue
		\global\@w@rname=\expandafter{\jobname_captions.tex}
		\typeout{Captions are written to \the\@w@rname}
}

\def\@p@@sprolog#1{\@prologfiletrue\def\@prologfileval{#1}}
\def\@p@@spostlog#1{\@postlogfiletrue\def\@postlogfileval{#1}}
\def\@cs@name#1{\csname #1\endcsname}
\def\@setparms#1=#2,{\@cs@name{@p@@s#1}{#2}}
\def\ps@init@parms{
		\@bbllxfalse \@bbllyfalse
		\@bburxfalse \@bburyfalse
		\@heightfalse \@widthfalse
		\@rheightfalse \@rwidthfalse
		\def\@p@sbbllx{}\def\@p@sbblly{}
		\def\@p@sbburx{}\def\@p@sbbury{}
		\def\@p@sheight{}\def\@p@swidth{}
		\def\@p@srheight{}\def\@p@srwidth{}
		\def\@p@sfile{}
		\def\@p@scost{10}
		\def\@sc{}
		\@prologfilefalse
		\@postlogfilefalse
		\@clipfalse
		\@rightfalse \@leftfalse
		\@boxfalse \@captionfalse
		\@toplinesfalse \@surroundfalse
		\@d@mextraheight=0pt
 		\@c@ttoplines=0
		\@pshape={} \def\@p@srheight@total{}
		\@captionwidthfalse \@d@mcaptionwidth=0pt
}
\def\parse@ps@parms#1{
	 	\@psdo\@psfiga:=#1\do
		   {\expandafter\@setparms\@psfiga,}}
\newif\ifno@bb
\newif\ifnot@eof
\newread\ps@stream
\newtoks\@linetok
\def\bb@missing{
	\typeout{psfig: searching \@p@sfile \space  for bounding box}
	\openin\ps@stream=\@p@sfile
	\no@bbtrue
	\not@eoftrue
	\catcode`\%=12
	\loop
		\read\ps@stream to \line@in
		\global\@linetok=\expandafter{\line@in}
		\ifeof\ps@stream \not@eoffalse \fi
		\@bbtest{\@linetok}
		\if@bbmatch\not@eoffalse\expandafter\bb@cull\the\@linetok\fi
	\ifnot@eof \repeat
	\catcode`\%=14
}	
\catcode`\%=12
\newif\if@bbmatch
\def\@bbtest#1{\expandafter\@a@\the#1
\long\def\@a@#1
     \ifx\@bbtest#2\@bbmatchfalse\else\@bbmatchtrue\fi}
\long\def\bb@cull#1 #2 #3 #4 #5 {
	\@d@mscratch=#2 bp\edef\@p@sbbllx{\number\@d@mscratch}
	\@d@mscratch=#3 bp\edef\@p@sbblly{\number\@d@mscratch}
	\@d@mscratch=#4 bp\edef\@p@sbburx{\number\@d@mscratch}
	\@d@mscratch=#5 bp\edef\@p@sbbury{\number\@d@mscratch}
	\no@bbfalse
}
\catcode`\%=14
\def\compute@bb{
		\no@bbfalse
		\if@bbllx \else \no@bbtrue \fi
		\if@bblly \else \no@bbtrue \fi
		\if@bburx \else \no@bbtrue \fi
		\if@bbury \else \no@bbtrue \fi
		\ifno@bb \bb@missing \fi
		\ifno@bb \typeout{FATAL ERROR: no bb supplied or found}
			\no-bb-error
		\fi
		\count203=\@p@sbburx
		\count204=\@p@sbbury
		\advance\count203 by -\@p@sbbllx
		\advance\count204 by -\@p@sbblly
		\edef\@bbw{\number\count203}
		\edef\@bbh{\number\count204}
}
\def\in@hundreds#1#2#3{\count240=#2 \count241=#3
		     \count100=\count240	
		     \divide\count100 by \count241
		     \count101=\count100
		     \multiply\count101 by \count241
		     \advance\count240 by -\count101
		     \multiply\count240 by 10
		     \count101=\count240	
		     \divide\count101 by \count241
		     \count102=\count101
		     \multiply\count102 by \count241
		     \advance\count240 by -\count102
		     \multiply\count240 by 10
		     \count102=\count240	
		     \divide\count102 by \count241
		     \count200=#1\count205=0
		     \count201=\count200
			\multiply\count201 by \count100
		     	\advance\count205 by \count201
		     \count201=\count200
			\divide\count201 by 10
		     	\multiply\count201 by \count101
			\advance\count205 by \count201
		     \count201=\count200
			\divide\count201 by 100
			\multiply\count201 by \count102
			\advance\count205 by \count201
		     \edef\@result{\number\count205}
}
\def\compute@wfromh{
		\in@hundreds{\@p@sheight}{\@bbw}{\@bbh}
		\edef\@p@swidth{\@result}
}
\def\compute@hfromw{
		\in@hundreds{\@p@swidth}{\@bbh}{\@bbw}
		\edef\@p@sheight{\@result}
}
\def\compute@handw{
		\if@height
			\if@width
			\else
				\compute@wfromh
			\fi
		\else
			\if@width
				\compute@hfromw
			\else
				\edef\@p@sheight{\@bbh}
				\edef\@p@swidth{\@bbw}
			\fi
		\fi
}
\def\compute@resv{
		\if@rheight \else \edef\@p@srheight{\@p@sheight} \fi
		\if@rwidth \else \edef\@p@srwidth{\@p@swidth} \fi
		\edef\@p@srheight@total{\@p@srheight}
}
\newtoks\@pshape
\def\@c@ttoplines{\count120}
\def\@c@theightcount{\count121}
\def\@c@tshapecount{\count122}
\newdimen\@d@mwidthshape
\newdimen\@d@mextraheight
\newdimen\@d@mscratch
\def\compute@parshape{
	\if@right
		\if@left
	   		\typeout{error: Can't have both left and right set}
			\@leftfalse
		\fi
	\fi
	\@d@mscratch=\@p@swidth truesp
	\divide \@d@mscratch by 19
	\multiply \@d@mscratch by 20
	\edef\@p@swidthdimen{\the\@d@mscratch}
	\@c@tshapecount=\@c@ttoplines
 	\@d@mscratch=\baselineskip
	\multiply \@d@mscratch by \@c@ttoplines
	\advance \@d@mscratch by .4\baselineskip
    	\edef\@p@stopdistance{\the\@d@mscratch }
	\@d@mscratch=\@p@sheight truesp
	\divide \@d@mscratch by 2
	\edef\@p@shalfboxheight{\the\@d@mscratch}
	\if@toplines
		\loop \@pshape=\expandafter{\the\@pshape 0pt \hsize}
		\advance\@c@ttoplines by -1
		\ifnum\@c@ttoplines>0 \repeat
	\fi
   	\@c@theightcount=\@p@srheight@total
	\advance \@c@theightcount by \@d@mextraheight
	\divide  \@c@theightcount by \baselineskip
	\advance \@c@theightcount by 1
    	\advance \@c@tshapecount by \@c@theightcount
	\advance \@c@theightcount by -1
	\@d@mwidthshape=\hsize
     	\advance \@d@mwidthshape by -\@p@swidthdimen
	\if@left
		\def\@moveshape{0pt}
		\ifnum\@c@theightcount>0
		      	\loop
			\@pshape=%
			\expandafter{\the\@pshape %
					\@p@swidthdimen \@d@mwidthshape}
			\advance \@c@theightcount by -1
			\ifnum\@c@theightcount>0 \repeat
		\else
			\advance \@c@tshapecount by 1
		\fi
	\fi
	\if@right
		\@d@mscratch=\hsize
		\advance \@d@mscratch by -\@p@swidth truesp
		\edef\@moveshape{\@d@mscratch}
		\ifnum\@c@theightcount>0
			\loop
			\@pshape=\expandafter{\the\@pshape 0pt \@d@mwidthshape}
			\advance \@c@theightcount by -1
			\ifnum\@c@theightcount>0 \repeat
		\else
			\advance \@c@tshapecount by 1
		\fi
	\fi
	\ifnum \@p@srheight=0
		\@pshape={}
		\@c@tshapecount = 0
	\else
	 	\@pshape=\expandafter{\the\@pshape 0pt \hsize}
	\fi
}
\def\@p@ssetsurroundboxes{
	\global\parshape=\count122 \the\@pshape		
 	\moveright\@moveshape
	\vbox to 0pt\bgroup\hskip0pt\vskip\@p@stopdistance
}
\newtoks\@captiontok
\newbox\@b@xcaption
\newdimen\@d@mcaptionwidth
\newdimen\@d@mcaptionheight
\newwrite\@w@rcaption
\newtoks\@w@rname
\def\setcaption#1{\@captiontok={#1}}
\def\@set@caption{
	\setbox\@b@xcaption\vbox{\hsize\@d@mcaptionwidth
	\tolerance=9000 \baselineskip=11.4pt
	\noindent\relax\the\@captiontok}
	\if@captionwrite
		\if@captionopen
		\else
			\global\@captionopentrue
			\immediate\openout\@w@rcaption=\the\@w@rname
		\fi
		\immediate\write\@w@rcaption{\the\@captiontok}
		\immediate\write\@w@rcaption{}
	\fi
}
\def\compute@caption{
	\if@captionwidth
	\else
		\@d@mcaptionwidth = \@p@swidth truesp
		\divide \@d@mcaptionwidth by 20
		\multiply \@d@mcaptionwidth by 17
	\fi
	\@set@caption
	\@d@mcaptionheight=\ht\@b@xcaption
	\if@rheight
	\else
		\count100=\@p@srheight
	   	\advance \count100 by \@d@mcaptionheight
	   	\advance \count100 by \bigskipamount
	   	\advance \count100 by \medskipamount
	   	\edef\@p@srheight@total{\number\count100}
	\fi
}
\newif\if@alreadyjtem \@alreadyjtemfalse
\def\newpar{\hfil\vadjust{\vskip\parskip}%
	{\count100=\parskip
	\count101=\baselineskip
	\divide\count101 by 10  \multiply\count101 by 3
	\advance \count100 by \count101
	\divide\count100 by \baselineskip
	\advance\count100 by \prevgraf
	\global\prevgraf=\count100}%
	\break\if@alreadyjtem\else\indent\fi%
}
\let\sav@par=\par
\def\jtem#1{%
    	\if@alreadyjtem\else\bgroup\fi
	\def\par{\sav@par\egroup\sav@par}
	\if@alreadyjtem\else\leftskip\parindent\fi
	\@alreadyjtemtrue
	\noindent\hskip0pt
	\llap{#1\ }\ignorespaces
}
\def\compute@sizes{%
	\compute@bb
	\compute@handw
  	\compute@resv
	\if@caption
		\compute@caption
	\fi
	\if@surround
		\compute@parshape
	\fi
}
\def\@p@sdospecials{
	\ifnum\@p@scost<\@psdraft
	       	\typeout{psfig: including \@p@sfile \space }
	\fi
	\special{ps::[begin] 	\@p@swidth \space \@p@sheight \space
			\@p@sbbllx \space \@p@sbblly \space
			\@p@sbburx \space \@p@sbbury \space
			startTexFig \space }
	\ifnum\@p@scost<\@psdraft
		\if@clip
			\typeout{(clip)}
			\special{ps:: \@p@sbbllx \space \@p@sbblly \space
				\@p@sbburx \space \@p@sbbury \space
			    	doclip \space }
		\fi
	\fi
	\if@box
		\typeout{(box)}
  		\special{ps:: \@p@sbbllx \space \@p@sbblly \space
			\@p@sbburx \space \@p@sbbury \space
		    	dobox \space }
	\fi
	\ifnum\@p@scost<\@psdraft
		\if@prologfile
	    		\special{ps: plotfile \@prologfileval \space }
		\fi
		\special{ps: plotfile \@p@sfile \space }
    		\if@postlogfile
			\special{ps: plotfile \@postlogfileval \space }
		\fi
	\fi
	\special{ps::[end] endTexFig \space }
}
\newif\if@putinvbox

\def\psfig#1{{%
	\ifhmode%
		\vbox\bgroup
		\@putinvboxtrue
	\else
		\@putinvboxfalse
	\fi
       	\ps@init@parms
	\parse@ps@parms{#1}
       	\compute@sizes
	\if@surround
		\psfig@for@surround
	\else
		\psfig@for@regular
	\fi
	\if@putinvbox
       		\egroup
	\fi
}}
\def\psfig@for@surround{%
	\@p@ssetsurroundboxes
	\ifnum\@p@scost<\@psdraft
		\@p@sdospecials
		\vbox to \@p@srheight true sp{\vss}
       	\else
		\if@box
			\@p@sdospecials
		\fi
		\vbox to \@p@srheight true sp{
			\vskip\@p@shalfboxheight
			\hbox to \@p@srwidth true sp{
				\hss
				\ifnum\@p@scost<\@psdraftspecial
					\@p@sfile
				\else
					\@p@sfake
				\fi
      				\hss
			}
		\vss
		}
	\fi
	\if@caption
		\medskip
		\hbox to \@p@srwidth true sp{
			\hss
			\box\@b@xcaption
			\hss
		}
 		\medskip
	\fi
	\vss\egroup
	\vskip-\parskip
}

\def\psfig@for@regular{%
	\if@putinvbox
	\else
		\vskip\parskip
	\fi
	\ifnum\@p@scost<\@psdraft
		\@p@sdospecials
		\vbox to \@p@srheight true sp{%
			\hbox to \@p@srwidth true sp{
			\hfil
			}
		\vfil
		}
       	\else
		\if@box
			\@p@sdospecials
		\fi
	    	\vbox to \@p@srheight true sp{
			\vss
			\hbox to \@p@srwidth true sp{
				\hss
				\ifnum\@p@scost<\@psdraftspecial
					\@p@sfile
				\else
					\@p@sfake
				\fi
				\hss
			}
		    	\vss
		}
	\fi
	\if@caption
		\medskip
		\hbox to \@p@srwidth true sp{
			\hss
			\box\@b@xcaption
			\hss
		}
		\bigskip
	\fi
	\if@putinvbox
	\else
		\vskip-\parskip
	\fi
}
\catcode`\@=12\relax

%% file: m-joyce.bbl
\newpage

{\footnotesize
\begin{thebibliography}{AAAAaa}
%
\bibitem[Ac1]{} B.S.\ Acharya,
 {\it $N=1$ heterotic/M-theory duality and Joyce manifolds},
 {\sl Nucl.\ Phys.}\ {\bf B475} (1996), pp.\ 579 - 596.

\bibitem[Ac2]{} --------,
 {\it On mirror symmetry for manifolds of exceptional holonomy},
 {\tt hep-th/9707186}.

\bibitem[As1]{} P.S.\ Aspinwall,
 {\it Some relationships between dualities in string theory},
 {\tt hep-th/9508154}.

\bibitem[As2]{} --------,
 {\it K3 surfaces and string duality}, {\tt hep-th/9611137}.

\bibitem[As3]{} --------,
 {\it M-theory versus F-theory pictures of the heterotic string},
 {\sl Adv.\ Theor.\ Math.\ Phys.}\ {\bf 1} (1998), pp.\ 127 - 147.

\bibitem[A-M]{} P.S.\ Aspinwall and D.R.\ Morrison,
 {\it String theory on K3 surfaces},
 in {\sl Mirror symmetry II}, B.R.\ Greene and S.-T.\ Yau eds.,
 pp.\ 703 - 716, Amer.\ Math.\ Soc./International Press, 1997.

\bibitem[Bo]{} C.\ Borcea,
 {\it K3 surfaces with involution and mirror pairs of Calabi-Yau
   manifolds},
 in {\sl Mirror symmetry II}, B.R.\ Greene and S.-T.\ Yau eds.,
 pp.\ 717 - 743, Amer.\ Math.\ Soc./International Press, 1997.

\bibitem[B-B-S]{} K.\ Becker, M.\ Becker, A.\ Strominger,
 {\it Fivebranes, menmbranes, and nonperturbative string theory},
 {\sl Nucl.\ Phys.}\ {\bf B456} (1995), pp.\ 130 - 152.

\bibitem[B-P-VV]{} W.\ Barth, C.\ Peters, and A.\ Van de Ven,
 {\sl Compact complex surfaces},
 Ser.\ Modern Surveys Math.\ 4, Springer-Verlag, 1984.

\bibitem[B-S]{} F.\ Bonahon and L.\ Siebenmann,
 {\it The classification of Seifert fibred $3$-orbifolds},
 in {\sl Low dimensional topology}, R.\ Fenn ed., pp.\ 19 - 85,
 London Math.\ Soc.\ Lect.\ Notes Ser.\ 95,
 Cambridge Univ.\ Press, 1985.

\bibitem[Di]{} A.\ Dimca,
 {\sl Singularities and topology of hypersurfaces},
 Springer-Verlag, 1992.

\bibitem[D-N-P]{} M.J.\ Duff, B.E.W.\ Nilsson, and C.N.\ Pope,
 {\it Compactification of $d=11$ supergravity on
      $\mbox{K3}\times{\footnotesizeBbb T}^3$},
 {\sl Phys.\ Lett.}\ {\bf 129B} (1983), pp.\ 39 - 42.

\bibitem[F-M]{} R.\ Friedman and J.W.\ Morgan,
 {\sl Smooth four-manifolds and complex surfaces},
 Springer-Verlag, 1994.

\bibitem[Gr]{} M.\ Gross,
 {\it Special Lagrangian fibrations I: Topology},
 {\tt alg-geom/9710006}.

\bibitem[G-H]{} P.\ Griffiths and J.\ Harris,
 {\sl Principles of algebraic geometry},
 John Wiley \& Sons, Inc., 1978.

\bibitem[G-W]{} M.\ Gross and P.M.H.\ Wilson,
 {\it Mirror symmetry via $3$-tori for a class of Calabi-Yau
   threefolds},
 {\tt alg-geom/9608004}.

\bibitem[Ha]{} F.R.\ Harvey,
 {\sl Spinors and calibrations}, Academic Press, 1990.

\bibitem[H\"{u}]{} T.\ H\"{u}bsch,
 {\sl Calabi-Yau manifolds -- a bestiary for physicists},
 World Scientific, 1992.

\bibitem[H-L]{} R.\ Harvey and H.B.\ Lawson,
 {\it Calibrated geometries},
 {\sl Acta Math.}\ {\bf 148} (1982), pp.\ 47 -157.

\bibitem[H-W1]{} P.\ Ho\v{r}ava and E.\ Witten,
 {\it Heterotic and type I string dynamics from eleven dimnsions},
 {\sl Nucl.\ Phys.}\ {\bf B460} (1996), pp.\ 506 - 524.

\bibitem[H-W2]{} --------,
 {\it Eleven-dimensional supergravity on a manifold with boundary},
 {\sl Nucl.\ Phys.}\ {\bf B475} (1996), pp.\ 94 - 114.

\bibitem[Jo1]{} D.D.\ Joyce,
 {\it Compact Riemannian $7$-manifolds with holonomy $G_2$. I \& II},
 {\sl J.\ Diff.\ Geom.}\ {\bf 43} (1996),
 pp.\ 291 - 328 and pp.\ 329 - 375.

\bibitem[Jo2]{} --------,
 {\it Compact $8$-manifolds with holonomy $\Spin(7)$},
 {\sl Invent.\ Math.}\ {\bf 123} (1996), pp.\ 507 - 552.

\bibitem[Jo3]{} --------,
 {\it On the topology of desingularizations of Calabi-Yau
      orbifolds},
 {\tt math.AG/9806146}.

\bibitem[Li]{} C.-H.\ Liu,
 {\it On the isolated singularity of a $7$-space obtained by
      rolling Calabi-Yau threefolds through extremal transitions},
 {\tt hep-th/9801175}.

\bibitem[McL]{} R.C.\ McLean,
 {\it Deformations and moduli of calibrated submanifolds},
 Ph.D thesis, Duke University, 1990.

\bibitem[Mo]{} J.M.\ Montesinos,
 {\it Classical tessellations and three-manifolds},
 Springer-Verlag, 1987.

\bibitem[Mor]{} D.R.\ Morrison,
 {\it The geometry underlying mirror symmetry},
 {\tt alg-geom/9608006}.

\bibitem[M-V]{} D.R.\ Morrison and C.\ Vafa,
 {\it Compactifications of F-theory on Calabi-Yau threefolds.\
      I $\&$ II},
 {\sl Nucl.\ Phys.}\ {\bf B473} (1996), pp.\ 74 - 92; and
 {\sl Nucl.\ Phys.}\ {\bf B476} (1996), pp.\ 437 - 469.

\bibitem[Ni]{} V.V.\ Nikulin,
 {\it Discrete reflection groups in Lobachevsky spaces and
   algebraic surfaces},
 {\sl Proc.\ Intern.\ Congress Math.\ Berkeley 1986},
 pp.\ 654 - 671.

\bibitem[Og1]{} K.\ Oguiso,
 {\it On algebraic fiber space structures on a Calabi-Yau
      $3$-fold},
 {\sl Intern.\ J.\ Math.}\ {\bf 4} (1993), pp.\ 439 - 465.

\bibitem[Og2]{} --------,
 {\it On certain rigid fibered Calabi-Yau threefolds},
 {\sl Math.\ Z.}\ {\bf 221} (1996), pp.\ 437 - 448.

\bibitem[Sc]{} P.\ Scott,
 {\it The geometry of $3$-manifolds},
 {\sl Bull.\ London Math.\ Soc.}\ {\bf 15} (1983), pp.\ 401 - 487.

\bibitem[Sh]{} I.R.\ Shafarevich,
 {\sl Basic algebraic geometry},
 {\sl vol I: varieties in projective space},
 {\sl vol II: schemes and complex manifolds},
 Springer-Verlag, 1994.

\bibitem[Sp]{} E.H.\ Spanier,
 {\it Algebraic topology},
 Springer-Verlag, 1966.

\bibitem[S-V]{} M.\ Schlessinger and C.\ Vafa,
 {\it Superstrings and manifolds of exceptional holonomy},
 {\sl Selecta Math.}\ {\bf 1} (1995), pp.\ 347 - 381.

\bibitem[S-Y-Z]{} A.\ Strominger, S.T.\ Yau, and E.\ Zaslow,
 {\it Mirror symmetry is T-duality},
 {\sl Nucl.\ Phys.}\ {\bf B479} (1996), pp.\ 243 - 259.

\bibitem[Th1]{} W.P.\ Thurston,
 {\sl The geometry and topology of three-manifolds},
 Princeton lecture notes, 1979. Revised version, 1991.

\bibitem[Th2]{} --------,
 {\sl Three-dimensional geometry and topology}, vol.\ 1,
 S.\ Levy ed., Princeton Univ.\ Press, 1997. 

\bibitem[To]{} P.K.\ Townsend,
 {\it Four lectures on M-theory},
 {\tt hep-th/9612121}.

\bibitem[Va]{} C.\ Vafa,
 {\it Lectures on strings and dualities},
 {\tt hep-th/9702201}.

\bibitem[Vo]{} C.\ Voisin,
 {\it Miroirs et involutions sur les surfaces K3},
 in {\sl Journe\'{e}s de G\'{e}om\'{e}trie alg\'{e}brique d'Orsay},
 {\sl Ast\'{e}risque} {\bf 218} (1993), pp.\ 273 - 323.

\bibitem[Wi1]{} E.\ Witten
 {\it String theory dynamics in various dimensions},
 {\sl Nucl.\ Phys.}\ {\bf B443} (1995), pp.\ 85 - 126.

\bibitem[Wi2]{} --------,
 {\it Phase transitions in M-theory and F-theory},
 {\sl Nucl.\ Phys.}\ {\bf B471} (1996), pp.\ 195 - 216.


\end{thebibliography}
}
